\documentclass[preprint,pra]{revtex4-2}

\usepackage{mathrsfs}
\usepackage{amsmath}
\usepackage{amssymb}

\begin{document}
%
%
\title{A Flow Equation Approach Striving towards an Energy-Separating Hamiltonian Unitary Equivalent to the Dirac Hamiltonian with Coupling to Electromagnetic Fields}
\author{N. Schopohl}
\email{nils.schopohl@uni-tuebingen.de}
\author{N.S. Cetin}
\date{\today}
\affiliation{Eberhard-Karls-Universit{\"a}t T{\"u}bingen\\ Institute for Theoretical Physics and CQ Center for Collective Quantum
Phenomena\\  Auf der Morgenstelle 14 , D-72076 T{\"u}bingen, Germany}
\begin{abstract}
The Dirac Hamiltonian $\mathsf{H}^{\left(D\right)}$ for relativistic charged fermions minimally coupled to (possibly time-dependent) electromagnetic fields
is transformed with a purpose-built flow equation method,  so that the result of that transformation is unitary equivalent to $\mathsf{H}^{\left(D\right)}$  $\emph{and}$  granted to strive towards a limiting value $\mathsf{H}^{\left(NW\right)}$ commuting with the Dirac $\beta$-matrix. Upon expansion of $\mathsf{H}^{\left(NW\right)}$ to order $\frac{v^2}{c^2}$ the nonrelativistic Hamiltonian $\mathsf{H}^{\left(SP\right)}$ of Schrödinger-Pauli quantum mechanics emerges as the leading order term adding to the rest energy   $mc^2$. All the  relativistic corrections to $\mathsf{H}^{\left(SP\right)}$ are explicitly taken into account in the guise of a Magnus type series expansion, the series coefficients generated to order $\left(\frac{v^{2}}{c^{2}}\right)^{n}$ for $n\geq2$ comprising partial sums of iterated commutators only.
In the special case of \emph{static} fields the equivalence of the flow equation method with the well known energy-separating unitary transformation of Eriksen  is established on the basis of an \emph{exact} solution of a reverse flow equation transforming the $\beta$-matrix into the energy-sign operator associated with $\mathsf{H}^{\left(D\right)}$. That way the identity $\mathsf{H}^{\left(NW\right)}=\beta\sqrt{\mathsf{H}^{\left(NW\right)}\mathsf{H}^{\left(NW\right)}}$ is established implying $\mathsf{H}^{\left(NW\right)}$ being determined \emph{unambiguously}.
In contrast to $\mathsf{H}^{\left(D\right)}$  it's unitary equivalent   $\mathsf{H}^{\left(NW\right)}$  generates the motion of electrons and positrons in the presence of weak external fields now as entirely separated wave packets carrying mass $m$, charge ${\pm{|e|}}$ and spin $\pm \frac{\hbar}{2}$ respectively, yet those wave packets being by construction bare of any "Zitterbewegung", akin to classical particles moving along individual trajectories under the influence of the Lorentz force.
 Upon expansion of $\mathsf{H}^{\left(NW\right)}$  to order $\left(\frac{v^{2}}{c^{2}}\right)^{n}$  for $n=1,2,3,...$  our results agree with results obtained recently by Silenko with a correction scheme developed for the original step-by-step FW-transformation method, the latter long-since known for not generating \emph{unambiguously} a unitary equivalent Hamiltonian being energy-separating.
\end{abstract}
\maketitle
\section*{Introduction}
Based on insight into the fundamental meaning of locality obtained by Newton and Wigner (NW) in 1949 for (freely moving) relativistic  fermions \cite{NewtonWigner}, Foldy and Wouthuysen (FW) established in 1950 with their unitary transformation of the free Dirac Hamiltonian an interpretation of concepts like position or velocity or spin or orbital angular momentum \cite{FoldyW50}, void of the well known paradoxical properties of operators in the Dirac representation (for instance the velocity operator having eigenvalues $\pm{c}$) and thus agreeable to the physical intuition ascribed to a moving single particle \cite{CostellaMcKellar95}.

For this reason we suggest to refer in the ensuing to the result $\mathsf{H}^{\left(NW\right)}$  of such a unitary transformation of the Dirac Hamiltonian $\mathsf{H}^{\left(D\right)}$  as the NW-Hamiltonian even if in $\mathsf{H}^{\left(D\right)}$ additional couplings to weak and slowly varying,  possibly time-dependent electromagnetic fields are taken into account.

For a relativistic charged fermion with minimal coupling to a magnetostatic field a generalization of the unitary FW-transformation in closed form has been discovered by Case \cite{Case54} and also, along a different line of reasoning,  by Eriksen \cite{Eriksen}, whose unitary transformation indeed applies as well to electrostatic and/or  magnetostatic fields superposed.

In the ensuing a purpose-built flow equation approach is introduced resulting in a limiting value $\mathsf{H}^{\left(NW\right)}$ equivalent to a unitary transformation of the Dirac-Hamiltonian for a charged fermion with coupling to \emph{time-dependent} electromagnetic fields, that in the special case of static fields turns out to be equivalent to the \emph{unambiguous} energy-separating result obtained with the Eriksen transformation.
The advantage of the new flow equation approach being that it provides in a lucid manner the expansion of $\mathsf{H}^{\left(NW\right)}$ in powers of $\frac{v}{c}$ to arbitrary order, and that in the special case of static fields that expansion is in full agreement with corresponding results obtained by the (arduous) expansion of Eriksen's transformation.

\section{The Flow Equation Method}
The so-called Hamiltonian flow equation method, originally developed
by Wegner  for problems in nonrelativistic many body physics \cite{Wegner94,Wegner2001},
pertains to a continuous unitary transformation of a given Hamiltonian
$\mathsf{H}$ via
\begin{eqnarray}
\mathsf{H}\left(s\right) & = & \mathsf{U}\left(s\right)\mathsf{H}\mathsf{U}^{\dagger}\left(s\right)\label{eq:continuous unitary transformation}\\
\frac{d}{ds}\mathsf{U}\left(s\right) & = & \eta\left(s\right)\mathsf{U}\left(s\right)\nonumber \\
\mathsf{U}\left(0\right) & = & \mathsf{1}_{4\times4}\nonumber \\
\eta^{\dagger}\left(s\right) & = & -\eta\left(s\right)\nonumber
\end{eqnarray}
Given an antisymmetric generator $\eta\left(s\right)$ for the flow,
the unitary transformation of an hermitean operator $\mathsf{H}$
is determined solving an initial value problem
\begin{eqnarray}
\frac{d}{ds}\mathsf{H}\left(s\right) & = & \left[\eta\left(s\right),\mathsf{H}\left(s\right)\right]\label{eq:flow equation}\\
\mathsf{H}\left(0\right) & = & \mathsf{H}\nonumber
\end{eqnarray}
Of course, the crux of the method is the choice of the generator $\eta\left(s\right)$,
with the essential point being, that $\eta\left(s\right)$ controls
the properties of the limiting value $\mathsf{H}\left(\infty\right)$.
A comprehensive review discussing various generators with a multitude
of applications of the Wegner flow equation method in nonrelativistic
many body physics has been given by Kehrein \cite{Kehrein}.
Let us choose a most simple generator that depends on a \emph{constant}
hermitean operator $\Gamma$ and consider a positive semi-definite
functional as introduced by Brockett \cite{Brockett91},\cite{DoubleBracketFlow}
\begin{equation}
\Phi(s)\equiv\frac{1}{2}\textrm{tr}\left(\left(\mathsf{H}\left(s\right)-\Gamma\right)^{2}\right)\geq0\label{eq:functional Phi(s)}
\end{equation}
Here $\mathsf{H}\left(s\right)$ is the solution of a flow with generator
$\eta\left(s\right)$, specified in terms of $\Gamma$ as
\begin{eqnarray}
\eta\left(s\right) & = & \left[\Gamma,\mathsf{H}\left(s\right)\right]\nonumber \\
\frac{d}{ds}\mathsf{H}\left(s\right) & = & \left[\eta\left(s\right),\mathsf{H}\left(s\right)\right]\label{eq:double bracket flow}\\
\mathsf{H}\left(0\right) & = & \mathsf{H}\nonumber
\end{eqnarray}
Exploiting the cyclic invariance of the trace one readily finds after
a sequence of elementary adjustments
\begin{eqnarray}
\frac{d}{ds}\Phi\left(s\right) & = & \textrm{tr}\left[\left(\mathsf{H}\left(s\right)-\Gamma\right)\frac{d}{ds}\mathsf{H}\left(s\right)\right]\label{eq:functional Phi(s) derivative}\\
 & = & \textrm{tr}\left[\left(\mathsf{H}\left(s\right)-\Gamma\right)\left[\eta\left(s\right),\mathsf{H}\left(s\right)\right]\right]\nonumber \\
 & = & \textrm{tr}\left[\left(\mathsf{H}\left(s\right)-\Gamma\right)\left[\left[\Gamma,\mathsf{H}\left(s\right)\right],\mathsf{H}\left(s\right)\right]\right]\nonumber \\
 & = & -\textrm{tr}\left[\left[\mathsf{H}\left(s\right),\Gamma\right]\left[\Gamma,\mathsf{H}\left(s\right)\right]\right]\nonumber \\
 & = & -\textrm{tr}\left[\eta^{\dagger}\left(s\right)\eta\left(s\right)\right]\nonumber \\
 & \leq & 0\nonumber
\end{eqnarray}
Because $\Phi(s)\geq0$ for all $s\geq0$ the only possible conclusion
being that
\begin{equation}
\lim_{s\rightarrow\infty}\frac{d}{ds}\Phi\left(s\right)=0\label{eq:d/ds Phi(s)->0}
\end{equation}
, i.e.
\begin{equation}
\lim_{s\rightarrow\infty}\eta\left(s\right)=\left[\Gamma,\mathsf{H}\left(\infty\right)\right]=\mathsf{0}_{4\times4}\label{eq:Commutator(Bgamma,H(infty))=0}
\end{equation}
So, a double bracket flow of the type (\ref{eq:double bracket flow}) indeed strives to a limiting value $\mathsf{H}\left(\infty\right)$
that commutes with the given hermitean operator $\Gamma$.

A while ago Bylev and Pirner (BP) \cite{Bylev1998} suggested
a flow equation approach to obtain a unitary transformation of the
Hamiltonian $\mathsf{H}^{\left(D\right)}$ for a Dirac particle moving in
an external \emph{static} electromagnetic potential, choosing as a generator
\begin{equation}
\eta\left(s\right)=\left[\beta,\mathsf{H}\left(s\right)\right]\label{eq:Bylev-Pirner Generator}
\end{equation}
and choosing as initial data at s=0
\begin{equation}
\mathsf{H}\left(0\right)=\mathsf{H}^{\left(D\right)}\label{eq:initial data Bylev-Pirner}
\end{equation}
The afore propounded argument in the special case $\Gamma=\beta$
at once reveals the limiting value $\mathsf{H}\left(\infty\right)$
of that flow being \emph{even}, i.e.
\begin{eqnarray}
\label{eq:Commutator(beta, H(infty))=0}
\mathsf{0}_{4\times4} & = & \eta\left(\infty\right)\,=\,\beta\mathsf{H}\left(\infty\right)-\mathsf{H}\left(\infty\right)\beta
\end{eqnarray}
Actually, the result $\mathsf{H}\left(\infty\right)$ being an even
operator was obtained in \cite{Bylev1998} merely for a special
case of \emph{perturbation theory}, akin (but not identical) with
the outcome of the \emph{perturbative} procedure of consecutive step
by step canonical transformations aiming at eliminating the odd terms
in the Dirac Hamiltonian due to Foldy and Wouthuysen (FW) \cite{FoldyW50}.
Comparing the results obtained with the BP-method with the ones obtained
by the FW-method a discrepancy arises in the $6^{th}$-order of perturbation
theory expanding in the small parameter $\kappa=\frac{v}{c}$  \cite{Schopohl2022}.
Mind because of the ambiguity in the definition of the sequence of operators
with such a step-by-step method, a blockdiagonal operator resulting from  \emph{several} successive unitary $4\times4$ -transformations is only guaranteed
to give results being equivalent up to a $2\times2$-blockdiagonal unitary transformation,
as earlier on was already emphasized by Pursey \cite{Pursey} and
by Eriksen and Kolsrud \cite{EriksenKolsrud}.
However, as has been concisely discussed in an elucidating article
by Costella and McKellar \cite{CostellaMcKellar95}, the crucial finding
of Foldy and Wouthuysen is not so much concerned with the (certainly
useful) perturbative step by step elimination of the odd terms in
a (possibly time-dependent) Dirac Hamiltonian, but is in the main
concerned with the unravelling of a problem of interpretation with
the four-component amplitude $\Psi_{\mu}^{\left(D\right)}\left(\mathbf{r},t\right)$
that solves the Dirac equation.
\section{A Problem of Interpretation with the Four-Component Dirac Amplitude}
In relativistic quantum mechanics a fermion with attributes mass
$m$, spin $\pm\frac{\hbar}{2}$ and charge $q_{e}=-|e|$, say moving in the presence of electrostatic and magnetostatic fields,
is described in terms of a four-component amplitude $\Psi_{\mu}^{\left(D\right)}\left(\mathbf{r},t\right)$
solving the Dirac equation
\footnote{As of now the summation convention always applies, if not suspended explicitely.}
:
\begin{eqnarray}
i\hbar\partial_{t}\Psi_{\mu}^{\left(D\right)}\left(\mathbf{r},t\right) & = & \mathsf{H}_{\mu,\mu'}^{\left(D\right)}\Psi_{\mu'}^{\left(D\right)}\left(\mathbf{r},t\right)\label{eq:Dirac equation}\\
\mu,\mu' & \epsilon & \left\{ 1,2,3,4\right\} \nonumber
\end{eqnarray}
Here
\begin{equation}
\mathsf{H}^{\left(D\right)}=mc^{2}\beta+c\alpha_{b}\varPi_{b}+q_{e}\varPhi\left(\mathsf{x}\right)\mathsf{1}_{4\times4}\label{eq:Dirac-Hamiltonian}
\end{equation}
denotes the relativistic Hamiltonian of the fermion with minimal coupling to electromagnetic fields in Dirac-Pauli
representation, with the well known $4\times4$ matrices $\beta$
and $\alpha_{b}$ anticommuting and being of square equal to unity,
see for instance \cite{Baym},\cite{Messiah},\cite{BjorkenDrell},\cite{IzyksonZuber}.
The magnetic induction field $\mathscr{\boldsymbol{B}}=\mathbf{rot}\mathcal{\boldsymbol{A}}$
is encoded in the gauge invariant derivative operator composed of
the Cartesian components of the conjugate momentum and position \emph{operators},
$\mathsf{p}_{b}$ and $\mathsf{x}_{a}$ , of fundamental quantum mechanics
\begin{eqnarray}
\varPi_{b}\left(\mathsf{p},\mathsf{x}\right)\equiv \varPi_b  & = & \mathsf{p}_{b}-q_{e}\mathcal{A}_{b}\left(\mathsf{x}\right)\label{eq:gauge invariant derivative}\\
\left[\mathsf{p}_{b},\mathsf{x}_{a}\right] & = & \frac{\hbar}{i}\delta_{a,b}\hat{1}\nonumber
\end{eqnarray}
According to the superposition principle the Dirac amplitude $\Psi_{\mu}^{\left(D\right)}\left(\mathbf{r},t\right)$
may be represented as a linear combination of a \emph{complete} system
of orthonormal \emph{four}-component eigenfunctions $U_{\mu}^{\left(D\right)}\left(\mathbf{r},k\right)$
and $V_{\mu}^{\left(D\right)}\left(\mathbf{r},\tilde{k}\right)$ of
$\mathsf{H}^{\left(D\right)}$ , entailing both, the positive and
the negative energy eigenvalues $E_{k}>0$ and $-E_{\tilde{k}}<0$ :
\begin{eqnarray}
\Psi_{\mu}^{\left(D\right)}\left(\mathbf{r},t\right) & = & \sum_{k}U_{\mu}^{\left(D\right)}\left(\mathbf{r},k\right)c_{k}e^{iE_{k}t}+\sum_{\tilde{k}}V_{\mu}^{\left(D\right)}\left(\mathbf{r},\tilde{k}\right)b_{\tilde{k}}e^{-iE_{\tilde{k}}t}\label{eq:expansion of Dirac amplitude in terms of eigenfunctions}
\end{eqnarray}
, whereby
\begin{eqnarray}
\mathsf{H}_{\mu,\mu'}^{\left(D\right)}U_{\mu'}^{\left(D\right)}\left(\mathbf{r},k\right) & = & E_{k}U_{\mu}^{\left(D\right)}\left(\mathbf{r},k\right)\label{eq:Dirac eigenvalue problem}\\
\mathsf{H}_{\mu,\mu'}^{\left(D\right)}V_{\mu'}^{\left(D\right)}\left(\mathbf{r},\tilde{k}\right) & = & -E_{\tilde{k}}V_{\mu}^{\left(D\right)}\left(\mathbf{r},\tilde{k}\right)\nonumber
\end{eqnarray}
In (\ref{eq:expansion of Dirac amplitude in terms of eigenfunctions})
the expansion coefficients, referred to as $c_{k}$ and $b_{\tilde{k}}$
, are c-numbers, obtainable from a prescribed amplitude $\Psi_{\mu}^{\left(D\right)}\left(\mathbf{r}\right)$
at an initial time $t=0$ making use of the orthonormality of those
eigenfunctions. Mind the labels $k$ and $\tilde{k}$ , counting the
positive-energy eigenmodes $U_{\mu}^{\left(D\right)}\left(\mathbf{r},k\right)$
or rather the negative-energy eigenmodes $V_{\mu}^{\left(D\right)}\left(\mathbf{r},\tilde{k}\right)$
of the Dirac Hamiltonian, have in the presence of an electrostatic
field $\mathscr{\boldsymbol{E}}\left(\mathbf{r}\right)=-\nabla\varPhi\left(\mathbf{r}\right)$
possibly different codomains (for instance if there exist bound states).
Because a complete set of orthonormal eigenfunctions of the Dirac
Hamiltonian comprises naturally the positive- \emph{and} the negative-energy
eigenstates \emph{jointly}, the position operator in the Dirac theory,
if defined by the operation of multiplication
\begin{equation}
\mathsf{x}_{a}\Psi_{\mu}^{\left(D\right)}\left(\mathbf{r},t\right)=r_{a}\Psi_{\mu}^{\left(D\right)}\left(\mathbf{r},t\right)\label{eq:position operator}
\end{equation}
, is \emph{not} an operator defined over well defined states of a
\emph{particle}, since in the evaluation of the expectation value
$\left\langle \Psi^{\left(D\right)}\right|$ $\mathsf{x}_{a}\left|\Psi^{\left(D\right)}\right\rangle $
of the position operator in a given Dirac state $\left|\Psi^{\left(D\right)}\right\rangle $
the positive-energy solutions \emph{interfere} with the (seemingly
unphysical) negative-energy solutions. The origin of this difficulty
is the assumed consent of the Dirac amplitude $\Psi_{\mu}^{\left(D\right)}\left(\mathbf{r},t\right)$
being a probability amplitude for \emph{particles} just like in nonrelativistic
Schr{\"o}dinger quantum mechanics, which is plainly wrong, as has been
first revealed by the analysis of the meaning of locality in quantum
mechanics by Newton and Wigner \cite{NewtonWigner}. For a brilliantly
witty discussion of this point, already elucidated in pioneering work
by Foldy and Wouthuysen \cite{FoldyW50}, we refer to Costella and
McKellar \cite{CostellaMcKellar95}.
Indeed interpreting $\Psi_{\mu}^{\left(D\right)}\left(\mathbf{r},t\right)$
as a probability amplitude gives cause to several well known absurdities,
for instance the components of the ``velocity'' operator in the
Heisenberg picture do not commute and have eigenvalues equal to $\pm c$
, see \cite{Baym},\cite{Messiah},\cite{Grainer}. Related to this
is the concession that the phenomenon of the highly oscillatory ``Zitterbewegung'',
sometimes discussed as an inevitable property of the relativistic electron,
is actually not a physical property of a moving \emph{particle} with attributes mass, charge and spin \cite{CostellaMcKellar95}.
\section{The Newton-Wigner Amplitude and the Notion of Energy Separation}
A physically correct probabilty amplitude, that in the style of the
discussion given by Costella and McKellar \cite{CostellaMcKellar95}
we refer to in what follows as the \emph{Newton-Wigner} amplitude
$\Psi_{\mu}^{\left(NW\right)}\left(\mathbf{r},t\right)$, can be constructed
from the four-component Dirac amplitude $\Psi_{\mu}^{\left(D\right)}\left(\mathbf{r},t\right)$
by a suitable \emph{unitary} transformation $\mathsf{T}$,
such that
\begin{equation}
\Psi_{\mu}^{\left(NW\right)}\left(\mathbf{r},t\right)=\mathsf{\mathsf{T}}_{\mu,\mu'}\Psi_{\mu'}^{\left(D\right)}\left(\mathbf{r},t\right)\label{eq:NW-amplitude}
\end{equation}
permits a meaningful expectation value in particular for the operators
of position, velocity, spin and also orbital angular momentum \cite{FoldyW50},\cite{CostellaMcKellar95}.

To this end one looks first for a unitary transformation $\mathsf{\mathsf{T}}$
requiring the transformed Dirac Hamiltonian
\begin{equation}
\mathsf{H}^{\left(NW\right)}\equiv\mathsf{\mathsf{T}}\mathsf{H}^{\left(D\right)}\mathsf{T}^{\dagger}\label{eq:H_(NW) I}
\end{equation}
be an \emph{even} operator, so that
\begin{equation}
\mathsf{\tilde{H}}^{\left(NW\right)}\beta\,=\,\beta\mathsf{\tilde{H}}^{\left(NW\right)}\label{eq:[beta, H_(NW)]=0}
\end{equation}
Taken by itself the criterion (\ref{eq:[beta, H_(NW)]=0})
only ensures, minding our choice $\beta=\textit{diag}\left\{1,1,-1,-1\right\}$  in Dirac-Pauli representation, that  $\mathsf{H}^{\left(NW\right)}$ assumes a block-diagonal guise,
but that doesn't warrant $\mathsf{H}^{\left(NW\right)}$ being \emph{energy-separating},
the  latter notion first introduced by Eriksen and Kolsrud \cite{EriksenKolsrud}.
Possibly the criterion is better understood introducing abstract bra-
and ket-notation, so that $U_{\mu}^{\left(D\right)}\left(\mathbf{r},k\right)=\left\langle \mathbf{r},\mu|U_{k}^{\left(D\right)}\right\rangle $
and $V_{\mu}^{\left(D\right)}\left(\mathbf{r},\tilde{k}\right)=\left\langle \mathbf{r},\mu|V_{\tilde{k}}^{\left(D\right)}\right\rangle $.
Introducing the spectral representations of the Dirac Hamiltonian
and it's associated energy-sign operator $\Lambda^{\left(D\right)}$
in the basis of Dirac eigenstates,
\begin{eqnarray}
\mathsf{H}^{\left(D\right)} & = & \sum_{k}E_{k}\left|U_{k}^{\left(D\right)}\right\rangle \left\langle U_{k}^{\left(D\right)}\right|+\sum_{\tilde{k}}\left(-E_{\tilde{k}}\right)\left|V_{\tilde{k}}^{\left(D\right)}\right\rangle \left\langle V_{\tilde{k}}^{\left(D\right)}\right|\label{eq:spectral representations}\\
\Lambda^{\left(D\right)} & = & \sum_{k}\left|U_{k}^{\left(D\right)}\right\rangle \left\langle U_{k}^{\left(D\right)}\right|-\sum_{\tilde{k}}\left|V_{\tilde{k}}^{\left(D\right)}\right\rangle \left\langle V_{\tilde{k}}^{\left(D\right)}\right|=\frac{\mathsf{H}^{\left(D\right)}}{\sqrt{\mathsf{H}^{\left(D\right)}\mathsf{H}^{\left(D\right)}}}\label{eq:energy sign operator}
\end{eqnarray}
, it follows at once in terms of the unitary transformed Dirac eigenstates
\begin{eqnarray}
\left|U_{k}^{\left(NW\right)}\right\rangle  & = & \mathsf{\mathsf{T}}\left|U_{k}^{\left(D\right)}\right\rangle \label{eq:NW-eigenstates}\\
\left|V_{\tilde{k}}^{\left(NW\right)}\right\rangle  & = & \mathsf{\mathsf{T}}\left|V_{\tilde{k}}^{\left(D\right)}\right\rangle \nonumber
\end{eqnarray}
 for the Dirac $\beta$-operator the representation
\begin{equation}
\beta=\mathsf{\mathsf{T}}\Lambda^{\left(D\right)}\mathsf{T}^{\dagger}=\frac{\mathsf{H}^{\left(NW\right)}}{\sqrt{\mathsf{H}^{\left(NW\right)}\mathsf{H}^{\left(NW\right)}}}\label{eq:representation Dirac-Beta}
\end{equation}
In reverse order, the\emph{ Newton-Wigner} Hamiltonian $\mathsf{H}^{\left(NW\right)}$
is defined as being one of a kind among all unitary transformed Hamiltonians
with block-diagonal guise, satisfying additionally the identity \cite{EriksenKolsrud}
\begin{eqnarray}
\mathsf{H}^{\left(NW\right)} & = & \beta\,\sqrt{\mathsf{H}^{\left(NW\right)}\mathsf{H}^{\left(NW\right)}}\label{eq:energy separation  identity H_(NW)}
\end{eqnarray}
As a direct consequence of (\ref{eq:energy separation  identity H_(NW)})
being true then
\begin{eqnarray}
\left(\mathsf{0}_{4\times1}\right)_{\mu} & = & \left(\mathsf{H}^{\left(NW\right)}-\beta\sqrt{\mathsf{H}^{\left(NW\right)}\mathsf{H}^{\left(NW\right)}}\right)_{\mu,\mu'}U_{\mu'}^{\left(NW\right)}\left(\mathbf{r},k\right)
\label{eq:energy separation identity H_(NW)  IIa}\\
 & = & E_{k}\left(\mathsf{\hat{1}}_{4\times4}-\beta\right)_{\mu,\mu'}U_{\mu'}^{\left(NW\right)}\left(\mathbf{r},k\right)\nonumber
\end{eqnarray}
\begin{eqnarray}
\left(\mathsf{0}_{4\times1}\right)_{\mu} & = & \left(\mathsf{H}^{\left(NW\right)}-\beta\sqrt{\mathsf{H}^{\left(NW\right)}\mathsf{H}^{\left(NW\right)}}\right)_{\mu,\mu'}V_{\mu'}^{\left(NW\right)}\left(\mathbf{r},\tilde{k}\right)\label{eq:energy separation identity H_(NW)  IIb}\\
 & = & \left(-E_{\tilde{k}}\right)\left(\mathsf{\hat{1}}_{4\times4}+\beta\right)_{\mu,\mu'}V_{\mu'}^{\left(NW\right)}\left(\mathbf{r},\tilde{k}\right)\nonumber
\end{eqnarray}
Assuming nonvanishing eigenvalues, $E_{k}\neq0$ and $E_{\tilde{k}}\neq0$,
this entails at once, minding $\beta$ being diagonal in Dirac-Pauli representation, that $U_{\mu}^{\left(NW\right)}\left(\mathbf{r},k\right)\equiv0$
regarding the lower components $\mu=3,4$ and $V_{\mu}^{\left(NW\right)}\left(\mathbf{r},\tilde{k}\right)\equiv0$
regarding the upper components $\mu=1,2$ . This distinguishing feature
indeed is the essence of the said energy-separating property of the
Newton-Wigner representation, it being the representation in which the operators position, velocity, orbital angular momentum and spin
 of the free theory are agreeable to physical intuition just like in classical physics \cite{FoldyW50,CostellaMcKellar95,LipingZou2020}.

Perhaps, the physical meaning of the concept of energy-separation
becomes more comprehensible if one applies the unitary transformation
$\mathsf{\mathsf{T}}$ to the Dirac-amplitude (\ref{eq:expansion of Dirac amplitude in terms of eigenfunctions})
and rewrites the Newton-Wigner amplitude (\ref{eq:NW-amplitude})
now with the transformed Dirac-eigenstates (\ref{eq:NW-eigenstates}) as
\begin{equation}
\Psi_{\mu}^{\left(NW\right)}\left(\mathbf{r},t\right)=\sum_{k}U_{\mu}^{\left(NW\right)}\left(\mathbf{r},k\right)c_{k}e^{iE_{k}t}+\sum_{\tilde{k}}V_{\mu}^{\left(NW\right)}\left(\mathbf{r},\tilde{k}\right)b_{\tilde{k}}e^{-iE_{\tilde{k}}t}
\label{eq:Newton-Wigner amplitude II}
\end{equation}
Expressely stated, $\Psi_{\mu}^{\left(NW\right)}\left(\mathbf{r},t\right)$
assumes in consequence of $U_{\mu}^{\left(NW\right)}\left(\mathbf{r},k\right)\equiv0$
for $\mu=3,4$ and $V_{\mu}^{\left(NW\right)}\left(\mathbf{r},k\right)\equiv0$
for $\mu=1,2$ the guise
\begin{equation}
\Psi_{\mu}^{\left(NW\right)}\left(\mathbf{r},t\right)=\left(\begin{array}{c}
\psi_{+}\left(\mathbf{r},t\right)\\
\psi_{-}\left(\mathbf{r},t\right)\\
0\\
0
\end{array}\right)_{\mu}+\left(\begin{array}{c}
0\\
0\\
\chi_{+}\left(\mathbf{r},t\right)\\
\chi_{-}\left(\mathbf{r},t\right)
\end{array}\right)_{\mu}\label{eq:Newton-Wigner amplitude III}
\end{equation}
Writing $\mathsf{H}^{\left(NW\right)}$ as well in explicit $2\times2$ block notation,
\begin{equation}
\mathsf{H}^{\left(NW\right)}\equiv\left(\begin{array}{cc}
\mathsf{H}_{2\times2}^{\left(e\right)} & ,\mathsf{0}_{2\times2}\\
\mathsf{0}_{2\times2} & ,-\mathsf{H}_{2\times2}^{\left(p\right)}
\end{array}\right)\label{eq:block decomposition H_NW}
\end{equation}
, the resulting equations of motion governing the time evolution of
the \emph{two}-component amplitudes $\psi_{\sigma}\left(\mathbf{r},t\right)$
and $\chi_{\sigma}\left(\mathbf{r},t\right)$ are now by construction
forever propagating\emph{ without }any interference of positive-  and
negative-energy states, i.e.
\begin{eqnarray}
i\hbar\partial_{t}\psi_{\sigma}\left(\mathbf{r},t\right) & = & \left(\mathsf{H}_{2\times2}^{\left(e\right)}\right)_{\sigma,\sigma'}\psi_{\sigma'}\left(\mathbf{r},t\right)\label{eq:equation of motion NW amplitude}\\
i\hbar\partial_{t}\chi_{\sigma}\left(\mathbf{r},t\right) & = & -\left(\mathsf{H}_{2\times2}^{\left(p\right)}\right)_{\sigma,\sigma'}\chi_{\sigma'}\left(\mathbf{r},t\right)\nonumber \\
\sigma,\sigma' & \epsilon & \left\{ +,-\right\} \nonumber
\end{eqnarray}
 These are Schr{\"o}dinger-Pauli type equations, whereas the amplitude
$\psi_{\sigma}\left(\mathbf{r},t\right)$ describing the electron
as a particle (wave packet) with attributes mass $m$, spin $\pm\frac{\hbar}{2}$
and charge $q_{e}=-|e|$ is composed \emph{exclusively}
of positive-energy eigenfunctions $U_{\mu}^{\left(NW\right)}\left(\mathbf{r},k\right)$.
Correlating with this, as the amplitude $\chi_{\sigma}\left(\mathbf{r},t\right)$
is composed solely of negative-energy eigenfunctions $V_{\mu}^{\left(NW\right)}\left(\mathbf{r},\tilde{k}\right)$,
it has been given a physical interpretation already by Dirac himself
via charge conjugation, so that $\chi_{-\sigma}^{\star}\left(\mathbf{r},t\right)$
is describing the positron as an ``anti- particle'' (wave packet)
with attributes mass $m$ , spin $\mp\frac{\hbar}{2}$ and opposite
charge $q_{p}=\left|e\right|$, albeit Dirac's hole picture implicitely
already involved a quantum field theory context. By this means the
specific unitary transformation $\mathsf{T}$ , that leads
from the Dirac Hamiltonian $\mathsf{H}^{\left(D\right)}$ to the energy-separating
Hamiltonian $\mathsf{H}^{\left(NW\right)}$, enables to bare the roots
of Dirac's discovery of antimatter.

To illustrate the concept of energy separation from a different point
of view, assume a certain unitary transformation $\mathsf{U}$ brought
the Dirac Hamiltonian $\mathsf{H}^{\left(D\right)}$ to a block-diagonal guise
\begin{equation}
\mathsf{H}^{\left(U\right)}=\mathsf{U}\mathsf{H}^{\left(D\right)}\mathsf{U}^{\dagger}=\left(\begin{array}{cc}
\mathsf{H}_{2\times2}^{\left(I\right)} & ,\mathsf{0}_{2\times2}\\
\mathsf{0}_{2\times2} & ,\mathsf{H}_{2\times2}^{\left(II\right)}
\end{array}\right)\label{eq:unitary transformation U applied to H_D}
\end{equation}
Now by construction $\mathsf{H}^{\left(U\right)}\beta=\beta\mathsf{H}^{\left(U\right)}$,
yet it is not ensured the upper block operator $\mathsf{H}_{2\times2}^{\left(I\right)}$
being positive definite and concurrently the lower block operator
$\mathsf{H}_{2\times2}^{\left(II\right)}$ being negative definite.
This entails that a \emph{four}-component amplitude $\Psi_{\mu}^{\left(D,I\right)}\left(\mathbf{r},t\right)$,
say generated by applying the inverse unitary transformation $\mathsf{U}^{\dagger}$
to a two-component amplitude $\psi_{\sigma}^{\left(I\right)}\left(\mathbf{r},t\right)$
being solely composed of the eigenfunctions of $\mathsf{H}_{2\times2}^{\left(I\right)}$
, i.e.
\begin{equation}
\Psi_{\mu}^{\left(D,I\right)}\left(\mathbf{r},t\right)=\left(\mathsf{U}^{\dagger}\right)_{\mu,\sigma}\psi_{\sigma}^{\left(I\right)}\left(\mathbf{r},t\right)\label{eq:back-transformed two component solution}
\end{equation}
, certainly represents a solution of the Dirac equation (\ref{eq:Dirac equation}).
Yet it (conceivably) comprises a linear combination\emph{ }of
positive energy \emph{and }negative energy eigenfunctions of $\mathsf{H}^{\left(D\right)}$
just like in (\ref{eq:expansion of Dirac amplitude in terms of eigenfunctions}).
Such a unitary transformation $\mathsf{U}$ is, of course, not energy-separating.

But even if  $\mathsf{H}_{2\times2}^{\left(I\right)}$ in (\ref{eq:unitary transformation U applied to H_D})
was positive definite and $\mathsf{H}_{2\times2}^{\left(II\right)}$
in (\ref{eq:unitary transformation U applied to H_D}) was negative
definite, there exists an ambiguity, as any \emph{additional} unitary
tranformation $\mathsf{N}$, taking a shape
\begin{equation}
\mathsf{N}=\left(\begin{array}{cc}
\mathsf{N}_{2\times2}^{\left(I\right)} & ,\mathsf{0}_{2\times2}\\
\mathsf{0}_{2\times2} & ,\mathsf{N}_{2\times2}^{\left(II\right)}
\end{array}\right)\label{eq:unitary transformation N}
\end{equation}
, now changes the $2\times2$-subblocks $\mathsf{H}_{2\times2}^{\left(I\right)}$
and $\mathsf{H}_{2\times2}^{\left(II\right)}$ in (\ref{eq:unitary transformation U applied to H_D})
into equivalent unitary transformed operators $\mathsf{N}_{2\times2}^{\left(I\right)}\mathsf{H}_{2\times2}^{\left(I\right)}\left(\mathsf{N}_{2\times2}^{\left(I\right)}\right)^{\dagger}$
and $\mathsf{N}_{2\times2}^{\left(II\right)}\mathsf{H}_{2\times2}^{\left(II\right)}\left(\mathsf{N}_{2\times2}^{\left(II\right)}\right)^{\dagger}$.
So then the question arises, how to remove
that ambiguity inherent to any such unitary transformation $\mathsf{V}=\mathsf{N}\mathsf{U}$ ?
In what follows we present a flow equation approach enabling to construct for the Dirac Hamiltonian it's unambiguous energy-separating unitary equivalent, the  Wigner-Newton Hamiltonian.

Last not least, while the Dirac Hamiltonian $\mathsf{H}^{\left(D\right)}$
is unique due to its linearity and minimal coupling to external fields,
the Newton-Wigner Hamiltonian $\mathsf{H}^{\left(NW\right)}$ is the
only one enabling a meaningful nonrelativistic limit in terms of particles
and anti-particles moving as completely separated entities along their
individual trajectories, as any bubble chamber track reveals. Unfortunately,
a disadvantage of this very useful property of $\mathsf{H}^{\left(NW\right)}$
is that locality of the Dirac-Hamiltonian $\mathsf{H}^{\left(D\right)}$,
as it comprises only first order differential operators, has been
traded off for a \emph{nonlocal} operator, with $\sqrt{\mathsf{H}^{\left(NW\right)}\mathsf{H}^{\left(NW\right)}}$
essentially being an integral kernel. Further, in marked contrast
to the Dirac equation, the unitary equivalent equations of motion
(\ref{eq:equation of motion NW amplitude}) based on the square root
operator (\ref{eq:block decomposition H_NW}) are obviously not covariant.
Damage of covariance regarding conformable square root operators has
been discussed before by several authors dealing, for instance, with
the easier problem of relativistic spin\emph{-$0$} particles \cite{Sucher_on_KG}\cite{Briegel_on_KG}\cite{Laemmerzahl1993}.
But in view of obtaining a suitable starting point for approximations
endeavouring to the quantum mechanics of nonrelativistic particles
that flaw has little concernment.
\section{External magnetostatic field}\label{sec:magnetostatic}
In the special case the Dirac particle is moving solely in the presence
of a magnetostatic field,
\begin{equation}
\mathsf{H}_{0}^{\left(D\right)}=mc^{2}\beta+c\alpha_{b}\varPi_{b}\label{eq:H_0_(D)}
\end{equation}
the unitary transformation $\mathsf{\mathsf{T}}_{0}$ transforming
$\mathsf{H}_{0}^{\left(D\right)}$ to $\mathsf{H}_{0}^{\left(NW\right)}$
is known exactly in terms of a straightforward generalization \cite{Case54,Eriksen}
of the result obtained afore by Foldy and Wouthuysen for the case
of a free (translational invariant) Dirac Hamiltonian \cite{FoldyW50}:
\begin{eqnarray}
\label{eq:unitary transformation T_0_(NW)}\\
\mathsf{\mathsf{T}}_{0} & = & \sqrt{\begin{array}{c}
\frac{1}{2}\left(\mathsf{1}_{4\times4}+\frac{mc^{2}}{\sqrt{\mathsf{H}_{0}^{\left(D\right)}\circ\mathsf{H}_{0}^{\left(D\right)}}}\right)\end{array}}+\beta\begin{array}{c}
\frac{\alpha_{b}\varPi_{b}}{\sqrt{\left(\alpha_{a}\varPi_{a}\right)^{2}}}\end{array}\sqrt{\begin{array}{c}
\frac{1}{2}\left(\mathsf{1}_{4\times4}-\frac{mc^{2}}{\sqrt{\mathsf{H}_{0}^{\left(D\right)}\circ\mathsf{H}_{0}^{\left(D\right)}}}\right)\end{array}}\nonumber
\end{eqnarray}
In this case the following explicit guise for the Newton-Wigner Hamiltonian
is obtained
\begin{eqnarray}
\mathsf{H}_{0}^{\left(NW\right)} & = & \mathsf{\mathsf{T}}_{0}\mathsf{H}_{0}^{\left(D\right)}\mathsf{\mathsf{T}}_{0}^{\dagger}=\beta\sqrt{\mathsf{H}_{0}^{\left(D\right)}\circ\mathsf{H}_{0}^{\left(D\right)}}\label{eq:H_0_(NW)}\\
 & = & \beta\,mc^{2}\,\cdot\sqrt{\mathsf{\mathrm{\mathsf{1}}}_{4\times4}+\frac{2}{mc^{2}}\left(\mathsf{1}_{2\times2}\otimes\mathsf{H}_{2\times2}^{\left(SP\right)}\right)}\nonumber
\end{eqnarray}
, with $\mathsf{H}_{2\times2}^{\left(SP\right)}$ the common Schr{\"o}dinger-Pauli Hamiltonian,
\begin{equation}
\label{eq:Schroedinger-Pauli Hamiltonian}
\mathsf{H}_{2\times2}^{\left(SP\right)}=\frac{\varPi_{b}\varPi_{b}}{2m}\mathsf{\mathrm{\mathsf{1}}}_{2\times2}-\frac{q\hbar}{2m}B_{b}^{\left(ext\right)}\sigma_{b}^{\left(P\right)}
\end{equation}
, now describing (in constant magnetic field) the Landau levels of
a nonrelativistic electron. From (\ref{eq:H_0_(NW)}) a meaningful
nonrelativistic Hamiltonian together with the lowest relativistic
correction on the energy scale of fine structure is readily obtained
considering the rest energy $mc^{2}$ of the electron as the predominant term:
\begin{equation}
\label{eq:nonrelativistic approximation to ( H_(NW)*H_(NW) )^(1/2)}\\
\beta\mathsf{H}_{0}^{\left(NW\right)}=\mathsf{\mathrm{\mathsf{1}}}_{2\times2}\otimes\left(mc^{2}\mathsf{\mathrm{\mathsf{1}}}_{2\times2}+\mathsf{H}_{2\times2}^{\left(SP\right)}-\frac{1}{2mc^{2}}\mathsf{H}_{2\times2}
^{\left(SP\right)}\mathsf{H}_{2\times2}^{\left(SP\right)}+...\right)
\end{equation}
In the expansion (\ref{eq:expansion of Dirac amplitude in terms of eigenfunctions})
of the four component Dirac amplitude, the positive-energy eigenmodes
$U_{\mu}\left(\mathbf{r},k\right)$, if regarded \emph{separately}
from the complementing negative-energy eigenmodes $V_{\mu}\left(\mathbf{r},\tilde{k}\right)$,
do \emph{not} represent a complete set of eigenfunctions, while the
NW-eigenmodes $U_{\mu}^{\left(NW\right)}\left(\mathbf{r},k\right)$
and $V_{\mu}^{\left(NW\right)}\left(\mathbf{r},\tilde{k}\right)$
of $\mathsf{H}_{0}^{\left(NW\right)}$ are directly connected to the
\emph{complete} and orthonormal set of the \emph{two-component} eigenmodes
$u_{\sigma}^{\left(SP\right)}\left(\mathbf{r},k\right)$ of $\mathsf{H}_{2\times2}^{\left(SP\right)}$.
Now restricting to the \emph{nonrelativistic} sector then the associated
eigenfunctions $u_{\sigma}^{\left(SP\right)}\left(\mathbf{r},k\right)$
of $\mathsf{H}_{2\times2}^{\left(SP\right)}$ and for that matter
the eigenfunctions $U_{\mu}^{\left(NW\right)}\left(\mathbf{r},k\right)$
and $V_{\mu}^{\left(NW\right)}\left(\mathbf{r},\tilde{k}\right)$
of $\mathsf{H}_{0}^{\left(NW\right)}$ will be slowly varying functions
on the scale of the Compton wavelength $\lambda_{C}$ , thus providing
an eminently suitable starting point for obtaining a nonrelativistic
approximation to matrix-elements originally build with \emph{four-component}
Dirac amplitudes.

As a final remark, given a complete system of eigenfunction of $\mathsf{H}_{2\times2}^{\left(SP\right)}$,
a corresponding complete system of eigenfunctions of $\mathsf{H}_{0}^{\left(D\right)}$ can be readily generated upon application
 of the inverse transformation $\mathsf{T}^{\dagger}$ applied to those eigenfunctions of $\mathsf{H}_{2\times2}^{\left(SP\right)}$.
\section{The beta-flow equation transforming the Dirac $\hspace{1mm}\beta\hspace{1mm} $ into the energy-sign operator $\Lambda^{\left(D\right)}$ for a Dirac Hamiltonian with coupling to electrostatic and magnetostatic fields}
If a charge carrying Dirac fermion moves in the presence of magnetostatic and electrostatic
fields superposed, it is generally accepted to be difficult \cite{Eriksen},\cite{EriksenKolsrud},\cite{Morpurgo},\cite{Osche},\cite{deVries}
obtaining the unitary transformation $\mathsf{\mathsf{T}}$, and in
this way the Newton-Wigner Hamiltonian $\mathsf{H}^{\left(NW\right)}$ from a Dirac Hamiltonian
$\mathsf{H}^{\left(D\right)}$ as stated in (\ref{eq:Dirac-Hamiltonian}).
Let us agree on terming operators $\mathcal{O}$ as being \emph{odd}
and operators $\mathcal{E}$ as being \emph{even}, iff
\begin{eqnarray}
\mathcal{O}\beta & = & -\beta\mathcal{O}\label{eq:def odd=000026even}\\
\mathcal{E}\beta & = & \beta\mathcal{E}\nonumber
\end{eqnarray}
We aim in this section at constructing a flow striving from a general Dirac Hamiltonian
\begin{eqnarray}
\mathsf{H}^{\left(D\right)} & = & \beta mc^{2}+\mathcal{O}+\mathcal{E}\label{eq:H_D}
\end{eqnarray}
towards the corresponding Newton-Wigner Hamiltonian $\mathsf{H}^{\left(NW\right)}$.
For example, a well known extension of the Dirac Hamiltonian in external
fields, as stated afore in (\ref{eq:Dirac-Hamiltonian}), takes into
account, besides mass $m$ and charge $q_{e}$, further phenomenological
attributes for the spin-$\frac{1}{2}$ ``particles'' like an intrinsic
\emph{anomalous} magnetic moment $\mu_{M}$ or even an intrinsic electric
dipole moment $d_{E}$, see for instance \cite{Barut}. In this case
\begin{eqnarray}
\mathcal{O} & = & c\alpha_{b}\varPi_{b}+i\beta\alpha_{b}\left(\frac{\mu_{M}}{c}E_{b}^{\left(ext\right)}-cd_{E}B_{b}^{\left(ext\right)}\right)\label{eq:odd operator   O  II-1}\\
\mathcal{E} & = & q_{e}\varPhi\left(\mathsf{x}\right)-\frac{\varPi_{b}}{mc}\left(\mu_{M}B_{b}^{\left(ext\right)}+d_{E}E_{b}^{\left(ext\right)}\right)\label{eq:even operator  E  II-1}
\end{eqnarray}
Of course, with $d_{E}\neq0$ then (spatial) parity is \emph{not} conserved
\cite{Pauli}. Whereas in standard QED for sure $d_{E}\equiv0$ \cite{Pauli},
instead in electroweak theory $d_{E}\neq0$ appears quite reasonable.
For a thorough discussion see \cite{Silenko2003},\cite{LipingZou2020}.

In order to determine the exact solution to the nonlinear flow equation
(\ref{eq:double bracket flow}) with generator $\eta\left(s\right)=\left[\beta,\mathsf{H}\left(s\right)\right]$,
the idea is to look for an operator $\mathsf{Z}\left(s\right)$ representing
a continuous unitary transformation of the Dirac matrix $\beta$ by solving the flow equation
\begin{eqnarray}
\frac{d}{ds}\mathsf{Z}\left(s\right) & = & \left[\omega\left(s\right),\mathsf{Z}\left(s\right)\right]\label{eq:beta-flow}\\
\mathsf{Z}\left(0\right) & = & \beta\nonumber
\end{eqnarray}
, and in particular to choose the generator $\omega\left(s\right)$
of that ``beta-flow'' in such a way, that the limiting value $\mathsf{Z}\left(\infty\right)$
commutes with the original Dirac Hamiltonian \footnote{Here $\mathsf{\tilde{H}}^{\left(D\right)}$ denotes a \emph{scaled} Dirac Hamiltonian.}
\begin{eqnarray}
\left[\mathsf{\tilde{H}}^{\left(D\right)},\mathsf{Z}\left(\infty\right)\right] & = & \mathsf{0}_{4\times4}\label{eq:[Z(infty), H_D]= 0}
\end{eqnarray}
Along the line of reasoning presented afore in (\ref{eq:double bracket flow})
a suitable antisymmetric generator of such a beta-flow emerges as
\begin{equation}
\omega\left(s\right)=\left[\mathsf{\tilde{H}}^{\left(D\right)},\mathsf{Z}\left(s\right)\right]\label{eq:generator omega}
\end{equation}
whereby
\begin{equation}
\mathsf{\tilde{H}}^{\left(D\right)}=\frac{1}{mc^{2}}\mathsf{H}^{\left(D\right)}=\beta+\mathcal{\tilde{E}}+\mathcal{\tilde{O}}\label{eq:scaled Hamiltonian}
\end{equation}
If (\ref{eq:beta-flow}) could be solved for $\mathsf{Z}\left(s\right)$,
then the generator $\omega\left(s\right)$ was known explicitely and
the unitary transformation of $\beta$ could be represented as
\begin{eqnarray}
\mathsf{Z}\left(s\right) & = & \mathsf{V}\left(s\right)\beta\,\mathsf{V}^{\dagger}\left(s\right)\label{eq:representation Z(s)=V*beta*V_(ad)}
\end{eqnarray}
, whereby the unitary transformation $\mathsf{V}\left(s\right)$ solves
\begin{eqnarray}
\frac{d}{ds}\mathsf{V}\left(s\right) & = & \omega\left(s\right)\mathsf{V}\left(s\right)\label{eq:ODE V(s)}\\
\mathsf{V}\left(0\right) & = & \mathsf{1}_{4\times4}\nonumber
\end{eqnarray}
And because the transformation $\mathsf{V}\left(s\right)$ is unitary, of course there holds
\begin{equation}
\beta\beta=\mathsf{Z}\left(s\right)\mathsf{Z}\left(s\right)=\mathsf{Z}\left(\infty\right)\mathsf{Z}\left(\infty\right)=\mathsf{1}_{4\times4}\label{eq:Z(s)^2=1}
\end{equation}

Consideration should be given to an ambiguity regarding the representation
of $\mathsf{Z}\left(s\right)$ in (\ref{eq:representation Z(s)=V*beta*V_(ad)})
with such a unitary transformation $\mathsf{V}\left(s\right)$. Indeed,
with $\mathsf{N}\left(s\right)$ another unitary operator with attributes
\begin{eqnarray}
\mathsf{N}\left(s\right)\beta\, & = & \beta\,\mathsf{N}\left(s\right)\label{eq:N(s)}
\end{eqnarray}
, then instead of (\ref{eq:representation Z(s)=V*beta*V_(ad)})
one finds for $\mathsf{Z}\left(s\right)$ as well the entirely equivalent representation
\begin{eqnarray}
\mathsf{Z}\left(s\right) & = & \mathsf{U}\left(s\right)\beta\,\mathsf{U}^{\dagger}\left(s\right)\label{eq:equivalent unitary transformation V(s)=U(s)N(s)}\\
\mathsf{V}\left(s\right) & = & \mathsf{U}\left(s\right)\mathsf{N}\left(s\right)\nonumber \\
\mathsf{U}\left(0\right)\mathsf{N}\left(0\right) & = & \mathsf{1}_{4\times4}\nonumber
\end{eqnarray}
So, even if $\mathsf{Z}\left(s\right)$ was known exactly, compliant with the representation (\ref{eq:representation Z(s)=V*beta*V_(ad)})
said unitary transformation $\mathsf{V}\left(s\right)$ in (\ref{eq:equivalent unitary transformation V(s)=U(s)N(s)})
cannot be determined any better up to an undetermined blockdiagonal factor $\mathsf{N}\left(s\right)$.

\section{Exact Solution of Beta-Flow Equation}\label{sec:ExactSolutionStaticBetaFlow }

With $\omega\left(s\right)$ as specified in (\ref{eq:generator omega})
the flow equation (\ref{eq:beta-flow}) determining $\mathsf{Z}\left(s\right)$ reads
\begin{eqnarray}
\label{eq:beta  flow II}
\frac{d}{ds}\mathsf{Z}\left(s\right) & = & \left[\left[\mathsf{\tilde{H}}^{\left(D\right)},\mathsf{Z}\left(s\right)\right],\mathsf{Z}\left(s\right)\right]\\
\mathsf{Z}\left(0\right) & = & \beta
\end{eqnarray}
Evaluation of the double commutator gives
\begin{equation}
\label{eq:double commutator Z}\\
\left[\left[\mathsf{\tilde{H}}^{\left(D\right)},\mathsf{Z}\left(s\right)\right],\mathsf{Z}\left(s\right)\right]=
\mathsf{\tilde{H}}^{\left(D\right)}\mathsf{Z}\left(s\right)\mathsf{Z}\left(s\right)
-2\mathsf{Z}\left(s\right)\mathsf{\tilde{H}}^{\left(D\right)}\mathsf{Z}\left(s\right)
+\mathsf{Z}\left(s\right)\mathsf{Z}\left(s\right)\mathsf{\tilde{H}}^{\left(D\right)}
\end{equation}
Because of (\ref{eq:Z(s)^2=1}) then (\ref{eq:double commutator Z})
simplifies and the differential equation (\ref{eq:beta flow II})
reads
\begin{eqnarray}
\label{eq:beta  flow III}
\frac{1}{2}\frac{d}{ds}\mathsf{Z}\left(s\right) & = & \mathsf{\tilde{H}}^{\left(D\right)}-\mathsf{Z}\left(s\right)\mathsf{\tilde{H}}^{\left(D\right)}\mathsf{Z}\left(s\right)\mathsf{Z}\left(0\right)
\end{eqnarray}
Serendipitously this being a matrix Riccati equation \cite{matrixRiccati},
we can find an \emph{exact }solution to this initial value problem in the guise
\begin{eqnarray}
\label{eq:exact solution Z(s)}
\mathsf{Z}\left(s\right) & = & \mathsf{W}\left(s\right)\beta\,\mathsf{W}^{-1}\left(s\right)
\end{eqnarray}
, with
\begin{eqnarray}
\label{eq:ingredients to solution Z(s)}
\mathsf{W}\left(s\right) & = & \mathsf{C}\left(s\right)\beta+\mathsf{S}\left(s\right)\\
\mathsf{C}\left(s\right) & = & \cosh\left(2s\,\mathsf{\tilde{H}}^{\left(D\right)}\right)\nonumber \\
\mathsf{S}\left(s\right) & = & \sinh\left(2s\,\mathsf{\tilde{H}}^{\left(D\right)}\right)\nonumber
\end{eqnarray}
Despite $\beta\mathsf{\tilde{H}}^{\left(D\right)}\neq\mathsf{\tilde{H}}^{\left(D\right)}\beta$
, there holds as an identity
\begin{equation}
\beta\mathsf{S}\left(s\right)\mathsf{S}\left(s\right)+\mathsf{C}\left(s\right)\mathsf{C}\left(s\right)\beta=\mathsf{S}\left(s\right)\mathsf{S}\left(s\right)\beta+\beta\mathsf{C}\left(s\right)\mathsf{C}\left(s\right)
\label{eq:beta*s^2+c^2*beta = beta*c^2 +s^2*beta}
\end{equation}
, and therefore
\begin{equation}
\label{eq:W_ad*W*beta=beta*W_ad*W}
\mathsf{W}^{\dagger}\left(s\right)\mathsf{W}\left(s\right)\beta=\beta\,\mathsf{W}^{\dagger}\left(s\right)\mathsf{W}\left(s\right)
\end{equation}
With that said
\begin{equation}
\mathsf{Z}\left(s\right)=\mathsf{Z}^{\dagger}\left(s\right)\label{eq:Z(s)=Z_(ad)(s)}
\end{equation}
, and in this way
\begin{equation}
\label{eq:Z(s)*Z_ad_(s)}
\mathsf{Z}\left(s\right)\mathsf{Z}^{\dagger}\left(s\right)=\mathsf{Z}^{\dagger}\left(s\right)\mathsf{Z}\left(s\right)
=\mathsf{Z}\left(s\right)\mathsf{Z}\left(s\right)=\mathsf{1}_{4\times4}
\end{equation}
, thus validating in accord with the representation (\ref{eq:representation Z(s)=V*beta*V_(ad)})
the operator $\mathsf{Z}\left(s\right)$ being unitary (and involutive as well).
With $\Lambda^{\left(D\right)}$ the energy-sign operator (\ref{eq:energy sign operator})
in the basis of Dirac eigenstates, there holds
\begin{eqnarray}
\label{eq:tanh( 2s (H_D*H_D)^(1/2) ) s->infty}
\lim_{s\rightarrow\infty}\mathsf{C}^{-1}\left(s\right)\mathsf{S}\left(s\right) & = \lim_{s\rightarrow\infty}
\frac{\mathsf{\tilde{H}}^{\left(D\right)}}{\sqrt{\mathsf{\tilde{H}}^{\left(D\right)}\mathsf{\tilde{H}}^{\left(D\right)}}}
\tanh\left(2s \sqrt{\mathsf{\tilde{H}}^{\left(D\right)}\mathsf{\tilde{H}}^{\left(D\right)}}\right) & = \Lambda^{\left(D\right)}
\end{eqnarray}
, and thus for $s\rightarrow\infty$ a meaningful limiting value of
$\mathsf{Z}\left(s\right)$ identical to the energy-sign operator $\Lambda^{\left(D\right)}$ exists:
\begin{eqnarray}
\mathsf{Z}\left(\infty\right) & = & \lim_{s\rightarrow\infty}\mathsf{Z}\left(s\right)\label{eq:Z(s->infty)}\\
 & = & \lim_{s\rightarrow\infty}\left(\mathsf{C}\left(s\right)\left(\beta+\mathsf{C}^{-1}\left(s\right)\mathsf{S}\left(s\right)\right)\beta\left(\beta+\mathsf{C}^{-1}\left(s\right)\mathsf{S}\left(s\right)\right)^{-1}\mathsf{C}^{-1}\left(s\right)\right)\nonumber \\
 & = & \lim_{s\rightarrow\infty}\left(\mathsf{C}\left(s\right)\left(\left(\beta+\Lambda^{\left(D\right)}\right)\beta\right)\left(\beta+\Lambda^{\left(D\right)}\right)^{-1}\mathsf{C}^{-1}\left(s\right)\right)\nonumber \\
 & = & \lim_{s\rightarrow\infty}\left(\mathsf{C}\left(s\right)\left(\Lambda^{\left(D\right)}\left(\beta+\Lambda^{\left(D\right)}\right)\right)\left(\beta+\Lambda^{\left(D\right)}\right)^{-1}\mathsf{C}^{-1}\left(s\right)\right)\nonumber \\
 & = & \Lambda^{\left(D\right)}\nonumber
\end{eqnarray}

\section{Construction of Unitary Transformation $\mathsf{V}\left(s\right)$}

Even though the operator $\mathsf{Z}\left(s\right)$ in (\ref{eq:exact solution Z(s)})
is the exact solution of the beta-flow (\ref{eq:beta  flow II}),
this doesn't concurrently determine the unitary transformation $\mathsf{V}\left(s\right)$
in the representation (\ref{eq:representation Z(s)=V*beta*V_(ad)}).
Of course, $\mathsf{W}^{\dagger}\left(s\right)\mathsf{W}\left(s\right)$
being positive definite then a unitary operator $\mathsf{U}^{\left(P\right)}\left(s\right)$
related to the polar decomposition of the operator $\mathsf{W}\left(s\right)$
can be identified in the guise \cite{polar}
\begin{equation}
\mathsf{U}^{\left(P\right)}\left(s\right)=\mathsf{W}\left(s\right)\left(\mathsf{W}^{\dagger}\left(s\right)\mathsf{W}\left(s\right)\right)^{-\frac{1}{2}}\label{eq:U_P_(s)  I}
\end{equation}
And with help of (\ref{eq:W_ad*W*beta=beta*W_ad*W}) it follows then
\begin{eqnarray}
\mathsf{Z}\left(s\right) & = & \mathsf{U}^{\left(P\right)}\left(s\right)\beta\,\left(\mathsf{U}^{\left(P\right)}\left(s\right)\right)^{\dagger}\label{eq:representation Z(s)= U_P*beta*U_(ad)_P}
\end{eqnarray}
, with $\mathsf{U}^{\left(P\right)}\left(s\right)$ (after re-arrangement of the square root term) given by
\begin{eqnarray}
\mathsf{U}^{\left(P\right)}\left(s\right) & = & \left(\mathsf{C}\left(s\right)\beta+\mathsf{S}\left(s\right)\right)\left(\left(\beta\:\mathsf{C}\left(s\right)+\mathsf{S}\left(s\right)\right)\left(\mathsf{C}\left(s\right)\beta+\mathsf{S}\left(s\right)\right)\right)^{-\frac{1}{2}}
\label{eq:U_P_(s)  II}
\end{eqnarray}

But now a complicacy arises, as the identification of (\ref{eq:U_P_(s)  II})
with the searched for unitary transformation $\mathsf{V}\left(s\right)$, setting
\begin{equation}
\mathsf{V}\left(s\right)\equiv\mathsf{U}^{\left(P\right)}\left(s\right)\beta\label{eq:identification V(s)=U_P_(s)*Beta}
\end{equation}
in view of the posed initial value (\ref{eq:ODE V(s)}) , actually is expedient only iff
\begin{eqnarray}
\left[\beta,\sqrt{\mathsf{\tilde{H}}^{\left(D\right)}\mathsf{\tilde{H}}^{\left(D\right)}}\right] & = & \mathsf{0}_{4\times4}\label{eq:special requirement}
\end{eqnarray}
This special constraint immediately implying $\mathsf{C}\left(s\right)\beta=\beta\,\mathsf{C}\left(s\right)$,
only then a meaningful limit of $\mathsf{U}^{\left(P\right)}\left(s\right)$
for $s\rightarrow\infty$ is readily obtained
\begin{equation}
\mathsf{U}^{\left(P\right)}\left(\infty\right)=\frac{\beta+\Lambda^{\left(D\right)}}{\sqrt{\left(\beta+\Lambda^{\left(D\right)}\right)^{2}}}\label{eq:U_P_(s -> infty)}
\end{equation}
Unfortunately, the restriction (\ref{eq:special requirement}) is
not fitting in at all, say, with the presence of an external electrostatic
potential $\varPhi\left(\mathbf{r}\right)\neq0$ in the Dirac Hamiltonian,
as the square $\mathsf{\tilde{H}}^{\left(D\right)}\mathsf{\tilde{H}}^{\left(D\right)}$
then also comprises odd terms (anti-commuting with $\beta$). Thus,
except in (special) cases when (\ref{eq:special requirement}) applies,
as concerns for instance the Dirac Hamiltonian $\mathsf{H}_{0}^{\left(D\right)}$stated
in (\ref{eq:H_0_(D)}), for a general Dirac Hamiltonian $\mathsf{\tilde{H}}^{\left(D\right)}$
the limiting value of $\mathsf{U}^{\left(P\right)}\left(s\right)$
indeed remains indeterminable due to our ignorance how to find the
limiting value of terms like $\mathsf{C}^{-1}\left(s\right)\beta\,\mathsf{S}\left(s\right)$
for $s\rightarrow\infty$. This symptom reveals the polar decomposition
(\ref{eq:U_P_(s)  I}) being pointless if (\ref{eq:special requirement})
ceases to be valid. Briefly speaking, that way we cannot find the
searched for unitary transformation for a general Dirac Hamiltonian,
even though the limiting value $\mathsf{Z}\left(s\right)$ for $s\rightarrow\infty$
has according to (\ref{eq:Z(s->infty)}) already an assigned value.

Progress comes, accepting (for the moment being) a simplified way
of writing $\mathsf{Z}\equiv\mathsf{Z}\left(s\right)$, from the observation
\begin{eqnarray}
\left(\beta+\mathsf{Z}\right)\beta & = & \mathsf{Z}\left(\beta+\mathsf{Z}\right)\label{eq:identity I}\\
\left[\beta,\left(\beta\mathsf{Z}+\mathsf{Z}\beta\right)\right] & = & \mathsf{0}_{4\times4}=\left[\mathsf{Z},\left(\beta\mathsf{Z}+\mathsf{Z}\beta\right)\right]\nonumber
\end{eqnarray}
, so that
\begin{eqnarray}
\mathsf{Z}\left(s\right) & = & \mathsf{Z}\left(\beta+\mathsf{Z}\right)\left(\beta+\mathsf{Z}\right)^{-1}\label{eq:identity II}\\
 & = & \left(\beta+\mathsf{Z}\right)\beta\left(\beta+\mathsf{Z}\right)^{-1}\nonumber \\
 & = & \left(\beta+\mathsf{Z}\right)\beta\left(\beta+\mathsf{Z}\right)^{-2}\left(\beta+\mathsf{Z}\right)\nonumber
\end{eqnarray}
The operator $\beta+\mathsf{Z}$ being hermitean (and excluding $zero$
as an eigenvalue of that operator), then for sure $\left(\beta+\mathsf{Z}\right)^{2}$
is positive definite, so that
\begin{eqnarray}
\left(\beta+\mathsf{Z}\right)^{-2} & = & \left(\sqrt{\left(\beta+\mathsf{Z}\right)^{2}}\sqrt{\left(\beta+\mathsf{Z}\right)^{2}}\right)^{-1}\label{eq:( square-root)^2-1}
\end{eqnarray}
On these grounds there follows now the representation
\begin{eqnarray}
\mathsf{Z}\left(s\right) & = & \mathsf{U}\left(s\right)\beta\,\mathsf{U}^{\dagger}\left(s\right)\label{eq:solution Z(s)  III}
\end{eqnarray}
, with $\mathsf{U}\left(s\right)$ being unitary and built-up in terms
of the exact solution $\mathsf{Z}\left(s\right)$ as given in (\ref{eq:exact solution Z(s)}):
\begin{eqnarray}
\mathsf{U}\left(s\right) & = & \frac{\beta+\mathsf{Z}\left(s\right)}{\sqrt{\left(\beta+\mathsf{Z}\left(s\right)\right)^{2}}\,}\label{eq:explicit unitary transformation U(s)}\\
\mathsf{U}\left(0\right) & = & \beta\label{eq:U(s->0)}
\end{eqnarray}
With the known limiting value (\ref{eq:Z(s->infty)}) then
\begin{eqnarray}
\mathsf{U}\left(\infty\right) & = & \frac{\beta+\Lambda^{\left(D\right)}}{\sqrt{\left(\beta+\Lambda^{\left(D\right)}\right)^{2}}}\label{eq:U(s->infty)}
\end{eqnarray}
Remarkably enough, even though the operator $\mathsf{\mathsf{U}^{\left(P\right)}}\left(s\right)$
introduced in (\ref{eq:U_P_(s)  I}) has only under the special
premise (\ref{eq:special requirement}) for $s\rightarrow\infty$
a definite limit (\ref{eq:U_P_(s -> infty)}), nonetheless that limiting
value coincides with the generally valid result (\ref{eq:U(s->infty)}).
Yet, in view of the posed initial value (\ref{eq:ODE V(s)}), the
searched for unitary transformation is \emph{not} $\mathsf{U}\left(s\right)$, but
\begin{eqnarray}
\mathsf{V}\left(s\right) & \equiv & \mathsf{U}\left(s\right)\beta=\frac{\beta+\mathsf{Z}\left(s\right)}{\sqrt{\left(\beta+\mathsf{Z}\left(s\right)\right)^{2}}\,}\beta\label{eq:identification V(s)=U(s)*Beta}
\end{eqnarray}
So in point of fact with $\mathsf{V}\left(s\right)$ given in (\ref{eq:identification V(s)=U(s)*Beta}) we have
\begin{eqnarray}
\mathsf{Z}\left(s\right) & = & \mathsf{V}\left(s\right)\beta\,\mathsf{V}^{\dagger}\left(s\right)\label{eq:representation Z(s)=V*beta*V_(ad) II}\\
\mathsf{V}\left(0\right) & = & \mathsf{1}_{4\times4}\nonumber \\
\mathsf{V}\left(s\right)\mathsf{V}^{\dagger}\left(s\right) & = & \mathsf{1}_{4\times4}=\mathsf{V}^{\dagger}\left(s\right)\mathsf{V}\left(s\right)\nonumber
\end{eqnarray}
Notably, based on the identities
\begin{eqnarray}
\mathsf{U}\left(s\right)\mathsf{U}\left(s\right) & = & \mathsf{1}_{4\times4}\label{eq:identities U(s)}\\
\beta\mathsf{U}\left(s\right) & = & \mathsf{U}\left(s\right)\mathsf{Z}\left(s\right)\nonumber
\end{eqnarray}
, there holds for this particular transformation (\ref{eq:identification V(s)=U(s)*Beta}) as well
\begin{eqnarray}
\mathsf{V}\left(s\right)\mathsf{V}\left(s\right) & = & \mathsf{U}\left(s\right)\beta\mathsf{U}\left(s\right)\beta\label{eq:V(s)V(s)}\\
 & = & \mathsf{U}\left(s\right)\mathsf{U}\left(s\right)\mathsf{Z}\left(s\right)\beta\nonumber \\
 & = & \mathsf{Z}\left(s\right)\beta\nonumber
\end{eqnarray}

\section{Retrieval of the Newton-Wigner Hamiltonian}

The basic cause of the unitary transformation $\mathsf{V}\left(s\right)$
stated in (\ref{eq:identification V(s)=U(s)*Beta}) being
for $s\rightarrow\infty$ in fact energy-separating, is the identity
\begin{equation}
\mathsf{V}\left(\infty\right)\beta\mathsf{V}^{\dagger}\left(\infty\right)=\mathsf{Z}\left(\infty\right)=\Lambda^{\left(D\right)}\label{eq:beta*U(infty)=U(infty)*Lambda}
\end{equation}
The seeked for unitary transformation $\mathsf{T}$ that maps the
Dirac Hamiltonian $\mathsf{H}^{\left(D\right)}$ to the Newton-Wigner
Hamiltonian $\mathsf{H}^{\left(NW\right)}$ we thus identify directly
from (\ref{eq:identification V(s)=U(s)*Beta}) as
\begin{equation}
\mathsf{T}\equiv\mathsf{V}^{\dagger}\left(\infty\right)=\beta\frac{\beta+\Lambda^{\left(D\right)}}{\sqrt{\left(\beta+\Lambda^{\left(D\right)}\right)^{2}}}\label{eq:unitary transformation T}
\end{equation}
Obviously there holds now
\begin{equation}
\beta\mathsf{T}=\mathsf{T}\Lambda^{\left(D\right)}\label{eq:beta*T=T*Lambda_D}
\end{equation}
Applying $\mathsf{T}$ to positive energy eigenstates $\left|U_{k}^{\left(D\right)}\right\rangle $
of $\mathsf{\tilde{H}}^{\left(D\right)}$, respectively applying $\mathsf{T}$
to negative energy eigenstates $\left|V_{\tilde{k}}^{\left(D\right)}\right\rangle $
of $\mathsf{\tilde{H}}^{\left(D\right)}$, it is manifest that the
unitary transformation $\mathsf{T}$ is as well energy-separating:
\begin{eqnarray}
\beta\left(\mathsf{T}\left|U_{k}^{\left(D\right)}\right\rangle \right) & = & \mathsf{T}\Lambda^{\left(D\right)}\left|U_{k}^{\left(D\right)}\right\rangle =+\mathsf{T}\left|U_{k}^{\left(D\right)}\right\rangle
\label{eq:energy-separation property}\\
\beta\left(\mathsf{T}\left|V_{\tilde{k}}^{\left(D\right)}\right\rangle \right) & = & \mathsf{T}\Lambda^{\left(D\right)}\left|V_{\tilde{k}}^{\left(D\right)}\right\rangle =-\mathsf{T}\left|V_{\tilde{k}}^{\left(D\right)}\right\rangle \nonumber
\end{eqnarray}
Indeed, due to (\ref{eq:Z(s->infty)}) we have
\begin{eqnarray}
\sqrt{\mathsf{H}^{\left(D\right)}\mathsf{H}^{\left(D\right)}} & = & \Lambda^{\left(D\right)}\mathsf{H}^{\left(D\right)}\label{eq:identity (H_D*H_D)^1/2}\\
 & = & \mathsf{Z}\left(\infty\right)\mathsf{H}^{\left(D\right)}\nonumber \\
 & = & \left(\mathsf{V}\left(\infty\right)\beta\mathsf{V}^{\dagger}\left(\infty\right)\right)\mathsf{H}^{\left(D\right)}\nonumber \\
 & = & \left(\mathsf{T}^{\dagger}\beta\,\mathsf{T}\right)\mathsf{H}^{\left(D\right)}\nonumber
\end{eqnarray}
, and so it follows
\begin{eqnarray}
\sqrt{\left(\mathsf{T}\mathsf{H}^{\left(D\right)}\mathsf{T}^{\dagger}\right)\left(\mathsf{T}\mathsf{H}^{\left(D\right)}\mathsf{T}^{\dagger}\right)} & = & \mathsf{T}\left(\sqrt{\mathsf{H}^{\left(D\right)}\mathsf{H}^{\left(D\right)}}\right)\mathsf{T}^{\dagger}\label{eq:unitary transformation (H_D*H_D)^(1/2)}\\
 & = & \mathsf{T}\left(\mathsf{T}^{\dagger}\beta\mathsf{T}\right)\mathsf{H}^{\left(D\right)}\mathsf{T}^{\dagger}\nonumber \\
 & = & \beta\left(\mathsf{T}\mathsf{H}^{\left(D\right)}\mathsf{T}^{\dagger}\right)\nonumber
\end{eqnarray}
Writing $\Lambda^{\left(D\right)}\mathsf{H}^{\left(D\right)}=\sqrt{\mathsf{H}^{\left(D\right)}\mathsf{H}^{\left(D\right)}}=\mathsf{H}^{\left(D\right)}\Lambda^{\left(D\right)}$,
then directly from (\ref{eq:identity (H_D*H_D)^1/2}) along the lines
indicated in (\ref{eq:unitary transformation (H_D*H_D)^(1/2)})
\begin{equation}
\beta\left(\mathsf{T}\mathsf{H}^{\left(D\right)}\mathsf{T}^{\dagger}\right)=\left(\mathsf{T}\mathsf{H}^{\left(D\right)}\mathsf{T}^{\dagger}\right)\beta\label{eq:T*H_(D)*T_ad   even}
\end{equation}
Comparing (\ref{eq:energy separation  identity H_(NW)}) with (\ref{eq:unitary transformation (H_D*H_D)^(1/2)})
we thus identify, among all unitary transformations producing merely
a $2\times2$ - block-diagonalization of the Dirac-Pauli Hamiltonian,
the even \emph{and} energy-separating Newton-Wigner Hamiltonian being
\begin{eqnarray}
\mathsf{H}^{\left(NW\right)} & = & \mathsf{T}\mathsf{H}^{\left(D\right)}\mathsf{T}^{\dagger}\label{eq:H_(NW)}
\end{eqnarray}
whereas $\mathsf{T}$ is the specific unitary transformation stated
in (\ref{eq:unitary transformation T}).

\section{Eriksen Transformation}

In pioneering work Eriksen \cite{Eriksen} fixed the unitary transformation
$\mathsf{U}_{E}$ bearing his name, by \emph{postulating}
\begin{equation}
\mathsf{U}_{E}^{\dagger}\beta=\beta\mathsf{U}_{E}\label{eq:Eriksen condition}
\end{equation}
From this he concluded
\begin{eqnarray}
\varLambda^{\left(D\right)}=\mathsf{U}_{E}^{\dagger}\beta\mathsf{U}_{E} & = & \beta\mathsf{U}_{E}^{2}\label{eq:Eriksen I}
\end{eqnarray}
, and (excluding $-1$ as an eigenvalue of $\beta\,\varLambda^{\left(D\right)}$)
obtained \cite{Eriksen}\cite{Vries1968}
\begin{eqnarray}
\mathsf{U}_{E} & = & \sqrt{\beta\,\varLambda^{\left(D\right)}}\equiv\frac{\frac{1}{2}\left(\mathsf{1}_{4\times4}+\beta\,\varLambda^{\left(D\right)}\right)}{\sqrt{\mathsf{1}_{4\times4}+\frac{1}{4}\left(\beta\,\varLambda^{\left(D\right)}+\varLambda^{\left(D\right)}\beta\,-2\times\mathsf{1}_{4\times4}\right)}}
\label{eq:Eriksen II}
\end{eqnarray}
Comparing $\mathsf{U}_{E}$ with our result (\ref{eq:unitary transformation T})
we readily confirm the limiting value $\mathsf{T}=\underset{s\rightarrow\infty}{\lim}\mathsf{V}^{\dagger}\left(s\right)$
being identical to Eriksen's transformation:
\begin{eqnarray}
\mathsf{T}=\beta\frac{\beta+\Lambda^{\left(D\right)}}{\sqrt{\left(\beta+\Lambda^{\left(D\right)}\right)^{2}}} & = & \mathsf{U}_{E}\label{eq:U(s->infty) II}
\end{eqnarray}

\section{A Link between the Hamiltonian Flow and the Beta-Flow}

As a general proposition, the solution $\mathsf{H}\left(s\right)$
to the (nonlinear) Hamiltonian flow equation (\ref{eq:double bracket flow})
with generator $\eta\left(s\right)=\left[\beta,\mathsf{H}\left(s\right)\right]$
enables a representation in the guise
\begin{equation}
\mathsf{H}\left(s\right)=\mathsf{V}^{\dagger}\left(s\right)\mathsf{\tilde{H}}^{\left(D\right)}\mathsf{V}\left(s\right)\label{eq:guise  H(s)}
\end{equation}
, whereby $\mathsf{V}\left(s\right)$ is the afore introduced \emph{specific}
unitary transformation solving the initial value problem (\ref{eq:ODE V(s)})
with the generator of the beta-flow (\ref{eq:beta-flow})
\begin{equation}
\omega\left(s\right)=\left[\mathsf{\tilde{H}}^{\left(D\right)},\mathsf{Z}\left(s\right)\right]\label{eq:generator omega_(s)  II}
\end{equation}
The validation of (\ref{eq:guise  H(s)}) follows readily calculating the derivative and minding subsequently (\ref{eq:ODE V(s)}),
(\ref{eq:generator omega_(s)  II}) and (\ref{eq:representation Z(s)=V*beta*V_(ad)}). Then
\begin{eqnarray}
\frac{d}{ds}\mathsf{H}\left(s\right) & = & \mathsf{V}^{\dagger}\left(s\right)\left[\mathsf{\tilde{H}}^{\left(D\right)},\omega\left(s\right)\right]\mathsf{V}\left(s\right)\label{eq:flow H(s) II}\\
 & = & \mathsf{V}^{\dagger}\left(s\right)\left(\left[\mathsf{\tilde{H}}^{\left(D\right)},\left[\mathsf{\tilde{H}}^{\left(D\right)},\mathsf{Z}\left(s\right)\right]\right]\right)\mathsf{V}\left(s\right)\nonumber \\
 & = & \mathsf{V}^{\dagger}\left(s\right)\left(\begin{array}{c}
\mathsf{\tilde{H}}^{\left(D\right)}\mathsf{\tilde{H}}^{\left(D\right)}\left(\mathsf{V}\left(s\right)\beta\mathsf{V}^{\dagger}\left(s\right)\right)\\
-2\mathsf{\tilde{H}}^{\left(D\right)}\left(\mathsf{V}\left(s\right)\beta\mathsf{V}^{\dagger}\left(s\right)\right)\mathsf{\tilde{H}}^{\left(D\right)}\\
+\left(\mathsf{V}\left(s\right)\beta\mathsf{V}^{\dagger}\left(s\right)\right)\mathsf{\tilde{H}}^{\left(D\right)}\mathsf{\tilde{H}}^{\left(D\right)}
\end{array}\right)\mathsf{V}\left(s\right)\nonumber \\
 & = & \mathsf{H}\left(s\right)\mathsf{H}\left(s\right)\beta-2\mathsf{H}\left(s\right)\beta\mathsf{H}\left(s\right)+\beta\mathsf{H}\left(s\right)\mathsf{H}\left(s\right)\nonumber \\
 & = & \left[\left[\beta,\mathsf{H}\left(s\right)\right],\mathsf{H}\left(s\right)\right]\nonumber
\end{eqnarray}
, or else
\begin{eqnarray}
\frac{d}{ds}\mathsf{H}\left(s\right) & = & \left[\eta\left(s\right),\mathsf{H}\left(s\right)\right]\label{eq:flow H(s)  III}\\
\eta\left(s\right) & = & \left[\beta,\mathsf{H}\left(s\right)\right]\nonumber \\
\mathsf{H}\left(0\right) & = & \mathsf{\tilde{H}}^{\left(D\right)}\nonumber
\end{eqnarray}
 The ODE obtained for $\mathsf{H}\left(s\right)$ this way, with (scaled)
initial data $\mathsf{H}\left(0\right)=\mathsf{\tilde{H}}^{\left(D\right)}$,
coincides with the Hamiltonian flow (\ref{eq:double bracket flow}),
thus substantiating the assertion (\ref{eq:guise  H(s)}). Not unsuspected
then, said generators, $\eta\left(s\right)$ for the Hamiltonian flow
(\ref{eq:flow H(s)  III}) and $\omega\left(s\right)$ for the associated
beta-flow (\ref{eq:beta-flow}), are mutually connected by the same
unitary transformation $\mathsf{V}\left(s\right)$, i.e. once $\omega\left(s\right)$
is known, then $\eta\left(s\right)$ is known and vice versa
\begin{equation}
\eta\left(s\right)=-\mathsf{V}^{\dagger}\left(s\right)\omega\left(s\right)\mathsf{V}\left(s\right)\label{eq:connection eta(s) and omega(s)}
\end{equation}
Indeed
\begin{eqnarray*}
\omega\left(s\right) & = & \left[\mathsf{H}^{\left(D\right)},\mathsf{Z}\left(s\right)\right]\\
 & = & \left[\mathsf{H}^{\left(D\right)},\mathsf{V}\left(s\right)\beta\mathsf{V}^{\dagger}\left(s\right)\right]\\
 & = & \mathsf{H}^{\left(D\right)}\mathsf{V}\left(s\right)\beta\mathsf{V}^{\dagger}\left(s\right)-\mathsf{V}\left(s\right)\beta\mathsf{V}^{\dagger}\left(s\right)\mathsf{H}^{\left(D\right)}\\
 & = & \mathsf{V}\left(s\right)\left(\mathsf{V}^{\dagger}\left(s\right)\mathsf{H}^{\left(D\right)}\mathsf{V}\left(s\right)\beta-\beta\mathsf{V}^{\dagger}\left(s\right)\mathsf{H}^{\left(D\right)}\mathsf{V}\left(s\right)\right)\mathsf{V}^{\dagger}\left(s\right)\\
 & = & \mathsf{V}\left(s\right)\left(\mathsf{H}\left(s\right)\beta-\beta\mathsf{H}\left(s\right)\right)\mathsf{V}^{\dagger}\left(s\right)\\
 & = & -\mathsf{V}\left(s\right)\left[\beta,\mathsf{H}\left(s\right)\right]\mathsf{V}^{\dagger}\left(s\right)\\
 & = & -\mathsf{V}\left(s\right)\eta\left(s\right)\mathsf{V}^{\dagger}\left(s\right)
\end{eqnarray*}
Notably, upon insertion of the afore obtained \emph{exact} solution
$\mathsf{Z}\left(s\right)$ stated in (\ref{eq:exact solution Z(s)})
then the generator $\omega\left(s\right)$ being a \emph{known} function
of the flow parameter $s$, this fact enables to write an explicit
(formal though) solution of the linear ODE (\ref{eq:ODE V(s)}) determining
the unitary transformation $\mathsf{V}\left(s\right)$ as an $s$-ordered exponential
\begin{equation}
\mathsf{V}\left(s\right)=T_{s}\exp\left[\int_{0}^{s}ds'\omega\left(s'\right)\right]\label{eq:Dyson series V(s)}
\end{equation}
or else known as the Dyson series \cite{IzyksonZuber}. Conversely,
using unitarity, (\ref{eq:ODE V(s)}) and reexpressing the generator
of the beta-flow $\omega\left(s\right)$ in terms of $\eta\left(s\right)$
using the identity (\ref{eq:connection eta(s) and omega(s)}), it follows at once
\begin{eqnarray*}
\mathsf{0} & = & \mathsf{V}^{\dagger}\left(s\right)\frac{d}{ds}\left(\underset{\mathsf{=1}}{\underbrace{\mathsf{V}\left(s\right)\mathsf{V}^{\dagger}\left(s\right)}}\right)\\
 & = & \mathsf{V}^{\dagger}\left(s\right)\left(\frac{d}{ds}\mathsf{V}\left(s\right)\right)\mathsf{V}^{\dagger}\left(s\right)+\frac{d}{ds}\mathsf{V}^{\dagger}\left(s\right)\\
 & = & \left(\mathsf{V}^{\dagger}\left(s\right)\omega\left(s\right)\mathsf{V}\left(s\right)\right)\mathsf{V}^{\dagger}\left(s\right)+\frac{d}{ds}\mathsf{V}^{\dagger}\left(s\right)\\
 & = & -\eta\left(s\right)\mathsf{V}^{\dagger}\left(s\right)+\frac{d}{ds}\mathsf{V}^{\dagger}\left(s\right)
\end{eqnarray*}
, i.e. $\mathsf{V}^{\dagger}\left(s\right)$ solves the ODE
\begin{eqnarray}
\frac{d}{ds}\mathsf{V}^{\dagger}\left(s\right) & = & \eta\left(s\right)\mathsf{V}^{\dagger}\left(s\right)\label{eq:ODE V_HC_(s)}\\
\mathsf{V}^{\dagger}\left(0\right) & = &  \mathsf{1}_{4\times4}\nonumber
\end{eqnarray}
This fact implicates an equivalent representation for the adjoint
(or else inverse) operator $\mathsf{V}^{\dagger}\left(s\right)$ in
the guise of the Dyson series constructed with the generator $\eta\left(s\right)$
of the Hamiltonian flow (\ref{eq:flow H(s)  III}),
\begin{equation}
\mathsf{V}^{\dagger}\left(s\right)=T_{s}\exp\left[\int_{0}^{s}ds'\eta\left(s'\right)\right]\label{eq:Dyson series V ad (s)}
\end{equation}
Alas, because $\left[\beta,\sqrt{\mathsf{\tilde{H}}^{\left(D\right)}\mathsf{\tilde{H}}^{\left(D\right)}}\right]\neq\mathsf{0}$
for a general Dirac Hamiltonian $\mathsf{\tilde{H}}^{\left(D\right)}$
then as well $\left[\omega\left(s_{1}\right),\omega\left(s_{2}\right)\right]\neq\mathsf{0}$
, alternatively $\left[\eta\left(s_{1}\right),\eta\left(s_{2}\right)\right]\neq\mathsf{0}$
, which feature as a rule prevents an elementary calculation of $\mathsf{V}\left(s\right)$
in closed form.

Be that as it may, with a generator $\omega\left(s\right)$
of known identity (\ref{eq:generator omega_(s)  II}) the explicit
solution $\mathsf{V}\left(s\right)$ of the \emph{linear} ODE (\ref{eq:ODE V(s)})
in the guise of the Dyson series (\ref{eq:Dyson series V(s)}) or
else the Dyson series (\ref{eq:Dyson series V ad (s)}) buildt
with $\eta\left(s\right)$, indeed provides an \emph{exact} solution
to the \emph{nonlinear} Hamiltonian flow equation (\ref{eq:flow H(s)  III}).
Mind however, that due to the afore mentioned ambiguity (\ref{eq:equivalent unitary transformation V(s)=U(s)N(s)}),
even though $\mathsf{H}\left(s\right)$ as a solution to a double-bracket
flow (\ref{eq:flow H(s)  III}) by construction strives for $s\rightarrow\infty$ to a limiting value
$\mathsf{H}\left(\infty\right)$ commuting with $\beta$, see (\ref{eq:Commutator(Bgamma,H(infty))=0}), this feature
taken by itself does \emph{not} warrant as well $\mathsf{H}\left(\infty\right)$ being energy-separating.
\section{The Newton-Wigner Hamiltonian for a Special Class of Dirac Hamiltonians }
The afore derived exact result (\ref{eq:H_(NW)}) for the Newton-Wigner
Hamiltonian is amenable to a substantial simplification for the particular
class of (scaled) Dirac-Hamiltonians $\mathsf{\tilde{H}}_{0}^{\left(D\right)}$
exhibiting the exceptional feature
\begin{eqnarray}
\left[\beta,\left(\mathsf{\tilde{H}}_{0}^{\left(D\right)}\mathsf{\tilde{H}}_{0}^{\left(D\right)}\right)\right] & = & \mathsf{0}_{4\times4}\label{eq:H_0*H_0 even}
\end{eqnarray}
, as applies for instance to the Dirac-Hamiltonian (\ref{eq:H_0_(D)})
for a relativistic particle moving solely in the presence of a static
external magnetic field and more generally to every (scaled) Dirac
Hamiltonians $\mathsf{\tilde{H}}_{0}^{\left(D\right)}=\beta+\mathcal{\tilde{O}}_{0}+\mathcal{\tilde{E}}_{0}$
with the property $\left\{ \mathcal{\tilde{O}}_{0},\mathcal{\tilde{E}}_{0}\right\} \equiv\mathsf{0}$
\cite{LipingZou2020}.
Replacing now everywhere in the exact expression ($\ref{eq:exact solution Z(s)}$)
the operator $\mathsf{\tilde{H}}^{\left(D\right)}$ by $\mathsf{\tilde{H}}_{0}^{\left(D\right)}$,
one readily obtains proceeding directly from (\ref{eq:exact solution Z(s)}) the result
\begin{eqnarray}
\mathsf{Z}_{0}\left(s\right) & = & \mathsf{t}_{0}\left(s\right)\Lambda_{0}^{\left(D\right)}\label{eq:exact solution Z_0_(s)}\\
\mathsf{t}_{0}\left(s\right) & = & \tanh\left(2s\sqrt{\mathsf{\tilde{H}}_{0}^{\left(D\right)}\mathsf{\tilde{H}}_{0}^{\left(D\right)}}+\textrm{artanh\ensuremath{\left(\beta\Lambda_{0}^{\left(D\right)}\right)}}\right)\nonumber
\end{eqnarray}
The associated generator $\omega_{0}\left(s\right)$ of the beta-flow assumes then the guise
\begin{eqnarray}
\omega_{0}\left(s\right) & = & \left[\mathsf{\tilde{H}}_{0}^{\left(D\right)},\mathsf{Z}_{0}\left(s\right)\right]\label{eq:generator omega_0_(s)}\\
 & = & \sqrt{\mathsf{\tilde{H}}_{0}^{\left(D\right)}\mathsf{\tilde{H}}_{0}^{\left(D\right)}}\left(\mathsf{t}_{0}^{\dagger}\left(s\right)-\mathsf{t}_{0}\left(s\right)\right)\nonumber
\end{eqnarray}
Besides being manifestly \emph{odd}
\begin{equation}
\beta\omega_{0}\left(s\right)\beta=-\omega_{0}\left(s\right)=\omega_{0}^{\dagger}\left(s\right).\label{eq:omega_0_(s)  odd}
\end{equation}
, that generator $\omega_{0}\left(s\right)$ has a \emph{vanishing}
commutator at different flow parameter values $s_{1}$, $s_{2}$ , so that
\begin{equation}
\left[\omega_{0}\left(s_{1}\right),\omega_{0}\left(s_{2}\right)\right]=\mathsf{0}_{4\times4}\label{eq:integrability omega_0_(s)}
\end{equation}
For details of the reasoning leading to (\ref{eq:exact solution Z_0_(s)})
and (\ref{eq:generator omega_0_(s)}) we refer to \cite{Schopohl2022}.

Because with (\ref{eq:exact solution Z_0_(s)}) the functional dependence
of $\omega_{0}\left(s\right)$ on the flow parameter $s$ is known,
now upon insertion of (\ref{eq:exact solution Z_0_(s)}) into (\ref{eq:generator omega_0_(s)})
an \emph{exact }analytical expression for the unitary transformation
$\mathsf{V}_{0}\left(s\right)$ can be given, so that
\begin{equation}
\mathsf{H}_{0}\left(s\right)=\mathsf{V}_{0}^{\dagger}\left(s\right)\mathsf{\tilde{H}}_{0}^{\left(D\right)}\mathsf{V}_{0}\left(s\right)\label{eq:unitary transformation H_0_(s)}
\end{equation}
This is because the ODE defining $\mathsf{V}_{0}\left(s\right)$ ,
\begin{eqnarray}
\frac{d}{ds}\mathsf{V}_{0}\left(s\right) & = & \omega_{0}\left(s\right)\mathsf{V}_{0}\left(s\right)\label{eq:ODE V_0_(s)}\\
\mathsf{V}_{0}\left(0\right) & = & \mathsf{1}_{4\times4}\nonumber
\end{eqnarray}
with $\omega_{0}\left(s\right)$ as stated in (\ref{eq:generator omega_0_(s)})
, can be solved due to (\ref{eq:integrability omega_0_(s)}) \emph{exactly}
\begin{eqnarray}
\mathsf{V}_{0}\left(s\right) & = & T_{s}\exp\left[\int_{0}^{s}ds'\omega_{0}\left(s'\right)\right]\equiv\exp\left[\int_{0}^{s}ds'\omega_{0}\left(s'\right)\right]\label{eq:exact solution ODE V_0_(s)  I}
\end{eqnarray}
Evaluation of the integral indeed gives \cite{Schopohl2022}
\begin{equation}
\mathsf{V}_{0}\left(s\right)=\sqrt{\mathsf{Z}_{0}\left(s\right)\beta}=\frac{\beta+\mathsf{Z}_{0}\left(s\right)}{\sqrt{\left(\beta+\mathsf{Z}_{0}\left(s\right)\right)^{2}}}\beta\label{eq:unitary transformation V_0_(s)}
\end{equation}
, which outcome is in full accordance with the afore derived general
result (\ref{eq:U(s->infty) II}), obtained by replacing in (\ref{eq:U(s->infty) II})
the operator $\mathsf{\tilde{H}}^{\left(D\right)}$ by $\mathsf{\tilde{H}}_{0}^{\left(D\right)}$.
The limiting value of $\mathsf{Z}_{0}\left(s\right)$ for $s\rightarrow\infty$
as determined from (\ref{eq:exact solution Z_0_(s)}) being now
\begin{equation}
\mathsf{Z}_{0}\left(\infty\right)=\Lambda_{0}^{\left(D\right)}=\frac{\mathsf{\tilde{H}}_{0}^{\left(D\right)}}{\sqrt{\mathsf{\tilde{H}}_{0}^{\left(D\right)}\mathsf{\tilde{H}}_{0}^{\left(D\right)}}}\label{eq:Z_0_(infty)}
\end{equation}
Defining $\mathsf{T}_{0}\equiv\mathsf{V}_{0}^{\dagger}\left(\infty\right)$, there follows using
\begin{eqnarray*}
\beta\mathsf{T}_{0} & = & \mathsf{T}_{0}\Lambda_{0}^{\left(D\right)}\\
\Lambda_{0}^{\left(D\right)}\mathsf{\tilde{H}}_{0}^{\left(D\right)} & = & \sqrt{\mathsf{\tilde{H}}_{0}^{\left(D\right)}\mathsf{\tilde{H}}_{0}^{\left(D\right)}}
\end{eqnarray*}
that (\ref{eq:H_0*H_0 even}) implies as well
\begin{equation}
\left[\mathsf{T}_{0},\sqrt{\mathsf{\tilde{H}}_{0}^{\left(D\right)}\mathsf{\tilde{H}}_{0}^{\left(D\right)}}\right]=\mathsf{0_{4\times4}}\label{eq:Commutator(T0,HD0^2)=0}
\end{equation}
It is then straightforward to show
\begin{eqnarray*}
\mathsf{\tilde{H}}_{0}^{\left(NW\right)} & = & \mathsf{T}_{0}\mathsf{\tilde{H}}_{0}^{\left(D\right)}\mathsf{T}_{0}^{\dagger}\\
 & = & \beta\left(\beta\mathsf{T}_{0}\right)\mathsf{\tilde{H}}_{0}^{\left(D\right)}\mathsf{T}_{0}^{\dagger}\\
 & = & \beta\left(\mathsf{T}_{0}\Lambda_{0}^{\left(D\right)}\right)\mathsf{\tilde{H}}_{0}^{\left(D\right)}\mathsf{T}_{0}^{\dagger}\\
 & = & \beta\mathsf{T}_{0}\left(\Lambda_{0}^{\left(D\right)}\mathsf{\tilde{H}}_{0}^{\left(D\right)}\right)\mathsf{T}_{0}^{\dagger}\\
 & = & \beta\mathsf{T}_{0}\sqrt{\mathsf{\tilde{H}}_{0}^{\left(D\right)}\mathsf{\tilde{H}}_{0}^{\left(D\right)}}\mathsf{T}_{0}^{\dagger}
\end{eqnarray*}
That way, with (\ref{eq:Commutator(T0,HD0^2)=0}) and $\mathsf{T}_{0}\mathsf{T}_{0}^{\dagger}=\mathsf{1}_{4\times4}$
the Newton-Wigner Hamiltonian $\mathsf{\tilde{H}}_{0}^{\left(NW\right)}$
associated with a Dirac Hamiltonian $\mathsf{\tilde{H}}_{0}^{\left(D\right)}$
obeying to (\ref{eq:H_0*H_0 even}) assumes the simplified guise
\begin{eqnarray}
\mathsf{\tilde{H}}_{0}^{\left(NW\right)} & = & \beta\:\sqrt{\mathsf{\tilde{H}}_{0}^{\left(D\right)}\mathsf{\tilde{H}}_{0}^{\left(D\right)}}\label{eq:NW-Hamiltonian for Commutator(beta H_D*H_D)=0}
\end{eqnarray}
This result coincides with findings obtained first without coupling
to external fields by Foldy and Wouthuysen \cite{FoldyW50}. It applies
as well in the presence of a magnetic induction field but excluding
any coupling to electric potentials, a result first obtained by Case
\cite{Case54} and using different methods by Eriksen \cite{Eriksen}.
A more general context where the feature (\ref{eq:NW-Hamiltonian for Commutator(beta H_D*H_D)=0})
applies is considered in \cite{Silenko2003},\cite{Silenko2013},\cite{Silenko2015},\cite{Silenko2016},\cite{LipingZou2020}.

\section{Manifestly even representation of the Newton-Wigner Hamiltonian as a Series of Iterated Commutators}
The ensuing considerations apply to a Dirac Hamiltonian incorporating
static external potentials and/or fields (\ref{eq:H_D}). According
to what has been said afore, the beta-flow equation (\ref{eq:beta  flow II}) transforms the operator $\beta$ into
\begin{eqnarray}
\mathsf{Z}\left(s\right) & = & \mathsf{V}\left(s\right)\beta\,\mathsf{V}^{\dagger}\left(s\right)\label{eq:representation Z(s)}
\end{eqnarray}
, whereas due to (\ref{eq:guise  H(s)}) that same unitary operator
$\mathsf{V}\left(s\right)$ concurrently serves as well to represent
the solution to the Hamiltonian flow equation (\ref{eq:flow H(s)  III})
\begin{eqnarray}
\mathsf{H}\left(s\right) & = & \mathsf{V}^{\dagger}\left(s\right)\mathsf{\tilde{H}}^{\left(D\right)}\mathsf{V}\left(s\right)\label{eq:representation H(s)}
\end{eqnarray}
Writing with anti-symmetric operators $\Omega_{u}\left(s\right)$ and $\Omega_{g}\left(s\right)$ now
\begin{equation}
\mathsf{V}\left(s\right)=e^{\Omega_{u}\left(s\right)}e^{\Omega_{g}\left(s\right)}\label{eq:representation V(s) with Omega_u and Omega_g I}
\end{equation}
, whereby $\Omega_{u}\left(s\right)$ is odd and $\Omega_{g}\left(s\right)$ is even, then
\begin{eqnarray}
\Omega_{u}^{\dagger}\left(s\right) & = & -\Omega_{u}\left(s\right)\label{eq:anti-fixed point}\\
\beta\Omega_{u}\left(s\right)\beta & = & -\Omega_{u}\left(s\right)\nonumber \\
\nonumber \\
\Omega_{g}^{\dagger}\left(s\right) & = & -\Omega_{g}\left(s\right)\nonumber \\
\beta\Omega_{g}\left(s\right)\beta & = & +\Omega_{g}\left(s\right)\label{eq:fixed-point}
\end{eqnarray}
It follows from (\ref{eq:representation H(s)})
\begin{eqnarray}
\mathsf{H}\left(s\right) & = & e^{-\Omega_{g}\left(s\right)}e^{-\Omega_{u}\left(s\right)}\mathsf{\tilde{H}}^{\left(D\right)}e^{\Omega_{u}\left(s\right)}e^{\Omega_{g}\left(s\right)}
\label{eq:representation V(s) with Omega_u and Omega_g II}
\end{eqnarray}
, whereas from (\ref{eq:representation Z(s)}) we obtain
\begin{equation}
\mathsf{Z}\left(s\right)=e^{\Omega_{u}\left(s\right)}e^{\Omega_{g}\left(s\right)}\beta\,e^{-\Omega_{g}\left(s\right)}e^{-\Omega_{u}\left(s\right)}=e^{2\Omega_{u}\left(s\right)}\beta
\label{eq:representation Z(s) with Omega_u}
\end{equation}
, which representation for $\mathsf{Z}\left(s\right)$ is inherently
consistent with (\ref{eq:V(s)V(s)}). That a factorization like (\ref{eq:representation V(s) with Omega_u and Omega_g I})
exists follows from a general result in Lie-group theory, understanding
the role of the operator $\beta$ in relations like (\ref{eq:fixed-point})
and (\ref{eq:anti-fixed point}) essentialy being equivalent to the
action of an ``involutive automorphism'' \cite{Zanna2004},\cite{A.Iserles2002}.

Because the limiting value $\mathsf{H}\left(\infty\right)$ of the
Hamiltonian flow (\ref{eq:double bracket flow}) obeys by construction
to $\left[\beta,\mathsf{H}\left(\infty\right)\right]=\mathsf{0}$,
now the searched for Newton-Wigner (NW) Hamiltonian arises in the guise
\begin{eqnarray}
\mathsf{\tilde{H}}^{\left(NW\right)} & \equiv & e^{+\Omega_{g}\left(\infty\right)}\mathsf{H}\left(\infty\right)e^{-\Omega_{g}\left(\infty\right)}\label{eq:representation H_NW   with Omega_g_(infty)}\\
 & = & e^{-\Omega_{u}\left(\infty\right)}\mathsf{\tilde{H}}^{\left(D\right)}e^{\Omega_{u}\left(\infty\right)}\label{eq:representation H_NW   with Omega_u_(infty) I}
\end{eqnarray}
If $\mathsf{H}\left(\infty\right)$ was obtained, for instance solving
the Hamiltonian flow (\ref{eq:flow H(s)  III}) perturbatively along
the lines indicated in \cite{Bylev1998}, then for sure $\mathsf{H}\left(\infty\right)$
is an even operator, but it is not guaranteed $\mathsf{H}\left(\infty\right)$
being energy-separating as well. Seen from another perspective, the
unitary transformation performed with the even operator $e^{+\Omega_{g}\left(\infty\right)}$
in (\ref{eq:representation H_NW   with Omega_g_(infty)}) can be
regarded, once $\Omega_{g}\left(\infty\right)$ is known, as a ``correction-scheme''
that converts the merely block-diagonal (even) limiting value $\mathsf{H}\left(\infty\right)$
of the Hamiltonian flow (\ref{eq:flow H(s)  III}) into the \emph{unique}
energy-separating Newton-Wigner Hamiltonian.

Alternatively, $\mathsf{\tilde{H}}^{\left(NW\right)}$ may be obtained
directly from the unitary transformation (\ref{eq:representation H_NW   with Omega_u_(infty) I})
performed solely with the odd operator $\Omega_{u}\left(\infty\right)$.
Once $\Omega_{u}\left(\infty\right)$ is known, then the common BCH-expansion
\cite{Wilcox} leads to a representation as a series of commutators
\begin{eqnarray}
\mathsf{\tilde{H}}^{\left(NW\right)} & = & \mathsf{\tilde{H}}^{\left(D\right)}-\left[\Omega_{u}\left(\infty\right),\mathsf{\tilde{H}}^{\left(D\right)}\right]+\frac{1}{2!}\left[\Omega_{u}\left(\infty\right),\left[\Omega_{u}\left(\infty\right),\mathsf{\tilde{H}}^{\left(D\right)}\right]\right]+....
\label{eq:representation H_NW   with Omega_u(infty) II}
\end{eqnarray}
Here it is important to realize, that a straightforward evaluation
of this series of iterated commutators is needlessly complicated,
because due to the first line in (\ref{eq:representation H_NW   with Omega_u_(infty) I})
the operator $\mathsf{\tilde{H}}^{\left(NW\right)}$ is unconditionally
guaranteed to be \emph{even}. A considerable simplification thus results
rewriting (\ref{eq:representation H_NW   with Omega_u_(infty) I}) in the equivalent guise
\begin{eqnarray}
\mathsf{\tilde{H}}^{\left(NW\right)} & = & \frac{1}{2}\left(e^{-\Omega_{u}\left(\infty\right)}\mathsf{\tilde{H}}^{\left(D\right)}e^{\Omega_{u}\left(\infty\right)}+\beta e^{-\Omega_{u}\left(\infty\right)}\mathsf{\tilde{H}}^{\left(D\right)}e^{\Omega_{u}\left(\infty\right)}\beta\right)\label{eq:representation H_NW   with Omega_u(infty) III}\\
\nonumber \\
 & = & \frac{1}{2}e^{-\Omega_{u}\left(\infty\right)}\mathsf{\tilde{H}}^{\left(D\right)}e^{\Omega_{u}\left(\infty\right)}
 +\frac{1}{2}e^{+\Omega_{u}\left(\infty\right)}\left(\beta\mathsf{\tilde{H}}^{\left(D\right)}\beta\right)e^{-\Omega_{u}\left(\infty\right)}\nonumber \\
\nonumber \\
 & = & \frac{1}{2}e^{-\Omega_{u}\left(\infty\right)}\left(\beta+\mathcal{\tilde{O}}+\mathcal{E}\right)e^{\Omega_{u}\left(\infty\right)}
 +\frac{1}{2}e^{+\Omega_{u}\left(\infty\right)}\left(\beta-\mathcal{\tilde{O}}+\mathcal{E}\right)e^{-\Omega_{u}\left(\infty\right)}\nonumber \\
\nonumber \\
 & = &
 \frac{1}{2}\begin{array}{c}\left(\exp\left(-\textrm{ad}_{\Omega_{u}\left(\infty\right)}\right)+\exp\left(+\textrm{ad}_{\Omega_{u}\left(\infty\right)}\right)\right)\end{array}\circ\left(\beta+\tilde{\mathcal{E}}\right)\nonumber\\
 & &-\frac{1}{2}\begin{array}{c}\left(\exp\left(\textrm{ad}_{\Omega_{u}\left(\infty\right)}\right)-\exp\left(-\textrm{ad}_{\Omega_{u}\left(\infty\right)}\right)\right)\end{array}\circ\tilde{\mathcal{O}}\nonumber
\end{eqnarray}
, whereby the symbol $\textrm{ad}_{X}$ denotes here a most useful
notation to describe iterated commutators as powers, see for instance
\cite{Iserles}. With given operators $X$ and $\mathcal{F}$ then
\begin{eqnarray}
\textrm{ad}_{X}\circ\mathcal{F} & = & \left[X,\mathcal{F}\right]\label{eq:symbol ad(Omega)}\\
\left(\textrm{ad}_{X}\right)^{2}\circ\mathcal{F} & = & \textrm{ad}_{X}\circ\left[X,\mathcal{F}\right]=\left[X,\left[X,\mathcal{F}\right]\right]\nonumber \\
 & ...\nonumber \\
\left(\textrm{ad}_{X}\right)^{n}\circ\mathcal{F} & = & \underset{\textrm{n-fold}}{\underbrace{\left[X,...\left[X,\left[X,\mathcal{F}\right]\right]\right]}}\nonumber
\end{eqnarray}
It follows introducing formal power series
\begin{eqnarray}
\exp\left(\textrm{ad}_{X}\right)\circ\mathcal{F} & = & \sum_{j=0}^{\infty}\frac{1}{j!}\left(\textrm{ad}_{X}\right)^{j}\circ\mathcal{F}\equiv e^{X}\mathcal{F}e^{-X}\nonumber \\
\cosh\left(\textrm{ad}_{X}\right)\circ\mathcal{F} & = & \sum_{j=0}^{\infty}\frac{1}{\left(2j\right)!}\left(\textrm{ad}_{X}\right)^{2j}\circ\mathcal{F}\label{eq:power series expansion hyperbolic functions}\\
\sinh\left(\textrm{ad}_{X}\right)\circ\mathcal{F} & = & \sum_{j=0}^{\infty}\frac{1}{\left(2j+1\right)!}\left(\textrm{ad}_{X}\right)^{2j+1}\circ\mathcal{F}\nonumber
\end{eqnarray}
That way an \emph{explicit} representation for the (scaled) energy-separating
Newton-Wigner Hamiltonian is obtained directly from (\ref{eq:representation H_NW   with Omega_u(infty) III})
in the guise of a series of iterated commutators composed solely with the odd operator
$\Omega_{u}\left(\infty\right)$, each term in this series expansion being\emph{ manifestly} even
\begin{eqnarray}
\mathsf{\tilde{H}}^{\left(NW\right)} & = & \cosh\left(\textrm{ad}_{\Omega_{u}\left(\infty\right)}\right)\circ\left(\beta+\tilde{\mathcal{E}}\right)-\sinh\left(\textrm{ad}_{\Omega_{u}\left(\infty\right)}\right)\circ\mathcal{\tilde{O}}
\label{eq:representation H_NW   with Omega_u(infty) IV}
\end{eqnarray}
Provided a perturbative expansion for the operator $\Omega_{u}\left(\infty\right)$
could be found, then the associated perturbative expansion for the
Newton-Wigner-Hamiltonian as represented by (\ref{eq:representation H_NW   with Omega_u(infty) IV})
is reduced to evaluating just a few commutators.

\section{A Perturbative Expansion in the Style of Magnus for the Operators $\Omega_{u}\left(s\right)$ and $\Omega_{g}\left(s\right)$  defining
 $\mathsf{V}{\left(s\right)}= e^{\Omega_{u}\left(s\right)} e^{\Omega_{g}\left(s\right)}$ }

Restriction of the Dirac Hamiltonian $\tilde{\mathsf{H}}^{\left(D\right)}$
in (\ref{eq:Dirac-Hamiltonian}) to the low energy sector of its
spectrum suggests a weighting of the respective contributions of the
kinetic energy and potential energy terms with regard to a small parameter
$\kappa=\frac{v}{c}$ (in what follows $\kappa$ serving as a formal
bookkeeping device, so that $\kappa=1$ at the end of the calculations).
Different from \cite{Bylev1998} though the  electric potential term  $\mathcal{E}$ is here considered in order of magnitude being comparable to the kinetic energy term $\mathcal{OO}$  for reasons of consistency with the nonrelativistic limit.
This implies (scaled units)
\begin{eqnarray}
\tilde{\mathsf{H}}^{\left(D\right)} & = & \beta+\kappa\mathcal{\tilde{O}}+\kappa^{2}\mathcal{\tilde{E}}\label{eq:Dirac-Hamiltonian with weighted kinetic and potential energy terms}
\end{eqnarray}
To obtain now a perturbation series expansion of the Newton-Wigner
Hamiltonian (\ref{eq:representation H_NW   with Omega_u(infty) IV})
in powers of $\kappa$ we first aim at obtaining a perturbation expansion
of the anti-hermitean operators $\Omega_{u}\left(s\right)$ and $\Omega_{g}\left(s\right)$,
those operators in fact being closely connected to the generator $\omega\left(s\right)$
of the beta-flow (\ref{eq:beta-flow}) or else to the generator $\eta\left(s\right)$
of the Hamiltonian flow (\ref{eq:flow H(s)  III}). To this end let
us rewrite the identity (\ref{eq:connection eta(s) and omega(s)}) in the guise
\begin{eqnarray}
e^{-\Omega_{u}\left(s\right)}\omega\left(s\right)e^{\Omega_{u}\left(s\right)} & = & -e^{\Omega_{g}\left(s\right)}\eta\left(s\right)e^{-\Omega_{g}\left(s\right)}\label{eq:connection eta(s) and omega(s) II}
\end{eqnarray}
As the generator $\eta\left(s\right)$ of the Hamiltonian flow is by construction an odd operator,
\begin{eqnarray}
\eta\left(s\right) & = & \left[\beta,\mathsf{H}\left(s\right)\right]\nonumber \\
\beta\eta\left(s\right)+\eta\left(s\right)\beta & = & \mathsf{0}_{4\times4}\label{eq:eta(s) odd}
\end{eqnarray}
and the commutator of an even operator with an odd operator is always
odd as well, then the even part of the left hand side in (\ref{eq:connection eta(s) and omega(s) II})
should vanish identically,
\begin{eqnarray}
\label{eq:connection eta(s) and omega(s) III}\\
e^{-\Omega_{u}\left(s\right)}\omega\left(s\right)e^{\Omega_{u}\left(s\right)}+\beta e^{-\Omega_{u}\left(s\right)}\omega\left(s\right)e^{\Omega_{u}\left(s\right)}\beta & = & -e^{\Omega_{g}\left(s\right)}\left(\eta\left(s\right)+\beta\eta\left(s\right)\beta\right)e^{-\Omega_{g}\left(s\right)}\equiv\mathsf{0}\nonumber
\end{eqnarray}
That way a (hidden) correlation between the even and odd parts of
the generator $\omega\left(s\right)$ of the beta-flow is revealed
\begin{equation}
\omega\left(s\right)=-e^{2\Omega_{u}\left(s\right)}\left(\beta\:\omega\left(s\right)\beta\right)e^{-2\Omega_{u}\left(s\right)}\label{eq:connection eta(s) and omega(s) IIIb}
\end{equation}
In consequence of $\omega\left(s\right)$ being (in general) composed
of even and odd terms,
\begin{equation}
\omega\left(s\right)=\omega_{g}\left(s\right)+\omega_{u}\left(s\right)\label{eq:decomposition omega(s)}
\end{equation}
, then
\begin{eqnarray}
\omega_{g}\left(s\right)+\omega_{u}\left(s\right) & = & e^{2\Omega_{u}\left(s\right)}\left(-\omega_{g}\left(s\right)+\omega_{u}\left(s\right)\right)e^{-2\Omega_{u}\left(s\right)}
\label{eq:connection eta(s) and omega(s) IVa}\\
 & = & \exp\left(2\mathsf{ad}_{\Omega_{u}\left(s\right)}\right)\circ\left(-\omega_{g}\left(s\right)+\omega_{u}\left(s\right)\right)\nonumber
\end{eqnarray}
, equivalently
\begin{equation}
\omega_{g}\left(s\right)=\tanh\left(\mathsf{ad}_{\Omega_{u}\left(s\right)}\right)\circ\omega_{u}\left(s\right)\label{eq:connection eta(s) and omega(s) IVb}
\end{equation}
From this insight a useful relation connecting the generator $\eta\left(s\right)$
solely with the odd part $\omega_{u}\left(s\right)$ of the generator
$\omega\left(s\right)$ emerges in the guise
\begin{eqnarray}
-e^{\Omega_{g}\left(s\right)}\eta\left(s\right)e^{-\Omega_{g}\left(s\right)} & = & e^{-\Omega_{u}\left(s\right)}\omega\left(s\right)e^{\Omega_{u}\left(s\right)}\label{eq:connection eta(s) and omega(s) V}\\
 & = & \exp\left(-\mathsf{ad}_{\Omega_{u}\left(s\right)}\right)\circ\left(\omega_{g}\left(s\right)+\omega_{u}\left(s\right)\right)\nonumber \\
 & = & \exp\left(-\mathsf{ad}_{\Omega_{u}\left(s\right)}\right)\left(\tanh\left(\mathsf{ad}_{\Omega_{u}\left(s\right)}\right)+\mathsf{1}\right)\circ\omega_{u}\left(s\right)\nonumber \\
 & = & \frac{1}{\mathsf{\cosh}\left(\mathsf{ad}_{\Omega_{u}\left(s\right)}\right)}\circ\omega_{u}\left(s\right)\nonumber
\end{eqnarray}
Consequently the ODE (\ref{eq:ODE V(s)}) together with (\ref{eq:connection eta(s) and omega(s) II}) leads to
\begin{eqnarray}
\frac{d}{ds}\mathsf{V}\left(s\right) & = & \frac{d}{ds}\left(e^{\Omega_{u}\left(s\right)}e^{\Omega_{g}\left(s\right)}\right)\label{eq:ODE exp(Omega(s)) II}\\
 & = & \left(\frac{d}{ds}e^{\Omega_{u}\left(s\right)}\right)e^{\Omega_{g}\left(s\right)}+e^{\Omega_{u}\left(s\right)}\left(\frac{d}{ds}e^{\Omega_{g}\left(s\right)}\right)\nonumber \\
 & \overset{!}{=} & \omega\left(s\right)\left(e^{\Omega_{u}\left(s\right)}e^{\Omega_{g}\left(s\right)}\right)\nonumber \\
 & = & -e^{\Omega_{u}\left(s\right)}e^{\Omega_{g}\left(s\right)}\eta\left(s\right)\nonumber
\end{eqnarray}
Employing the well known formula for the derivative of an exponential
$e^{\Omega\left(s\right)}$ in case $\left[\Omega\left(s_{1}\right),\Omega\left(s_{2}\right)\right]\neq\mathsf{0}$
, for instance \cite{Wilcox},\cite{Iserles}, now
\begin{eqnarray}
\frac{d}{ds}e^{\Omega\left(s\right)} & = & \int_{0}^{1}d\tau e^{\tau\,\Omega\left(s\right)}\frac{d}{ds}\Omega\left(s\right)e^{-\tau\,\Omega\left(s\right)}e^{\Omega\left(s\right)}
\label{eq:derivative of an exponential}\\
 & = & \left(\int_{0}^{1}d\tau e^{\tau\,\mathsf{ad}_{\Omega\left(s\right)}}\circ\frac{d}{ds}\Omega\left(s\right)\right)e^{\Omega\left(s\right)}\nonumber \\
 & = & \left(\frac{\exp\left(\mathsf{ad}_{\Omega\left(s\right)}\right)-\mathsf{1}}{\mathsf{ad}_{\Omega\left(s\right)}}\circ\frac{d}{ds}\Omega\left(s\right)\right)e^{\Omega\left(s\right)}\nonumber
\end{eqnarray}
Up next the ODE (\ref{eq:ODE exp(Omega(s)) II}) along
with (\ref{eq:connection eta(s) and omega(s) V}) is readily shown to be equivalent to
\begin{eqnarray}
-e^{\Omega_{g}\left(s\right)}\eta\left(s\right)e^{-\Omega_{g}\left(s\right)} & = & \frac{1}{\mathsf{\cosh}\left(\mathsf{ad}_{\Omega_{u}\left(s\right)}\right)}\circ\omega_{u}\left(s\right)
\label{eq:ODE exp(Omega(s)) III}\\
 & = & \frac{\mathsf{1}-\exp\left(-\mathsf{ad}_{\Omega_{u}\left(s\right)}\right)}{\mathsf{ad}_{\Omega_{u}\left(s\right)}}\circ\frac{d}{ds}\Omega_{u}\left(s\right)+\frac{\exp\left(\mathsf{ad}_{\Omega_{g}\left(s\right)}\right)-\mathsf{1}}{\mathsf{ad}_{\Omega_{g}\left(s\right)}}\circ\frac{d}{ds}\Omega_{g}\left(s\right)\nonumber
\end{eqnarray}
The left hand side in (\ref{eq:ODE exp(Omega(s)) III})
being manifestly odd, the \emph{first} term on the right hand side
in (\ref{eq:ODE exp(Omega(s)) III}) decomposes into even and odd parts
\begin{eqnarray}
 &  & \frac{\mathsf{1}-\exp\left(-\mathsf{ad}_{\Omega_{u}\left(s\right)}\right)}{\mathsf{ad}_{\Omega_{u}\left(s\right)}}\circ\frac{d}{ds}\Omega_{u}\left(s\right)\label{eq:decomposition d_exp  u}\\
\nonumber \\
 & = & \left(\frac{\mathsf{1}-\cosh\left(\mathsf{ad}_{\Omega_{u}\left(s\right)}\right)}{\mathsf{ad}_{\Omega_{u}\left(s\right)}}+\frac{\sinh\left(\mathsf{ad}_{\Omega_{u}\left(s\right)}\right)}{\mathsf{ad}_{\Omega_{u}\left(s\right)}}\right)\circ\frac{d}{ds}\Omega_{u}\left(s\right)\nonumber \\
\nonumber \\
 & = & \underset{\textrm{even}}{\underbrace{\frac{\mathsf{1}-\cosh\left(\mathsf{ad}_{\Omega_{u}\left(s\right)}\right)}{\mathsf{ad}_{\Omega_{u}\left(s\right)}}\circ\frac{d}{ds}\Omega_{u}\left(s\right)}}+\underset{\textrm{odd}}{\underbrace{\frac{\sinh\left(\mathsf{ad}_{\Omega_{u}\left(s\right)}\right)}{\mathsf{ad}_{\Omega_{u}\left(s\right)}}\circ\frac{d}{ds}\Omega_{u}\left(s\right)}}\nonumber
\end{eqnarray}
, whereas the \emph{second} term on the right hand side in (\ref{eq:ODE exp(Omega(s)) III}) is manifestly even
\begin{eqnarray}
\underset{\textrm{even}}{\underbrace{\frac{\exp\left(\mathsf{ad}_{\Omega_{g}\left(s\right)}\right)-\mathsf{1}}{\mathsf{ad}_{\Omega_{g}\left(s\right)}}\circ\frac{d}{ds}\Omega_{g}\left(s\right)}} & = & \sum_{n=0}^{\infty}\frac{1}{\left(n+1\right)!}\mathsf{ad}_{\Omega_{g}\left(s\right)}^{n}\circ\frac{d}{ds}\Omega_{g}\left(s\right)\label{eq:decomposition d_exp  g-1}
\end{eqnarray}
A decomposition like (\ref{eq:decomposition d_exp  u}) applies,
because the commutator of an even operator with an even operator or
else the commutator of an odd operator with an odd operator always
results in an operator being \emph{even}, wheras the commutator of
an even operator with an odd operator and vice versa always results
in an operator being \emph{odd}.

Equating the even parts and the odd parts on either side of (\ref{eq:ODE exp(Omega(s)) III})
regarded separately one obtains instead of (\ref{eq:ODE exp(Omega(s)) III})
now two equations for $\frac{d}{ds}\Omega_{u}\left(s\right)$ and
$\frac{d}{ds}\Omega_{g}\left(s\right)$ :
\begin{eqnarray}
\frac{\sinh\left(\mathsf{ad}_{\Omega_{u}\left(s\right)}\right)}{\mathsf{ad}_{\Omega_{u}\left(s\right)}}\circ\frac{d}{ds}\Omega_{u}\left(s\right) & = & \frac{1}{\mathsf{\cosh}\left(\mathsf{ad}_{\Omega_{u}\left(s\right)}\right)}\circ\omega_{u}\left(s\right)\label{eq:ODE exp(Omega(s)) VI a}
\end{eqnarray}
\begin{equation}
\frac{\mathsf{1}-\cosh\left(\mathsf{ad}_{\Omega_{u}\left(s\right)}\right)}{\mathsf{ad}_{\Omega_{u}\left(s\right)}}\circ\frac{d}{ds}\Omega_{u}\left(s\right)+\frac{\exp\left(\mathsf{ad}_{\Omega_{g}\left(s\right)}\right)-\mathsf{1}}{\mathsf{ad}_{\Omega_{g}\left(s\right)}}\circ\frac{d}{ds}\Omega_{g}\left(s\right)=\mathsf{0}
\label{eq:ODE exp(Omega(s)) VIb}
\end{equation}
Obviously, the first equation (\ref{eq:ODE exp(Omega(s)) VI a})
directly determines $\frac{d}{ds}\Omega_{u}\left(s\right)$, as it
is\emph{ decoupled} from the equation for $\frac{d}{ds}\Omega_{g}\left(s\right)$,
whereas the determination of $\frac{d}{ds}\Omega_{g}\left(s\right)$
with the second equation (\ref{eq:ODE exp(Omega(s)) VIb})
requires prior knowledge of $\frac{d}{ds}\Omega_{u}\left(s\right)$.
Fortunately though, the determination of $\Omega_{g}\left(s\right)$
is for the purpose of determining the Newton-Wigner Hamiltonian $\mathsf{H}^{\left(NW\right)}$
redundant, as the representation (\ref{eq:representation H_NW   with Omega_u(infty) IV})
reveals at one glance.

Solving finally (\ref{eq:ODE exp(Omega(s)) VI a})
for $\frac{d}{ds}\Omega_{u}\left(s\right)$ an equivalent ODE determining
$\Omega_{u}\left(s\right)$ is obtained
\begin{eqnarray}
\frac{d}{ds}\Omega_{u}\left(s\right) & = & \frac{2\mathsf{ad}_{\Omega_{u}\left(s\right)}}{\sinh\left(2\mathsf{ad}_{\Omega_{u}\left(s\right)}\right)}\circ\omega_{u}\left(s\right)
\label{eq:ODE  exp(Omega_u_(s)) VII}\\
\Omega_{u}\left(0\right) & = & 0\nonumber
\end{eqnarray}
This being at first sight a scary nonlinear problem, in what follows
it proves otherwise, noting that integration of (\ref{eq:ODE  exp(Omega_u_(s)) VII})
with respect to $s$ leads to the integral equation
\begin{eqnarray}
\Omega_{u}\left(s\right) & = & \int_{0}^{s}ds'\frac{2\mathsf{ad}_{\Omega_{u}\left(s'\right)}}{\sinh\left(2\mathsf{ad}_{\Omega_{u}\left(s'\right)}\right)}\circ\omega_{u}\left(s'\right)\label{eq:IGL  Omega_u_(s) I}
\end{eqnarray}
Provided the odd part $\omega_{u}\left(s\right)$ of the generator
$\omega\left(s\right)$ of the beta-flow can be considered as being
small, then that integral equation is amenable to a perturbative solution
for the odd operator $\Omega_{u}\left(s\right)$ by the method of
Picard iteration. Clearly, the described perturbative solution method
is a close relative to the well known Magnus series expansion, for
a comprehensive review see \cite{Iserles}.

\section{Perturbation Series for Generator $\omega\left(s\right)$ of Beta-Flow}

Unfortunately, it is difficult to obtain for a general Dirac Hamiltonian
(\ref{eq:Dirac-Hamiltonian with weighted kinetic and potential energy terms})
an expansion of the generator
\begin{eqnarray}
\omega\left(s\right) & = & \left[\mathsf{\tilde{H}}^{\left(D\right)},\mathsf{Z}\left(s\right)\right]=\sum_{n=1}^{\infty}\kappa^{n}\omega^{\left(n\right)}\left(s\right)\label{eq:perturbation series omega(s)}
\end{eqnarray}
 with regard to the parameter $\kappa$, even though the exact solution
$\mathsf{Z}\left(s\right)$ of the beta-flow has been obtained in
(\ref{eq:exact solution Z(s)}).
To find the perturbation terms $\omega^{\left(n\right)}\left(s\right)$
in (\ref{eq:perturbation series omega(s)}) it is convenient to consider the operators
\begin{eqnarray}
\mathsf{Q}\left(s\right) & = & \mathsf{Z}\left(s\right)\mathsf{\tilde{H}}^{\left(D\right)}\label{eq:Definition Q(s)}\\
\mathsf{Q}^{\dagger}\left(s\right) & = & \mathsf{\tilde{H}}^{\left(D\right)}\mathsf{Z}\left(s\right)\nonumber
\end{eqnarray}
, so that
\begin{eqnarray}
\omega\left(s\right) & = & \mathsf{Q}^{\dagger}\left(s\right)-\mathsf{Q}\left(s\right)\label{eq:omega(s) = Q_ad_(s) - Q(s)}
\end{eqnarray}
From the beta-flow (\ref{eq:beta  flow III}) we readily confirm
that $\mathsf{Q}\left(s\right)$ solves the ODE
\begin{eqnarray}
\frac{1}{2}\frac{d}{ds}\mathsf{Q}\left(s\right) & = & \mathsf{\tilde{H}}^{\left(D\right)}\mathsf{\tilde{H}}^{\left(D\right)}-\mathsf{Q}\left(s\right)\mathsf{Q}\left(s\right)\label{eq:ODE  Q(s)}\\
\mathsf{Q}\left(0\right) & = & \beta\mathsf{\tilde{H}}^{\left(D\right)}= \mathsf{1}_{4\times4}+\kappa\beta\mathcal{\tilde{O}}+\kappa^{2}\beta\mathcal{\tilde{E}}\nonumber
\end{eqnarray}
Now a perturbation series expansion of the solution $\mathsf{Q}\left(s\right)$
to this ODE (\ref{eq:ODE  Q(s)}) is searched for in the guise
\begin{equation}
\mathsf{Q}\left(s\right)= \mathsf{1}_{4\times4}+\sum_{n=1}^{\infty}\kappa^{n}\mathsf{Q}^{\left(n\right)}\left(s\right)\label{eq:perturbation expansion Q(s)}
\end{equation}
Once the operators $\mathsf{Q}^{\left(n\right)}\left(s\right)$ are found, then
\begin{eqnarray}
\omega\left(s\right) & = & \sum_{n=1}^{\infty}\kappa^{n}\omega^{\left(n\right)}\left(s\right)\label{eq:perturbation series expansion generator omega(s)}\\
\omega^{\left(n\right)}\left(s\right) & = & \left(\mathsf{Q}^{\left(n\right)}\left(s\right)\right)^{\dagger}-\mathsf{Q}^{\left(n\right)}\left(s\right)\nonumber
\end{eqnarray}
Insertion of the ansatz (\ref{eq:perturbation expansion Q(s)}) into (\ref{eq:ODE  Q(s)}) gives
\[
\frac{d}{ds}\mathsf{Q}\left(s\right)=\kappa\frac{d}{ds}\mathsf{Q}^{\left(1\right)}\left(s\right)+\sum_{n=2}^{\infty}\kappa^{n}\frac{d}{ds}\mathsf{Q}^{\left(n\right)}\left(s\right)
\]
, and minding
\begin{eqnarray}
\mathcal{\tilde{O}}\beta & = & -\beta\mathcal{\tilde{O}}\label{eq:even-odd relations}\\
\mathcal{\tilde{E}}\beta & = & \beta\tilde{\mathcal{E}}\nonumber \\
\beta\beta & = &  \mathsf{1}_{4\times4}\nonumber
\end{eqnarray}
 the inhomogenous term in the ODE (\ref{eq:ODE  Q(s)}) reads
\begin{eqnarray}
\label{eq:H_D*H_D expansin kappa}\\
\mathsf{\tilde{H}}^{\left(D\right)}\mathsf{\tilde{H}}^{\left(D\right)} & = & \left(\beta+\kappa\tilde{\mathcal{O}}+\kappa^{2}\tilde{\mathcal{E}}\right)^{2}=\mathsf{1}_{4\times4}+\kappa^{2}\left(2\beta\tilde{\mathcal{E}}
+\mathcal{\tilde{O}}^{2}\right)+\kappa^{3}\left(\tilde{\mathcal{E}}\tilde{\mathcal{O}}+\tilde{\mathcal{O}}\tilde{\mathcal{E}}\right)+\kappa^{4}\tilde{\mathcal{E}}^{2}\nonumber \\
 & \equiv & \mathsf{1}_{4\times4}+\sum_{n=2}^{\infty}\kappa^{n}\mathsf{R}^{\left(n\right)}\nonumber \\
\mathsf{R}^{\left(2\right)} & = & 2\beta\tilde{\mathcal{E}}+\tilde{\mathcal{O}}^{2}\nonumber \\
\mathsf{R}^{\left(3\right)} & = & \mathcal{\tilde{E}}\tilde{\mathcal{O}}+\tilde{\mathcal{O}}\tilde{\mathcal{E}}\nonumber \\
\mathsf{R}^{\left(4\right)} & = & \mathcal{\tilde{E}}^{2}\nonumber \\
n & > & 4\nonumber \\
\mathsf{R}^{\left(n\right)} & = & \mathsf{0}_{4\times4}\nonumber
\end{eqnarray}
, where as the quadratic term in (\ref{eq:ODE  Q(s)}) assumes the guise
\begin{eqnarray}
\label{eq:series expansion Q(s)^2}\\
\mathsf{Q}\left(s\right)\mathsf{Q}\left(s\right) & = & \mathsf{1}_{4\times4}+2\kappa\mathsf{Q}^{\left(1\right)}\left(s\right)+\sum_{n=2}^{\infty}\kappa^{n}\left(2\mathsf{Q}^{\left(n\right)}\left(s\right)
+\sum_{j=1}^{n-1}\mathsf{Q}^{\left(j\right)}\left(s\right)\mathsf{Q}^{\left(n-j\right)}\left(s\right)\right)\nonumber
\end{eqnarray}
Comparing coefficients of $\kappa^{n}$ for $n=1,2,3,...$ on either
side of (\ref{eq:ODE  Q(s)}) then the following set of linear differential
equations for the determination of the operators $\mathsf{Q}^{\left(n\right)}\left(s\right)$ is obtained
\begin{eqnarray}
\frac{1}{2}\frac{d}{ds}\mathsf{Q}^{\left(1\right)}\left(s\right) & = & -2\mathsf{Q}^{\left(1\right)}\left(s\right)\nonumber \\
n & \geq & 2\label{eq:ODE  Q_n_(s)}\\
\frac{1}{2}\frac{d}{ds}\mathsf{Q}^{\left(n\right)}\left(s\right) & = & -2\mathsf{Q}^{\left(n\right)}\left(s\right)+\mathsf{R}^{\left(n\right)}-\sum_{j=1}^{n-1}\mathsf{Q}^{\left(j\right)}\left(s\right)\mathsf{Q}^{\left(n-j\right)}\left(s\right)\nonumber
\end{eqnarray}
To be consistent with the initial data posed at $s=0$ in (\ref{eq:ODE  Q(s)}) it is required
\begin{eqnarray}
\mathsf{Q}^{\left(1\right)}\left(0\right) & = & \beta\tilde{\mathcal{O}}\nonumber \\
\mathsf{Q}^{\left(2\right)}\left(0\right) & = & \beta\mathcal{\tilde{E}}\label{eq:initial data Q_n_(s=0)}\\
\textrm{for }n & > & 2\textrm{ }\nonumber \\
\mathsf{Q}^{\left(n\right)}\left(0\right) & = & \mathsf{0}_{4\times4}\nonumber
\end{eqnarray}
so that
\begin{eqnarray}
\mathsf{Q}\left(0\right) & = & \mathsf{1}_{4\times4}+\sum_{n=1}^{\infty}\kappa^{n}\mathsf{Q}^{\left(n\right)}\left(0\right)\label{eq:initial data Q(s=0)}\\
 & = & \mathsf{1}_{4\times4}+\kappa\beta\mathcal{\tilde{O}}+\kappa^{2}\beta\mathcal{\tilde{E}}\nonumber
\end{eqnarray}
The retained inhomogeneous linear differential equations (\ref{eq:ODE  Q_n_(s)})
are readily integrated and enable now a recursive determination of
the operators $\mathsf{Q}^{\left(n\right)}\left(s\right)$ as follows
\begin{eqnarray}
\mathsf{Q}^{\left(1\right)}\left(s\right) & = & e^{-4s}\beta\mathcal{O}\nonumber \\
n & \geq & 2\label{eq:recursion Q_n_(s)}\\
\mathsf{Q}^{\left(n\right)}\left(s\right) & = & e^{-4s}\mathsf{Q}^{\left(n\right)}\left(0\right)+2\int_{0}^{s}ds'e^{-4\left(s-s'\right)}\left(\mathsf{R}^{\left(n\right)}
-\sum_{j=1}^{n-1}\mathsf{Q}^{\left(j\right)}\left(s'\right)\mathsf{Q}^{\left(n-j\right)}\left(s'\right)\right)\nonumber
\end{eqnarray}
A straightforward analysis of this recursion up to and including the
terms of fourth order $\mathsf{Q}^{\left(4\right)}\left(s\right)$
leads on the basis of (\ref{eq:perturbation series expansion generator omega(s)}) to the following results
\begin{eqnarray}
\omega^{\left(1\right)}\left(s\right) & = & -2e^{-4s}\beta\mathcal{\tilde{O}}\label{eq:omega_1_(s)}\\
\omega^{\left(2\right)}\left(s\right) & = & \mathsf{0}_{4\times4}\label{eq:omega_2_(s)}\\
\omega^{\left(3\right)}\left(s\right) & = & -4e^{-4s}s\left[\mathcal{\tilde{O}},\mathcal{\tilde{E}}\right]+\left(e^{-4s}\left(-\frac{1}{2}+4s\right)+\frac{e^{-12s}}{2}\right)\beta\mathcal{\tilde{O}}^{3}\label{eq:omega_3_(s)}\\
\omega^{\left(4\right)}\left(s\right) & = & \left(e^{-4s}\left(-\frac{1}{2}+2s\right)+\frac{e^{-8s}}{2}\right)\beta\left[\mathcal{\tilde{O}},\left(\tilde{\mathcal{E}}\tilde{\mathcal{O}}+\tilde{\mathcal{O}}\tilde{\mathcal{E}}\right)\right]\label{eq:omega_4_(s)}
\end{eqnarray}
The result for $\omega^{\left(5\right)}\left(s\right)$ is available
in the complemental material \cite{Schopohl2022}.

Quite generally, the recursion (\ref{eq:recursion Q_n_(s)}) reveals
the even-numbered terms $\omega^{\left(2n\right)}\left(s\right)$
are \emph{even} operators whereas the odd-numbered terms $\omega^{\left(2n+1\right)}\left(s\right)$ are \emph{odd} operators:
\begin{eqnarray}
\beta\omega^{\left(2n\right)}\left(s\right)\beta & = & \omega^{\left(2n\right)}\left(s\right)\label{eq:omega_2n_(s)  even}\\
\beta\omega^{\left(2n+1\right)}\left(s\right)\beta & = & -\omega^{\left(2n+1\right)}\left(s\right)\label{eq:omega_(2n+1)_(s)  odd}
\end{eqnarray}
As a result of this the series expansion (\ref{eq:perturbation series omega(s)})
representing the generator $\omega\left(s\right)$ decomposes into
\begin{eqnarray}
\omega\left(s\right) & = & \omega_{g}\left(s\right)+\omega_{u}\left(s\right)\nonumber \\
\omega_{g}\left(s\right) & = & \sum_{n=2}^{\infty}\kappa^{2n}\omega^{\left(2n\right)}\left(s\right)\label{eq:Connection omega_g_(s) and omega_u_(s)}\\
\omega_{u}\left(s\right) & = & \sum_{n=0}^{\infty}\kappa^{2n+1}\omega^{\left(2n+1\right)}\left(s\right)\nonumber
\end{eqnarray}
Note that because $\omega^{\left(2\right)}\left(s\right)\equiv\mathsf{0}$
the series determining the even part $\omega_{g}\left(s\right)$ of
the generator $\omega\left(s\right)$ is small of order $O(\kappa^{4})$.

\section{Perturbation Series for operator $\Omega_{u}\left(s\right)$}

With
\begin{equation}
\omega_{u}\left(s\right)=\sum_{n=0}^{\infty}\kappa^{2n+1}\omega^{\left(2n+1\right)}\left(s\right)\label{eq:expansion generator omega_u_(s)}
\end{equation}
being odd it is natural adopting a corresponding approach for a perturbative
series expansion of $\Omega_{u}\left(s\right)$
\begin{equation}
\Omega_{u}\left(s\right)=\sum_{n=0}^{\infty}\kappa^{2n+1}\Omega^{\left(2n+1\right)}\left(s\right)\label{eq:Ansatz operator Omega_u_(s)}
\end{equation}
Writing
\begin{eqnarray}
\frac{2z}{\sinh\left(2z\right)} & = & 1-\frac{2}{3}z^{2}+\frac{14}{45}z^{4}-\frac{124}{945}z^{6}+O\left(z^{8}\right)\label{eq:power series V}
\end{eqnarray}
there follows in place of (\ref{eq:IGL  Omega_u_(s) I})
\begin{eqnarray}
\label{eq:IGL  Omega_u_(s) II}\\
\Omega_{u}\left(s\right) & = & \int_{0}^{s}ds'\left(1-\frac{2}{3}\left(\mathsf{ad}_{\Omega_{u}\left(s'\right)}\right)^{2}+\frac{14}{45}\left(\mathsf{ad}_{\Omega_{u}\left(s'\right)}\right)^{4}-\frac{124}{945}\left(\mathsf{ad}_{\Omega_{u}\left(s'\right)}\right)^{6}+O\left(\kappa^{8}\right)\right)\circ\omega_{u}\left(s'\right)\nonumber
\end{eqnarray}
Inserting the series expansions (\ref{eq:expansion generator omega_u_(s)})
and (\ref{eq:Ansatz operator Omega_u_(s)}) into (\ref{eq:IGL  Omega_u_(s) II})
one finds in a straightforward manner comparing coefficients of $\kappa^{2n+1}$on
either side of (\ref{eq:IGL  Omega_u_(s) II}) a recursion relation
determining the searched for terms $\Omega^{\left(2n+1\right)}\left(s\right)$.
Obviously the first order term being
\begin{equation}
\Omega^{\left(1\right)}\left(s\right)=\int_{0}^{s}ds'\omega^{\left(1\right)}\left(s'\right)=-\frac{1-e^{-4s}}{2}\beta\mathcal{\tilde{O}}\label{eq:Omega_1_(s)}
\end{equation}
, this immediately implies the vanishing of the commutator
\begin{equation}
\left[\Omega^{\left(1\right)}\left(s'\right),\omega^{\left(1\right)}\left(s'\right)\right]=\mathsf{0}\label{eq:Comutator(Omega_1, omega_1)=0}
\end{equation}
Consequently all the commutator terms $\left(\mathsf{ad}_{\Omega_{u}\left(s'\right)}\right)^{2n}\circ\omega_{u}\left(s'\right)$
in (\ref{eq:IGL  Omega_u_(s) II}) are at least small of order $O\left(\kappa^{2n+3}\right)$.
With that said the third order term is
\begin{eqnarray}
\Omega^{\left(3\right)}\left(s\right) & = & \int_{0}^{s}ds'\omega^{\left(3\right)}\left(s'\right)\label{eq:Omega_3_(s)}\\
 & = & \left(-\frac{1}{4}+e^{-4s}\left(\frac{1}{4}+s\right)\right)\left[\mathcal{\tilde{O}},\tilde{\mathcal{E}}\right]
 +\left(\frac{1}{6}-e^{-4s}\left(\frac{1}{8}+s\right)-\frac{e^{-12s}}{24}\right)\beta\mathcal{\tilde{O}}^{3}\nonumber
\end{eqnarray}
 , whereas the fifth-order term is then determined by
\begin{eqnarray}
\label{eq:Omega_5_(s)}\\
\Omega^{\left(5\right)}\left(s\right) & = & \begin{cases}
\int_{0}^{s}ds'\omega^{\left(5\right)}\left(s'\right)\\
\\
-\frac{2}{3}\int_{0}^{s}ds'\left(\left[\Omega^{\left(1\right)}\left(s'\right),\left[\Omega^{\left(3\right)}\left(s'\right),\omega^{\left(1\right)}\left(s'\right)\right]\right]+\left[\Omega^{\left(1\right)}\left(s'\right),\left[\Omega^{\left(1\right)}\left(s'\right),\omega^{\left(3\right)}\left(s'\right)\right]\right]\right)
\end{cases}\nonumber
\end{eqnarray}
An explicit evaluation of (\ref{eq:Omega_5_(s)}) is given in \cite{Schopohl2022}.
As we show in the ensuing, $\Omega^{\left(5\right)}\left(s\right)$
will appear in the calculations for the first time if the expansion
of $\mathsf{\tilde{H}}^{\left(D\right)}$ according to the lines indicated
in (\ref{eq:representation H_NW   with Omega_u(infty) IV}) aims
at an accuracy \emph{better} than $\kappa^{6}$.

Note we did not adress the issue of the \emph{convergence} of the
Magnus-type expansion (\ref{eq:Ansatz operator Omega_u_(s)}). Guided
by physical intuition, with operators $\tilde{\mathcal{E}}$ and $\mathcal{\tilde{O}}$
assigned to a Dirac Hamiltonian in external potentials like in (\ref{eq:Dirac-Hamiltonian}),
the described expansion method is conjectured being convergent for
all electric field strengths $\mathscr{E}=\left\Vert \frac{i}{\hbar}\left[\mathcal{\tilde{O}},\tilde{\mathcal{E}}\right]\right\Vert $
vastly below the Schwinger critical field $\mathscr{E}_{S}=\frac{m}{\left|e\right|}\frac{c^{2}}{\lambda_{C}}\simeq1.3\times10^{18}\left[\frac{V}{m}\right]$.
Perhaps one could find along the lines discussed in \cite{Iserles}
a sharp estimate for the radius of convergence of that expansion (\ref{eq:Ansatz operator Omega_u_(s)}), establishing that way, for instance, a stability criterion
for the existence of the Newton-Wigner Hamiltonian of the relativistic electron in the presence of strong electrostatic fields.

\section{The relativistic corrections to the Schr{\"o}dinger Pauli Hamiltonian $\mathsf{H}^{\left(SP\right)}$ as a series progressing in powers of $\frac{v^{2}}{c^{2}}$ } \label{sec:Perturbative H_NW Hamiltonian static}

The remaining task is now to evaluate the afore given\emph{ explicit}
representation (\ref{eq:representation H_NW   with Omega_u(infty) IV})
for the Newton-Wigner Hamiltonian as a series of iterated commutators
by inserting the obtained series expansion for $\Omega_{u}\left(s\right)$
in (\ref{eq:Ansatz operator Omega_u_(s)}) and to find that way the
searched for perturbative terms $\Omega^{\left(2n+1\right)}\left(s\right)$.
Let us agree in the ensuing on abbreviating the limiting value of
the operator $\Omega_{u}\left(s\right)$ for $s\rightarrow\infty$ as
\begin{equation}
\Omega_{u}\left(\infty\right)\equiv\Omega_{u}=\kappa\Omega^{\left(1\right)}+\kappa^{3}\Omega^{\left(3\right)}+\kappa^{5}\Omega^{\left(5\right)}+...\label{eq:perturbation series Omega_u_(infty)}
\end{equation}
, whereby according to (\ref{eq:Omega_1_(s)}), (\ref{eq:Omega_3_(s)}),
(\ref{eq:Omega_5_(s)}) the terms $\Omega^{\left(2n+1\right)}\equiv\Omega^{\left(2n+1\right)}\left(\infty\right)$
are given by
\begin{eqnarray}
\Omega^{\left(1\right)} & = & -\frac{1}{2}\beta\mathcal{\tilde{O}}\label{eq:perturbation terms Omega_(2n+1)}\\
\nonumber \\
\Omega^{\left(3\right)} & = & \frac{1}{6}\beta\mathcal{\tilde{O}}^{3}-\frac{1}{4}\left[\mathcal{\tilde{O}},\tilde{\mathcal{E}}\right]\nonumber \\
\nonumber \\
\Omega^{\left(5\right)} & = & -\frac{1}{10}\beta\mathcal{\tilde{O}}^{5}+\frac{1}{9}\left[\tilde{\mathcal{O}}^{3},\tilde{\mathcal{E}}\right]
+\frac{5}{144}\left[\tilde{\mathcal{O}},\left[\tilde{\mathcal{O}},\left[\tilde{\mathcal{O}},\tilde{\mathcal{E}}\right]\right]\right]
-\frac{1}{8}\beta\left[\left[\tilde{\mathcal{O}},\tilde{\mathcal{E}}\right],\tilde{\mathcal{E}}\right]\nonumber
\end{eqnarray}
, and so on.

Restricting the expansion of $\mathsf{\tilde{H}}^{\left(NW\right)}$ to accuracy
$O\left(\kappa^{8}\right)$ it follows directly from (\ref{eq:representation H_NW   with Omega_u(infty) IV})
\begin{eqnarray}
\label{eq:series expansion  H_NW  I}\\
\mathsf{\tilde{H}}^{\left(NW\right)} & = & \begin{cases}
\left(\beta+\kappa^{2}\tilde{\mathcal{E}}\right)\\
\\
+\frac{1}{2}\left(\textrm{ad}_{\Omega_{u}}\right)^{2}\circ\left(\beta+\kappa^{2}\tilde{\mathcal{E}}\right)
+\frac{1}{24}\left(\textrm{ad}_{\Omega_{u}}\right)^{4}\circ\left(\beta+\kappa^{2}\tilde{\mathcal{E}}\right)
+\frac{1}{720}\left(\textrm{ad}_{\Omega_{u}}\right)^{6}\circ\left(\beta+\kappa^{2}\tilde{\mathcal{E}}\right)\\
\\
-\left(\textrm{ad}_{\Omega_{u}}\right)\circ\kappa\mathcal{\tilde{O}}
-\frac{1}{6}\left(\textrm{ad}_{\Omega_{u}}\right)^{3}\circ\kappa\mathcal{\tilde{O}}
-\frac{1}{120}\left(\textrm{ad}_{\Omega_{u}}\right)^{5}\circ\kappa\mathcal{\tilde{O}}\\
\\
+O\left(\kappa^{8}\right)
\end{cases}\nonumber
\end{eqnarray}
At first sight the indicated accuracy $O\left(\kappa^{8}\right)$
holds true with the operator $\Omega_{u}$ being expanded up to \emph{and}
including the fifth order term $\Omega^{\left(5\right)}$, because with the expansion
\begin{equation}
\textrm{ad}_{\Omega_{u}}\circ\mathcal{F}=\sum_{j=0}^{\infty}\kappa^{2j+1}\textrm{ad}_{\Omega^{\left(2j+1\right)}}\circ\mathcal{F}\label{eq:expansion ad_(Omega_u)}
\end{equation}
then
\begin{eqnarray}
\left(\textrm{ad}_{\Omega_{u}}\right)^{6}\circ\beta & = & \kappa^{6}\left(\textrm{ad}_{\Omega^{\left(1\right)}}\right)^{6}\circ\beta+O\left(\kappa^{8}\right)\label{eq:expansion ad_(Omega_u) II}\\
\left(\textrm{ad}_{\Omega_{u}\left(\infty\right)}\right)^{5}\circ\left(\kappa\mathcal{\tilde{O}}\right) & = & \kappa^{6}\left(\textrm{ad}_{\Omega^{\left(1\right)}}\right)^{5}\circ\tilde{O}+O\left(\kappa^{8}\right)\nonumber \\
\left(\textrm{ad}_{\Omega_{u}}\right)^{4}\circ\left(\kappa^{2}\tilde{\mathcal{E}}\right) & = & \kappa^{6}\left(\textrm{ad}_{\Omega^{\left(1\right)}}\right)^{4}\circ\tilde{\mathcal{E}}+O\left(\kappa^{8}\right)\nonumber
\end{eqnarray}
Yet the commutator terms at order $\kappa^{2n}$ in the expansion
(\ref{eq:series expansion  H_NW  I}) involving the operators $\Omega^{\left(2n-1\right)}$
\emph{cancel} for $n\geq2$ , because of the identity
\begin{equation}
\frac{1}{2}\textrm{ad}_{\Omega^{\left(1\right)}}\circ\textrm{ad}_{\Omega^{\left(2n-1\right)}}\circ\beta+\frac{1}{2}\textrm{ad}_{\Omega^{\left(2n-1\right)}}\circ\textrm{ad}_{\Omega^{\left(1\right)}}\circ\beta=\textrm{ad}_{\Omega^{\left(2n-1\right)}}\circ\mathcal{\tilde{O}}
\label{eq:cancel Omega_(2n+1) identity}
\end{equation}
This is fortunate, as it implies, that in order to achieve accuracy
$O\left(\kappa^{2n}\right)$ for $n\geq2$ only the operators $\Omega^{\left(1\right)},\Omega^{\left(3\right)},...,\Omega^{\left(2n-3\right)}$are
required! So we obtain from (\ref{eq:series expansion  H_NW  I})
now the following expansion for the (scaled) Newton-Wigner Hamiltonian
in the guise (here we set $\kappa=1$, as no book keeping is required anymore)
\begin{eqnarray}
\mathsf{\tilde{H}}^{\left(NW\right)} & = & \beta+\tilde{\mathsf{h}}^{\left(2\right)}+\tilde{\mathsf{h}}^{\left(4\right)}+\tilde{\mathsf{h}}^{\left(6\right)}+...
\label{eq:series expansion  H_NW  II}
\end{eqnarray}
whereas
\begin{eqnarray}
\tilde{\mathsf{h}}^{\left(2\right)} & = & \tilde{\mathcal{E}}+\frac{1}{2}\left[\Omega^{\left(1\right)},\left[\Omega^{\left(1\right)},\beta\right]\right]-\left[\Omega^{\left(1\right)},\mathcal{\tilde{O}}\right]
\label{eq:h_2  I}\\
\nonumber \\
\tilde{\mathsf{h}}^{\left(4\right)} & = & \left(\begin{array}{c}
+\frac{1}{24}\left[\Omega^{\left(1\right)},\left[\Omega^{\left(1\right)},\left[\Omega^{\left(1\right)},\left[\Omega^{\left(1\right)},\mathcal{\beta}\right]\right]\right]\right]\\
+\frac{1}{2}\left[\Omega^{\left(1\right)},\left[\Omega^{\left(1\right)},\tilde{\mathcal{E}}\right]\right]\\
-\frac{1}{6}\left[\Omega^{\left(1\right)},\left[\Omega^{\left(1\right)},\left[\Omega^{\left(1\right)},\mathcal{\tilde{O}}\right]\right]\right]
\end{array}\right)\label{eq:h_4  I}
\end{eqnarray}
\begin{eqnarray}
\label{eq:h_6  I}\\
\tilde{\mathsf{h}}^{\left(6\right)} & = &
 \begin{cases}
\frac{1}{2}\left[\Omega^{\left(3\right)},\left[\Omega^{\left(3\right)},\beta\right]\right]\\
+\frac{1}{2}\left[\Omega^{\left(1\right)},\left[\Omega^{\left(3\right)},\tilde{\mathcal{E}}\right]\right]+\frac{1}{2}\left[\Omega^{\left(3\right)},\left[\Omega^{\left(1\right)},\tilde{\mathcal{E}}\right]\right]\\
\\
+\frac{1}{24}\left[\Omega^{\left(1\right)},\left[\Omega^{\left(1\right)},\left[\Omega^{\left(1\right)},\left[\Omega^{\left(3\right)},\beta\right]\right]\right]\right]+\frac{1}{24}\left[\Omega^{\left(1\right)},\left[\Omega^{\left(1\right)},\left[\Omega^{\left(3\right)},\left[\Omega^{\left(1\right)},\beta\right]\right]\right]\right]\\
+\frac{1}{24}\left[\Omega^{\left(1\right)},\left[\Omega^{\left(3\right)},\left[\Omega^{\left(1\right)},\left[\Omega^{\left(1\right)},\beta\right]\right]\right]\right]+\frac{1}{24}\left[\Omega^{\left(3\right)},\left[\Omega^{\left(1\right)},\left[\Omega^{\left(1\right)},\left[\Omega^{\left(1\right)},\beta\right]\right]\right]\right]\\
+\frac{1}{24}\left[\Omega^{\left(1\right)},\left[\Omega^{\left(1\right)},\left[\Omega^{\left(1\right)},\left[\Omega^{\left(1\right)},\tilde{\mathcal{E}}\right]\right]\right]\right]\\
+\frac{1}{720}\left[\Omega^{\left(1\right)},\left[\Omega^{\left(1\right)},\left[\Omega^{\left(1\right)},\left[\Omega^{\left(1\right)},\left[\Omega^{\left(1\right)},\left[\Omega^{\left(1\right)},\beta\right]\right]\right]\right]\right]\right]\\
\\
-\frac{1}{6}\left[\Omega^{\left(1\right)},\left[\Omega^{\left(1\right)},\left[\Omega^{\left(3\right)},\mathcal{\tilde{O}}\right]\right]\right]-\frac{1}{6}\left[\Omega^{\left(1\right)},\left[\Omega^{\left(3\right)},\left[\Omega^{\left(1\right)},\mathcal{\tilde{O}}\right]\right]\right]-\frac{1}{6}\left[\Omega^{\left(3\right)},\left[\Omega^{\left(1\right)},\left[\Omega^{\left(1\right)},\mathcal{\tilde{O}}\right]\right]\right]\\
-\frac{1}{120}\left[\Omega^{\left(1\right)},\left[\Omega^{\left(1\right)},\left[\Omega^{\left(1\right)},\left[\Omega^{\left(1\right)},\left[\Omega^{\left(1\right)},\mathcal{\tilde{O}}\right]\right]\right]\right]\right]
\end{cases}
\nonumber
\end{eqnarray}
, and so on.

A straightforward analysis \cite{Schopohl2022} of these expression leads to
\begin{eqnarray}
\tilde{\mathsf{h}}^{\left(2\right)} & = & \tilde{\mathcal{E}}+\frac{1}{2}\beta\mathcal{\tilde{O}}^{2}\label{eq:h_2  II}\\
\nonumber \\
\tilde{\mathsf{h}}^{\left(4\right)} & = & -\frac{1}{8}\beta\mathcal{\tilde{O}}^{4}-\frac{1}{8}\left[\mathcal{\mathcal{\tilde{O}}},\left[\mathcal{\tilde{O}},\tilde{\mathcal{E}}\right]\right]\label{eq:h_4  II}\\
\nonumber
\end{eqnarray}
\begin{eqnarray}
\tilde{\mathsf{h}}^{\left(6\right)} & = &
 \begin{cases}
\frac{1}{16}\beta\mathcal{\tilde{O}}^{6}\\
\\
\frac{1}{32}\left[\mathcal{\tilde{O}}^{3},\left[\mathcal{\tilde{O}},\tilde{\mathcal{E}}\right]\right]
+\frac{1}{64}\left[\mathcal{\mathcal{\tilde{O}}},\left(\mathcal{O}^{2}\left[\tilde{\mathcal{O}},\mathcal{\tilde{E}}\right]+\left[\tilde{\mathcal{O}},\mathcal{\tilde{E}}\right]\mathcal{O}^{2}\right)\right]\\
+\frac{1}{128}\left[\mathcal{\mathcal{\tilde{O}}},\left[\mathcal{\mathcal{\tilde{O}}},\left[\mathcal{\mathcal{\tilde{O}}},\left[\tilde{\mathcal{O}},\mathcal{\tilde{E}}\right]\right]\right]\right]\\
\\
+\frac{1}{16}\beta\left(\mathcal{\mathcal{\tilde{O}}}\left[\left[\tilde{\mathcal{O}},\mathcal{\tilde{E}}\right],\tilde{\mathcal{E}}\right]
+\left[\left[\tilde{\mathcal{O}},\mathcal{\tilde{E}}\right],\tilde{\mathcal{E}}\right]\mathcal{\tilde{O}}\right)
\end{cases}\label{eq:h_6  II}
\end{eqnarray}
The term $\tilde{\mathsf{h}}^{\left(2\right)}$ in (\ref{eq:h_2  II})
is the Hamiltonian of Schr{\"o}dinger Pauli quantum mechanics (scaled
units). The term $\tilde{\mathsf{h}}^{\left(4\right)}$ in (\ref{eq:h_4  II})
coincides with the standard outcome for the relativistic correction
to the kinectic energy, the Darwin term and the spin-orbit
interaction term with the (static) external electric field, in agreement
with the (laborious) step-by-step FW-transformation method \cite{FoldyW50}.

Beyond order $\kappa^{4}$ results obtained by the FW-transformation method
are not energy-separating. This has been realized already early on by
Eriksen and Kolsrud \cite{EriksenKolsrud}.  An additional unitary
transformation is required to generate suitable correction terms,
so that the results obtained then coincide with results obtained by
the Eriksen method \cite{Eriksen}. A general scheme that provides for the
original FW-transformation to every order $\kappa^{2n}$ the required correction terms,
so that it coincides with the result of the (tedious) Eriksen transformation method expanded to that same order $\kappa^{2n}$,
 has been developed by Silenko \cite{Silenko2016}.

It should be emphasized that our result for $\tilde{\mathsf{h}}^{\left(6\right)}$ in (\ref{eq:h_6  II}) is by construction energy-separating.
The result is in accord with results obtained by Silenko's correction scheme \cite{Silenko2016} applied to the original FW-transformation,
and it also coincides with the table
provided by deVries and Jonker \cite{Vries1968}, who obtained (though in a less practicable guise)
their expansion in powers of $\kappa=\frac{v}{c}$ with the Pauli-Achieser-Berestezki elimination method
(using computer algebra) and provided a proof of equivalence of their approach with the unitary transformation method of Eriksen.

The introduced expansion method, see (\ref{eq:representation H_NW   with Omega_u(infty) IV})
together with (\ref{eq:IGL  Omega_u_(s) I}), constitutes the central
result of this article. It fully implements a convenient energy-separating scheme to
reconstruct the expansion in powers of $\frac{v}{c}$ for the Newton-Wigner Hamiltonian
 of the relativistic electron moving in static external potentials.

\section{Hamiltonian Flow Equations for Time-Dependent Electromagnetic Fields}\label{sec:Time-Dependent-Hamiltonian}

We consider now a (scaled) \emph{time-dependent} Dirac Hamiltonian, composed of even and odd parts
\begin{equation}
\mathsf{\tilde{H}}^{\left(D\right)}\left(t\right)=\beta+\mathcal{\tilde{E}}\left(t\right)+\mathcal{\tilde{O}}\left(t\right)\label{eq:time dependent Dirac Hamiltonian I}
\end{equation}
, agreeing in what follows to choose $\beta$ and $\alpha_{b}$ in Dirac-Pauli representation, always minding $\beta$ being then a diagonal matrix (for details and a discussion of other unitary equivalent representations see the useful appendix A-2 in ref. \cite{IzyksonZuber}).
For instance, taking into account external (c-number valued) electromagnetic
fields with prescribed (parametric) time-dependence,
\begin{eqnarray}
\mathscr{E}_{b}\left(\mathbf{r},t\right) & = & -\frac{\partial\Phi\left(\mathbf{r},t\right)}{\partial r_{b}}-\frac{\partial\mathcal{A}_{b}\left(\mathbf{r},t\right)}{\partial t}\label{eq:EM-fields time-dependent}\\
\mathscr{B}_{a}\left(r,t\right) & = & \varepsilon_{abc}\frac{\partial}{\partial r_{b}}\mathcal{A}_{c}\left(\mathbf{r},t\right)\nonumber
\end{eqnarray}
, then the even and odd operators in (\ref{eq:time dependent Dirac Hamiltonian I}) are, respectively
\begin{eqnarray}
\mathcal{\tilde{E}}\left(t\right) & = & \frac{q_{e}}{mc^{2}}\Phi\left(\mathsf{x},t\right)\mathsf{1}_{4\times4}\label{eq:operator E}\\
\mathcal{\tilde{O}}\left(t\right) & = & \alpha_{b}\frac{\Pi_{b}}{mc}\label{eq:operator O}\\
\Pi_{b}& \equiv & \Pi_{b}\left(\mathsf{p},\mathsf{x},t\right)  =  \mathsf{p}_{b}-q_{e}\mathcal{A}_{b}\left(\mathsf{x},t\right) \nonumber
\end{eqnarray}

In the ensuing a flow equation based scheme is searched for to find
a time-dependent unitary transformation $\mathsf{U}\left(t,s\right)$
of the four-component Dirac Amplitude $\Psi_{\mu}^{\left(D\right)}\left(\mathbf{r},t\right)$
that strives for $s\rightarrow\infty$ to an amplitude
\[
\Psi_{\mu}^{\left(U\right)}\left(\mathbf{r},t\right)=\mathsf{U}_{\mu,\mu'}\left(t,\infty\right)\Psi_{\mu'}^{\left(D\right)}\left(\mathbf{r},t\right)
\]
in the guise (\ref{eq:Newton-Wigner amplitude III}), so that also
in the presence of spatiotemporal electromagnetic fields \emph{separate}
equations of motion govern the time development of the two-component
amplitudes $\psi_{\sigma}\left(\mathbf{r},t\right)$ and $\chi_{\sigma}\left(\mathbf{r},t\right)$
for \emph{all }times.

Specifically, if  at time $t=0$  there holds $\Psi_{\mu}^{\left(U\right)}\left(\mathbf{r},0\right) = 0$ for $\mu=3,4$
then at  \emph{all}   later times $t>0$ it ought to be as well $\Psi_{\mu}^{\left(U\right)}\left(\mathbf{r},t\right) = 0$  for $\mu =3,4$.
Ditto, if  at time $t=0$  there holds $\Psi_{\mu}^{\left(U\right)}\left(\mathbf{r},0\right) = 0$ for $\mu=1,2$
 then at \emph{all}  later times $t>0$ it ought to be as well $\Psi_{\mu}^{\left(U\right)}\left(\mathbf{r},t\right) = 0$  for $\mu =1,2$.

This not being the only distinctive feature of the unitary transformation
$\mathsf{U}\left(t,\infty\right)$ it is concurrently required that
returning to the special case of \emph{static} fields the unitary
transformation $\mathsf{U}\left(t,s\right)$ should coincide with
the afore obtained unitary transformation (\ref{eq:identification V(s)=U(s)*Beta}), the latter converging for $s\to\infty$  to the unitary transformation $\mathsf{T}$ in (\ref{eq:U(s->infty) II}) equivalent to the Eriksen transformation.

Now choosing as initial value of the flow at $s=0$ the (scaled) time
dependent Dirac Hamiltonian $\mathsf{\tilde{H}}^{\left(D\right)}\left(t\right)$
defined in (\ref{eq:time dependent Dirac Hamiltonian I}), a unitary
transformed time-dependent Hamiltonian $\mathsf{\tilde{H}}^{\left(U\right)}\left(s,t\right)$ arises
, that governs the time evolution of the unitary transformed Dirac amplitude
\[
\Psi_{\mu}^{\left(U\right)}\left(\mathbf{r},t;s\right)=\mathsf{U}_{\mu,\mu'}\left(t,s\right)\Psi_{\mu'}^{\left(D\right)}\left(\mathbf{r},t;s\right)
\]
 at fixed value $s$ according to
\begin{eqnarray}
\mathsf{\tilde{H}}^{\left(U\right)}\left(t,s\right) & = & \mathsf{U}\left(t,s\right)\circ\left(\mathsf{\tilde{H}}^{\left(D\right)}\left(t\right)-i\hat{\partial}_{t}\right)\circ\mathsf{U}^{\dagger}\left(t,s\right)+i\hat{\partial}_{t}\label{eq:definition H_(U)}
\end{eqnarray}
, so that instead of
\begin{eqnarray}
i\frac{\hbar}{mc^{2}}\frac{\partial}{\partial t}\Psi_{\mu}^{\left(D\right)}\left(\mathbf{r},t\right) & = & \mathsf{\tilde{H}}_{\mu,\mu'}^{\left(D\right)}\left(t\right)\Psi_{\mu'}^{\left(D\right)}\left(\mathbf{r},t\right)\label{eq:time development Dirac amplitude}
\end{eqnarray}
 we have now
\begin{eqnarray}
i\frac{\hbar}{mc^{2}}\frac{\partial}{\partial t}\Psi_{\mu}^{\left(U\right)}\left(\mathbf{r},t;s\right) & = & \mathsf{\tilde{H}}_{\mu,\mu'}^{\left(U\right)}\left(t,s\right)\Psi_{\mu'}^{\left(U\right)}\left(\mathbf{r},t;s\right)
\label{eq:time development unitary transformed amplitude}\\
\Psi_{\mu}^{\left(U\right)}\left(\mathbf{r},t;0\right) & = & \Psi_{\mu}^{\left(D\right)}\left(\mathbf{r},t\right)\nonumber
\end{eqnarray}
For convenience here we introduced the symbol $\hat{\partial}_{t}$
to denote a $\emph{scaled}$ time derivative \emph{operator} so that
\begin{equation}
\left[\hat{\partial}_{t}\:,\mathsf{U}^{\dagger}\left(t,s\right)\right]=\frac{\hbar}{mc^{2}}\frac{\partial\mathsf{U}^{\dagger}\left(t,s\right)}{\partial t}\label{eq:derivative of function}
\end{equation}
Note the unitary transformation $\mathsf{U}\left(t,s\right)$ depends
on the flow parameter $s$ \emph{and} on time $t$ in consequence
of the time dependence of the initial data of the flow at $s=0$ given by
\begin{eqnarray}
\mathsf{U}\left(t,0\right) & = &  \mathsf{1}_{4\times4}\label{eq:initial data for the time dependent Hamiltonian flow}\\
\mathsf{\tilde{H}}^{\left(U\right)}\left(t,0\right) & = & \mathsf{\tilde{H}}^{\left(D\right)}\left(t\right)\nonumber
\end{eqnarray}
Because the time derivative \emph{operator} $i\hat{\partial}_{t}$
in the Dirac equation (\ref{eq:time development Dirac amplitude})
necessarily gets affected too by any time-dependent unitary transformation, as emphasized in
\cite{FoldyW50}, at first sight the method(s) described in the previous
sections for static external fields are not likely to apply.

Progress comes introducing the \emph{operator}
\begin{equation}
\mathsf{K}\left(t,s\right)\equiv\mathsf{\tilde{H}}^{\left(U\right)}\left(t,s\right)-i\hat{\partial}_{t}
=\mathsf{U}\left(t,s\right)\left(\mathsf{\tilde{H}}^{\left(D\right)}\left(t\right)-i\hat{\partial}_{t}\right)\mathsf{U}^{\dagger}\left(t,s\right)
\label{eq:operator K(s,t) I}
\end{equation}
Given a suitable generator $\eta\left(t,s\right)=-\eta^{\dagger}\left(t,s\right)$
of the Hamiltonian flow then the associated unitary transformation is determined by
\begin{eqnarray}
\frac{\partial}{\partial s}\mathsf{U}\left(t,s\right) & = & \eta\left(t,s\right)\mathsf{U}\left(t,s\right)\label{eq:unitary transformation with generator eta}\\
\mathsf{U}\left(t,0\right) & = & \mathsf{1}_{4\times4}\nonumber
\end{eqnarray}
, and with that said the ODE determining $\mathsf{K}\left(t,s\right)$
and taking into account the initial data (\ref{eq:initial data for the time dependent Hamiltonian flow}) reads
\begin{eqnarray}
\frac{\partial}{\partial s}\mathsf{K}\left(t,s\right) & = & \left[\eta\left(t,s\right),\mathsf{K}\left(t,s\right)\right]\label{eq:flow equation operator K(s,t)  I}\\
\mathsf{K}\left(t,0\right) & = & \mathsf{\tilde{H}}^{\left(D\right)}\left(t\right)-i\hat{\partial}_{t}\nonumber
\end{eqnarray}

A \emph{specific} flow equation striving towards a \emph{time-dependent}
Newton-Wigner representation,  that we now introduce in full analogy
to the previous (static) flow equation approach (\ref{eq:flow H(s)  III})
, employs as a time-dependent generator
\begin{equation}
\eta\left(t,s\right)=\left[\beta,\mathsf{K}\left(t,s\right)\right]\label{eq:time-dependent BP-generator}
\end{equation}
The proof of the wanted property $\mathsf{K}\left(t,\infty\right)$ being even (block-diagonal), i.e.
\begin{equation}
\lim_{s\rightarrow\infty}\eta\left(t,s\right)=\mathsf{0}_{4\times4}\label{eq:vanishing of generator eta_(s,t) for s->infty}
\end{equation}
, is readily transferred from it's static prefiguration adjusting
the functional (\ref{eq:functional Phi(s)}) and it's derivative
(\ref{eq:functional Phi(s) derivative}) to the time-dependent case.

Combining now the generator (\ref{eq:time-dependent BP-generator})
with (\ref{eq:flow equation operator K(s,t)  I}) thereby emerges
the time-dependent flow equation
\begin{eqnarray}
\frac{\partial}{\partial s}\mathsf{K}\left(t,s\right) & = & \left[\left[\beta,\mathsf{K}\left(t,s\right)\right],\mathsf{K}\left(t,s\right)\right]\label{eq:flow equation operator K(t,s)  II}\\
\mathsf{K}\left(t,0\right) & = & \beta+\mathcal{\tilde{O}}\left(t\right)+\mathcal{\tilde{E}}\left(t\right)-i\hat{\partial}_{t}\nonumber
\end{eqnarray}
, with the limiting value $\mathsf{K}\left(t,\infty\right)$ of that
flow for $s\rightarrow\infty$ by construction being even, i.e. $\left[\beta,\mathsf{K}\left(t,\infty\right)\right]=\mathsf{0}_{4\times4}$.

Introducing a splitting of the operator $\mathsf{K}\left(t,s\right)$
into an \emph{even} operator $\mathsf{K}_{g}\left(t,s\right)$ and
an \emph{odd }operator\emph{ }$\mathsf{K}_{u}\left(t,s\right)$,
\begin{eqnarray}
\mathsf{K}\left(t,s\right) & = & \frac{1}{2}\left(\mathsf{K}\left(t,s\right)+\beta\mathsf{K}\left(t,s\right)\beta\right)+\frac{1}{2}\left(\mathsf{K}\left(t,s\right)-\beta\,\mathsf{K}\left(t,s\right)\beta\right)
\label{eq:splitting of operator K(t,s)}\\
 & \equiv & \mathsf{K}_{g}\left(t,s\right)+\mathsf{K}_{u}\left(t,s\right)\nonumber \\
\beta\mathsf{K}_{g}\left(t,s\right) & = & \mathsf{K}_{g}\left(t,s\right)\beta\nonumber \\
\beta\mathsf{K}_{u}\left(t,s\right) & = & -\mathsf{K}_{u}\left(t,s\right)\beta\nonumber
\end{eqnarray}
, the generator $\eta\left(t,s\right)$ of the flow is indeed the
generalization of the time-independent generator used by Bylev and
Pirner \cite{Bylev1998} to the time-dependent case:
\begin{equation}
\eta\left(t,s\right)=\left[\beta,\mathsf{K}\left(t,s\right)\right]=2\beta\mathsf{K}_{u}\left(t,s\right)\label{eq:time-dependent  generator of flow}
\end{equation}
Consequently the flow equation (\ref{eq:flow equation operator K(t,s)  II})
determining the operator $\mathsf{K}\left(t,s\right)$ is equivalent
to \emph{two} coupled equations determining $\mathsf{K}_{g}\left(t,s\right)$
and $\mathsf{K}_{u}\left(t,s\right)$ :
\begin{eqnarray}
\label{eq:flow equation  K_u and K_g}
\frac{d}{ds}\mathsf{K}_{g}\left(t,s\right) & = & \left[\left[\beta,\mathsf{K}_{u}\left(t,s\right)\right],\mathsf{K}_{u}\left(t,s\right)\right]=4\beta\mathsf{K}_{u}\left(t,s\right)\mathsf{K}_{u}\left(t,s\right) \\
\frac{d}{ds}\mathsf{K}_{u}\left(t,s\right) & = & \left[\left[\beta,\mathsf{K}_{u}\left(t,s\right)\right],\mathsf{K}_{g}\left(t,s\right)\right]=2\beta\left[\mathsf{K}_{u}\left(t,s\right),\mathsf{K}_{g}\left(t,s\right)\right]\nonumber
\end{eqnarray}
, with the associated initial data at $s=0$ being
\begin{eqnarray}
\mathsf{K}_{g}\left(t,0\right) & = & \beta+\mathcal{\tilde{E}}\left(t\right)-i\hat{\partial}_{t}\label{eq:initial data E(t,0) and O(t,0)}\\
\mathsf{K}_{u}\left(t,0\right) & = & \mathcal{\tilde{O}}\left(t\right)\nonumber
\end{eqnarray}
In general we don't expect to find an exact solution to (\ref{eq:flow equation operator K(t,s)  II}).
But if we restrict to the \emph{nonrelativistic} sector and to weak
field strengths (far below the Schwinger critical field) with slow
time-dependence compared with the fast time scale set by $\frac{mc^{2}}{\hbar}$,
a series expansion of the sought solutions $\mathsf{K}_{g}\left(t,s\right)$
and $\mathsf{K}_{u}\left(t,s\right)$ progressing in powers of the
small parameter $\kappa=\frac{v}{c}$ is adequate:
\begin{eqnarray}
\mathsf{K}_{g}\left(t,s\right) & = & \beta+\sum_{j=1}^{\infty}\kappa^{2j}\mathsf{K}^{\left(2j\right)}\left(t,s\right)\label{eq:perturbation series expansion  E(t,s) and O(t,s)}\\
\mathsf{K}_{u}\left(t,s\right) & = & \sum_{j=0}^{\infty}\kappa^{2j+1}\mathsf{K}^{\left(2j+1\right)}\left(t,s\right)\nonumber
\end{eqnarray}
In view of our aim obtaining from the Dirac Hamiltonian the nonrelativistic Schr{\"o}dinger-Pauli quantum
mechanics together with the relativistic corrections progressing in
powers of $\kappa$ now for a time-dependent Dirac Hamiltonian here
we rank the term $\mathcal{\tilde{E}}\left(t\right)-i\hat{\partial}_{t}$
on equal footing with $\mathcal{\tilde{O}}\left(t\right)\mathcal{\tilde{O}}\left(t\right)$.
Of course, if strong(er) electric fields should be considered, like
during the passage of an electron near by an atomic nucleus in a scattering
experiment, then a ranking of $\mathcal{\tilde{E}}\left(t\right)-i\hat{\partial}_{t}$
on equal footing with $\mathcal{\tilde{O}}\left(t\right)$ should be preferable.

For consistency with (\ref{eq:initial data E(t,0) and O(t,0)})
the initial data of those series elements $\mathsf{K}^{\left(j\right)}\left(t,s\right)$
at $s=0$ are
\begin{eqnarray}
\mathsf{K}^{\left(1\right)}\left(t,0\right) & = & \mathcal{\tilde{O}}\left(t\right)\label{eq:initial data E_j_(t,0) and O_j_(t,0)}\\
\mathsf{K}^{\left(2\right)}\left(t,0\right) & = & \mathcal{\tilde{E}}\left(t\right)-i\hat{\partial}_{t}\nonumber \\
j & \geq & 3\nonumber \\
\mathsf{K}^{\left(j\right)}\left(t,0\right) & = & \mathsf{0}_{4\times4}\nonumber
\end{eqnarray}
Insertion of the series representations (\ref{eq:perturbation series expansion  E(t,s) and O(t,s)})
into the non linear differential equations (\ref{eq:flow equation  K_u and K_g})
a \emph{linear} system of coupled ordinary differential equations
emerges enabling the recursive determination of the sequence of operators
$\mathsf{K}^{\left(j\right)}\left(0,t\right)$ in full analogy to
the afore addressed (simpler) case of a \emph{time-independent} Hamiltonian
\cite{Bylev1998}. That way with the prescribed initial data at
$s=0$ the recursion relation obtained reads
\begin{eqnarray}
\label{eq:recursion  E_j_(s,t) and O_j_(s,t)}\\
\mathsf{K}^{\left(2n\right)}\left(t,s\right) & = & \mathsf{K}^{\left(2n\right)}\left(t,0\right)+4\beta\sum_{j=0}^{n-1}\int_{0}^{s}ds'\mathsf{K}^{\left(2j+1\right)}\left(t,s'\right)\mathsf{K}{}^{\left(2n-2j-1\right)}\left(t,s'\right)\nonumber \\
\mathsf{K}^{\left(2n+1\right)}\left(t,s\right) & = & e^{-4s}\mathsf{K}^{\left(2n+1\right)}\left(t,0\right)+2\beta\sum_{j=0}^{n-1}\int_{0}^{s}ds'e^{-4\left(s-s'\right)}\left[\mathsf{K}^{\left(2j+1\right)}\left(t,s'\right),\mathsf{K}{}^{\left(2n-2j\right)}\left(t,s'\right)\right]\nonumber
\end{eqnarray}
All odd numbered terms $\mathsf{K}^{\left(2n+1\right)}$ manifestly
vanishing in the limit $s\rightarrow\infty$ the searched for unitary
transformed operator $\mathsf{K}\left(t,s\right)$ emerges in the
limit $s\rightarrow\infty$ as a series of terms progressing in powers
of $\kappa^{2}$ given by
\begin{eqnarray}
\mathsf{K}\left(t,\infty\right) & = & \beta+\sum_{j=1}^{\infty}\kappa^{2j}\mathsf{K}^{\left(2j\right)}\left(t,\infty\right)\label{eq:unitary transformed QED Hamiltonian I}\\
\mathsf{\tilde{H}}^{\left(U\right)}\left(t,\infty\right) & = & i\hat{\partial}_{t}+\mathsf{K}\left(t,\infty\right)\nonumber
\end{eqnarray}
For convenience let us adopt compact notation \cite{Silenko2016}
\begin{eqnarray*}
\mathcal{\tilde{F}} & = & \mathcal{\tilde{E}}\left(t\right)-i\hat{\partial}_{t}\\
\tilde{O} & = & \tilde{O}\left(t\right)\\
\mathcal{\tilde{E}} & = & \mathcal{\tilde{E}}\left(t\right)
\end{eqnarray*}
With the details of the calculations all recorded in \cite{Schopohl2022},
let us summarize our results obtained from straightforward perturbation
theory up to and including the sixth order terms $\kappa^{6}$:
\begin{equation}
\mathsf{\tilde{H}}^{\left(U\right)}\left(t,\infty\right)=\beta+\kappa^{2}\mathsf{\tilde{h}}^{\left(U,2\right)}\left(t\right)
+\kappa^{4}\mathsf{\tilde{h}}^{\left(U,4\right)}\left(t\right)+\kappa^{6}\mathsf{\tilde{h}}^{\left(U,6\right)}\left(t\right)+...
\label{eq:result time dependent flow}
\end{equation}
whereas (scaled units)
\begin{eqnarray}
\mathsf{\tilde{h}}^{\left(U,2\right)}\left(t\right) & = & \mathcal{\tilde{E}}
+\beta\frac{\mathcal{\tilde{O}}^{2}}{2}\label{eq:result h2_h4_h6}\\
\nonumber \\
\mathsf{\tilde{h}}^{\left(U,4\right)}\left(t\right) & = & -\frac{1}{8}\beta\mathcal{\tilde{O}}^{4}
-\frac{1}{8}\left[\mathcal{\tilde{O}},\left[\mathcal{\tilde{O}},\mathcal{\tilde{F}}\right]\right]\nonumber \\
\nonumber
\end{eqnarray}
\begin{eqnarray}
\mathsf{\tilde{h}}^{\left(U,6\right)}\left(t\right) & = &
\begin{cases}
\begin{array}{c}
+\frac{1}{16}\beta\mathcal{\tilde{O}}^{6}+\frac{1}{16}\beta\left(\mathcal{\tilde{O}}^{2}\mathcal{\tilde{F}}^{2}
+\mathcal{\tilde{F}}^{2}\mathcal{\tilde{O}}^{2}-2\mathcal{\tilde{O}}\mathcal{\tilde{F}}\mathcal{\tilde{O}}\mathcal{\tilde{F}}
-2\mathcal{\tilde{F}}\mathcal{\tilde{O}}\mathcal{\tilde{F}}\mathcal{\tilde{O}}+2\mathcal{\tilde{O}}\mathcal{\tilde{F}}^{2}\mathcal{\tilde{O}}\right)\\
\\
+\frac{7}{128}\left(\mathcal{\tilde{O}}^{4}\mathcal{\tilde{F}}+
\mathcal{\tilde{F}}\mathcal{\tilde{O}}^{4}\right)-\frac{3}{32}\left(\mathcal{\tilde{O}}^{3}\mathcal{\tilde{F}}\mathcal{\tilde{O}}
+\mathcal{\tilde{O}}\mathcal{\tilde{F}}\mathcal{\tilde{O}}^{3}\right)+\frac{5}{64}\mathcal{\tilde{O}}^{2}\mathcal{\tilde{F}}\mathcal{\tilde{O}}^{2}\\
\\
-\frac{1}{32}\beta\left[\mathcal{\tilde{F}},\left[\mathcal{\tilde{F}},\mathcal{\tilde{O}}^{2}\right]\right]
+\frac{1}{64}\left[\mathcal{\tilde{O}}^{2},\left[\mathcal{\tilde{O}}^{2},\mathcal{\tilde{F}}\right]\right]
\end{array}
\end{cases}\nonumber
\end{eqnarray}

The term $\mathsf{\tilde{h}}^{\left(U,2\right)}\left(t\right)$ is the (expected)
nonrelativistic Hamiltonian of Schr{\"o}dinger-Pauli quantum mechanics
in the presence of time-dependent electromagnetic fields.

 The fourth order term $\mathsf{\tilde{h}}^{\left(U,4\right)}\left(t\right)$ adds to
$\mathsf{h}^{\left(U,2\right)}\left(t\right)$ the leading order relativistic
corrections. It comprises,  besides the correction to the kinetic energy $-\frac{1}{8}\beta\mathcal{\tilde{O}}^{4}$,
the Darwin term \emph{and} the spin-orbit interaction, both terms encoded in the double commutator $-\frac{1}{8}\left[\mathcal{\tilde{O}},\left[\mathcal{\tilde{O}},\mathcal{\tilde{F}}\right]\right]$, but
now with the time-dependent \emph{total} electric field (longitudinal
\emph{and} transversal). This is readily seen evaluating first, with $\mathcal{\tilde{O}}$
and $\mathcal{\tilde{E}}$ as given in (\ref{eq:operator O}), (\ref{eq:operator E}),
the commutator
\begin{eqnarray}
\left[\mathcal{\tilde{O}},\mathcal{\tilde{F}}\right] & = & \left[\alpha_{b}\tilde{\Pi}_{b},\left(\mathcal{\tilde{E}}-i\hat{\partial}_{t}\right)\right]\label{eq:commutator giving total electric field}\\
 & = & \left(\frac{1}{mc}\right)\left(\frac{\hbar q_{e}}{m c^{2}}\right)i\left(-\frac{\partial\Phi\left(\mathbf{r},t\right)}{\partial\mathsf{r}_{b}}-\frac{\partial\mathcal{A}_{b}\left(\mathbf{r},t\right)}{\partial t}\right)\alpha_{b}\nonumber \\
 & = & \left(\frac{1}{mc^{2}}\frac{\hbar q_{e}}{mc}\right) i\mathscr{E}_{b}\left(\mathbf{r},t\right)\alpha_{b}\nonumber \\
 & \equiv & i\mathscr{\tilde{E}}_{b}\left(\mathbf{r},t\right)\alpha_{b}\nonumber
\end{eqnarray}
, whereby $\mathscr{\tilde{E}}_{b}\left(\mathbf{r},t\right)$ denotes
now the (scaled) \emph{total} electric field including both contributions,
longitudinal \emph{and} transversal. Subsequently one finds
\begin{eqnarray}
\left[\mathcal{\tilde{O}},\left[\mathcal{\tilde{O}},\mathcal{\tilde{F}}\right]\right] \label{eq:Darwin + SO}
& = & \left[\tilde{\Pi}_{a}\alpha_{a},i\mathscr{\tilde{E}}_{b}\left(\mathbf{r},t\right)\alpha_{b}\right]\\
 & = & \frac{1}{2}\left(\left[\tilde{\Pi}_{a},i\mathscr{\tilde{E}}_{b}\right]2\delta_{a,b}\mathsf{1}_{4\times4}+i\left(\tilde{\Pi}_{a}\mathscr{\tilde{E}}_{b}
 +\mathscr{\tilde{E}}_{b}\tilde{\Pi}_{a}\right)2i\varepsilon_{abb'}\sigma_{b'}\right)\nonumber\\
 & = &
 \begin{array}{c}
\frac{1}{mc}\left(\frac{1}{mc^{2}}\frac{\hbar}{mc}q_{e}\right)
\end{array}
\left( \hbar\textrm{div}\mathscr{E}-\boldsymbol{\Pi}\wedge\mathscr{E}+\mathscr{E}\wedge\boldsymbol{\Pi} \right)_{b'}\sigma_{b'}\nonumber
\end{eqnarray}
Note the contribution of the transversal electric field to the spin-orbit
interaction is missed out in the flow equation approach \cite{Bylev1998}, as it applies only to \emph{static} external potentials.

The term $\mathsf{\tilde{h}}^{\left(U,6\right)}\left(t\right)$ represents
the corrections of order $\kappa^{6}$ as obtained by the above presented
Hamiltonian flow equation method. This term we stated here for clarification
and for comparison only. Specializing to \emph{time-independent} fields,
then in all commutator terms we have $\tilde{\mathcal{F}}\equiv\mathcal{\tilde{E}}$,
and comparing with our afore obtained energy-separating result (\ref{eq:h_6  II}) there is indeed a discrepancy
\begin{eqnarray}
\mathsf{\tilde{h}}^{\left(U,6\right)}&=&\mathsf{\tilde{h}}^{\left(6\right)}
-\frac{1}{32}\beta\left[\mathcal{\tilde{E}},\left[\mathcal{\tilde{E}},\mathcal{\tilde{O}}^{2}\right]\right]
+\frac{1}{64}\left[\mathcal{\tilde{O}}^{2},\left[\mathcal{\tilde{O}}^{2},\mathcal{\tilde{E}}\right]\right]\nonumber\\
&=&\frac{1}{32}\left[\beta\left[\mathcal{\tilde{E}},\mathcal{\tilde{O}}^{2}\right],\mathsf{\tilde{h}}^{\left(2\right)}\right]
\label{eq:discrepancy sixth order term h_6}
\end{eqnarray}
That the established step-by-step Foldy-Wouthuysen method actually disagrees
in the sixth order term $\mathsf{\tilde{h}}^{\left(FW,6\right)}$ with results
obtained by the (tedious) perturbative Eriksen method, has been discovered already early on for minimal coupling to static fields \cite{EriksenKolsrud}.
So it comes as no surprise learning that
our results obtained for $\mathsf{\tilde{H}}^{\left(U\right)}\left(t,\infty\right)$ with
the afore discussed time-dependent Hamiltonian flow equation method (\ref{eq:flow equation operator K(t,s)  II}),
if expanded beyond order $\kappa^{4}$ , now disagree as well with
the afore obtained contribution (\ref{eq:h_6  II}) to the Newton-Wigner
Hamiltonian.

As regards the additional discrepancy $\mathsf{\tilde{h}}^{\left(FW,6\right)}\neq\mathsf{\tilde{h}}^{\left(U,6\right)}$
, this is reflecting merely the afore mentioned ambiguity (\ref{eq:unitary transformation N})
regarding unitary transformations $\mathsf{U}$ and also  $\mathsf{N}\mathsf{U}$
both transforming the Dirac Hamiltonian to a block-diagonal guise.

\section{Construction of the Newton-Wigner Hamiltonian for a Dirac Fermion with  minimal coupling to time-dependent electromagnetic fields}

In this section we first introduce a generalization of the afore discussed
static beta-flow equation method (\ref{eq:flow equation operator K(t,s)  II})
in order to derive now the time-dependent Newton-Wigner Hamiltonian
assiciated with a time-dependent Dirac Hamiltonian. The proposition
being, that all is required to lift the previous result for static
fields (\ref{eq:representation H_NW   with Omega_u(infty) IV}) now
to time-dependent fields, lies in the replacement \cite{EriksenKolsrud}

\[
\mathcal{\tilde{E}}\rightarrow\mathcal{\tilde{F}}\left(t\right)=\mathcal{\tilde{E}}\left(t\right)-i\hat{\partial}_{t}
\]
in all \emph{commutators} defining the Newton-Wigner Hamiltonian (\ref{eq:representation H_NW   with Omega_u(infty) IV}),
as generated with the expansions (\ref{eq:representation H_NW   with Omega_u(infty) IV})
and (\ref{eq:IGL  Omega_u_(s) I}), repectively (\ref{eq:IGL  Omega_u_(s) II}).

To provide a proof we introduce to this end now a \emph{time-dependent} beta-flow
\begin{eqnarray}
\frac{\partial}{\partial s}\mathsf{Z}\left(t,s\right) & = & \left[\omega\left(t,s\right),\mathsf{Z}\left(t,s\right)\right]\label{eq:beta-flow-time-dependent I}\\
\mathsf{Z}\left(t,0\right) & = & \beta\nonumber
\end{eqnarray}
In contrast to the afore discussed (static) beta-flow in (\ref{eq:beta-flow})
here we adopt the generator $\omega\left(t,s\right)$ in such a way,
that the operator
\begin{equation}
\tilde{\mathsf{K}}\left(t\right)=\mathsf{\tilde{H}}^{\left(D\right)}\left(t\right)-i\hat{\partial}_{t}\label{eq:operator K(t)}
\end{equation}
obeys to
\begin{eqnarray}
\left[\tilde{\mathsf{K}}\left(t\right),\mathsf{Z}\left(t,\infty\right)\right] & = & \mathsf{0}_{4\times4}\label{eq:Commutator(Z(infty,t), H_D)=0}
\end{eqnarray}
Along the line of reasoning presented afore in (\ref{eq:double bracket flow})
a suitable antisymmetric generator of such a time-dependent beta-flow emerges as
\begin{equation}
\omega\left(t,s\right)=\left[\tilde{\mathsf{K}}\left(t\right),\mathsf{Z}\left(t,s\right)\right]\label{eq:generator omega(s,t)}
\end{equation}
Every bit said afore in the static case applies now in the time-dependent case with the operator $\tilde{\mathsf{K}}\left(t\right)$
in place of $\mathsf{\tilde{H}}^{\left(D\right)}$. With that said
the exact solution to (\ref{eq:beta-flow}) reads
\begin{eqnarray}
\mathsf{Z}\left(t,s\right) & = & \mathsf{W}\left(t,s\right)\beta\,\mathsf{W}^{-1}\left(t,s\right)\label{eq:exact solution Z(t,s)}
\end{eqnarray}
, whereas
\begin{eqnarray}
\mathsf{W}\left(t,s\right) & = & \mathsf{C}\left(t,s\right)\beta+\mathsf{S}\left(t,s\right)\label{eq:ingredients to solution Z(t,s)}\\
\mathsf{C}\left(t,s\right) & = & \cosh\left(2s\,\tilde{\mathsf{K}}\left(t\right)\right)=\cosh\left(2s\,\sqrt{\tilde{\mathsf{K}}\left(t\right)\tilde{\mathsf{K}}\left(t\right)}\right)\nonumber \\
\mathsf{S}\left(t,s\right) & = & \sinh\left(2s\,\tilde{\mathsf{K}}\left(t\right)\right)=\frac{\tilde{\mathsf{K}}\left(t\right)}{\sqrt{\tilde{\mathsf{K}}\left(t\right)\tilde{\mathsf{K}}\left(t\right)}}\sinh\left(2s\,\sqrt{\tilde{\mathsf{K}}\left(t\right)\tilde{\mathsf{K}}\left(t\right)}\right)\nonumber
\end{eqnarray}
As follows in reference to the exact solution (\ref{eq:exact solution Z(t,s)}) one readily confirms
\begin{eqnarray}
\mathsf{Z}\left(t,s\right) & = & \mathsf{Z}^{\dagger}\left(t,s\right)\label{eq:properties of Z(t,s)}\\
\nonumber \\
\left[\beta,\left(\beta\mathsf{Z}\left(t,s\right)+\mathsf{Z}\left(t,s\right)\beta\right)\right] & = & \mathsf{0}_{4\times4}=\left[\left[\mathsf{Z}\left(t,s\right),\left(\beta\,\mathsf{Z}\left(t,s\right)+\mathsf{Z}\left(t,s\right)\beta\right)\right]\right]\nonumber \\
\nonumber \\
\mathsf{Z}\left(t,s\right)\mathsf{Z}^{\dagger}\left(t,s\right) & = & \mathsf{Z}^{\dagger}\left(t,s\right)\mathsf{Z}\left(t,s\right)\nonumber \\
\nonumber \\
\mathsf{Z}\left(t,s\right)\mathsf{Z}\left(t,s\right) & = & \mathsf{1}_{4\times4}\nonumber
\end{eqnarray}
, thus validating the operator $\mathsf{Z}\left(t,s\right)$ being
unitary (and involutive as well). Furthermore the exact solution (\ref{eq:exact solution Z(t,s)})
has the limiting value
\begin{equation}
\lim_{s\rightarrow\infty}\mathsf{Z}\left(t,s\right)=\mathsf{Z}\left(t,\infty\right)=\frac{\tilde{\mathsf{K}}\left(t\right)}{\sqrt{\tilde{\mathsf{K}}\left(t\right)\tilde{\mathsf{K}}\left(t\right)}}
\label{eq:Z(t, s->infty )}
\end{equation}
In full analogy to the afore discussed time-independent case there
follows, introducing the unitary operator
\[
\mathsf{V}\left(t,s\right)\equiv\frac{\beta+\mathsf{Z}\left(t,s\right)}{\sqrt{\left(\beta+\mathsf{Z}\left(t,s\right)\right)^{2}}\,}\beta
\]
, now the representation
\begin{eqnarray}
\mathsf{Z}\left(t,s\right) & = & \mathsf{V}\left(t,s\right)\beta\,\mathsf{V}^{\dagger}\left(t,s\right)
\end{eqnarray}
And just like in the time-independent case there holds on the basis
of the commutator relation stated in (\ref{eq:properties of Z(t,s)}).
\begin{eqnarray}
\mathsf{V}\left(t,s\right)\mathsf{V}\left(t,s\right) & = & \mathsf{Z}\left(t,s\right)\beta
\end{eqnarray}
In view of the close analogy between the static case discussed afore
in (\ref{sec:ExactSolutionStaticBetaFlow }) and the here treated time-dependent case,
it seems natural there exists the solution $\mathsf{K}\left(t,s\right)$
to the (nonlinear) Hamiltonian flow equation (\ref{eq:flow equation operator K(t,s)  II})
with time-dependent generator (\ref{eq:time-dependent BP-generator})
in the guise
\begin{equation}
\mathsf{K}\left(t,s\right)=\mathsf{V}^{\dagger}\left(t,s\right)\mathsf{\tilde{K}}\left(t\right)\mathsf{V}\left(t,s\right)\label{eq:guise  K(t,s)}
\end{equation}
, with $\mathsf{V}\left(t,s\right)$ now a\emph{ specific} unitary
transformation solving the initial value problem
\begin{eqnarray}
\frac{\partial}{\partial s}\mathsf{V}\left(t,s\right) & = & \omega\left(t,s\right)\mathsf{V}\left(t,s\right)\label{eq:ODE V(t,s)}\\
\mathsf{V}\left(t,0\right) & = &  \mathsf{1}_{4\times4}\nonumber
\end{eqnarray}
 , and $\omega\left(t,s\right)$ being the antisymmetric generator
of the time-dependent beta-flow (\ref{eq:beta-flow})
\begin{equation}
\omega\left(t,s\right)=\left[\mathsf{\tilde{K}}\left(t\right),\mathsf{Z}\left(t,s\right)\right]\label{eq:generator omega_(t,s)  II}
\end{equation}
The validation of (\ref{eq:guise  K(t,s)}) follows readily directly
from (\ref{eq:ODE V(t,s)}) and leads akin to the time-independent case  (\ref{eq:flow H(s)  III}) to
\begin{eqnarray}
\frac{\partial}{\partial s}\mathsf{K}\left(t,s\right) & = & \left[\left[\beta,\mathsf{K}\left(t,s\right)\right],\mathsf{K}\left(t,s\right)\right]\label{eq:flow K(t,s) II}
\end{eqnarray}
, or else
\begin{eqnarray}
\frac{\partial}{\partial s}\mathsf{K}\left(t,s\right) & = & \left[\eta\left(t,s\right),\mathsf{K}\left(t,s\right)\right]\label{eq:flow K(t,s)  III}\\
\eta\left(t,s\right) & = & \left[\beta,\mathsf{K}\left(t,s\right)\right]\nonumber \\
\mathsf{K}\left(t,0\right) & = & \mathsf{\tilde{K}}\left(t\right)\nonumber
\end{eqnarray}
The flow equation obtained for $\mathsf{K}\left(t,s\right)$ this
way coincides with the time-dependent flow (\ref{eq:flow equation operator K(t,s)  II}),
thus substantiating the assertion (\ref{eq:guise  K(t,s)}).

The now time-dependent generators, $\eta\left(t,s\right)$ and $\omega\left(t,s\right)$,
with $\eta\left(t,s\right)$ being odd and $\omega\left(t,s\right)=\omega_{u}\left(t,s\right)+\omega_{g}\left(t,s\right)$
being decomposed into an even and an odd part (just like in the time-independent
case), are mutually connected by the same unitary transformation $\mathsf{V}\left(t,s\right)$,
i.e. once $\omega\left(t,s\right)$ is known, then $\eta\left(t,s\right)$
is known and vice versa
\begin{equation}
\eta\left(t,s\right)=-\mathsf{V}^{\dagger}\left(t,s\right)\omega\left(t,s\right)\mathsf{V}\left(t,s\right)\label{eq:connection eta(t,s) and omega(t,s)}
\end{equation}
Representing next, like before in the static case (\ref{sec:Perturbative H_NW Hamiltonian static}),
the unitary transformation as
\begin{eqnarray}
\mathsf{V}\left(t,s\right) & = & e^{\Omega_{g}\left(t,s\right)}e^{\Omega_{u}\left(t,s\right)}\label{eq:representation V(t,s) with Omega_u_(t,s) and Omega_g_(t,s)}
\end{eqnarray}
, with $\Omega_{g}\left(t,s\right)$ being even and $\Omega_{u}\left(t,s\right)$
being odd, then (\ref{eq:connection eta(t,s) and omega(t,s)}) implies
\[
e^{-\Omega_{u}\left(t,s\right)}\omega\left(t,s\right)e^{\Omega_{u}\left(t,s\right)}=-e^{\Omega_{g}\left(t,s\right)}\eta\left(t,s\right)e^{-\Omega_{g}\left(t,s\right)}
\]
This important fact engenders that everything said afore regarding
the result for the perturbation expansion in the \emph{time-independent}
case, applies as well in the time-dependent case, so that we have
in full analogy to the derivation of (\ref{eq:IGL  Omega_u_(s) I}) now
\begin{eqnarray}
\Omega_{u}\left(t,s\right) & = & \int_{0}^{s}ds'\frac{2\mathsf{ad}_{\Omega_{u}\left(t,s'\right)}}{\sinh\left(2\mathsf{ad}_{\Omega_{u}\left(t,s'\right)}\right)}\circ\omega_{u}\left(t,s'\right)
\label{eq:IGL  Omega_u_(t,s) I}
\end{eqnarray}
Identifying the limiting value of the time-dependent flow determining $\mathsf{\tilde{H}}^{\left(U\right)}\left(t,s\right)$ as
\begin{eqnarray}
\mathsf{\tilde{H}}^{\left(U\right)}\left(t,\infty\right)-i\hat{\partial}_{t} & \equiv & \mathsf{K}\left(t,\infty\right)
\label{eq:K(t,infty) }\\
 & = & \lim_{s\rightarrow\infty}\mathsf{V}^{\dagger}\left(t,s\right)\mathsf{\tilde{K}}\left(t\right)\mathsf{V}\left(t,s\right)\nonumber \\
 & = & \lim_{s\rightarrow\infty}e^{-\Omega_{u}\left(t,s\right)}e^{-\Omega_{g}\left(t,s\right)}\mathsf{\tilde{K}}\left(t\right)e^{\Omega_{g}\left(t,s\right)}e^{\Omega_{u}\left(t,s\right)}\nonumber
\end{eqnarray}
, and because the limiting value $\mathsf{K}\left(t,\infty\right)$
of the flow (\ref{eq:flow K(t,s)  III}) obeys by construction to
$\left[\beta,\mathsf{K}\left(t,\infty\right)\right]=\mathsf{0}$,
now the searched for time-dependent Newton-Wigner (NW) Hamiltonian arises in the guise
\begin{eqnarray}
\mathsf{\tilde{H}}^{\left(NW\right)}\left(t\right)-i\hat{\partial}_{t} & \equiv & e^{+\Omega_{g}\left(t,\infty\right)}\mathsf{K}\left(t,\infty\right)e^{-\Omega_{g}\left(t,\infty\right)}
\label{eq:time-dependent  H_NW  I}\\
 & = & e^{-\Omega_{u}\left(t,\infty\right)}\mathsf{\tilde{K}}\left(t\right)e^{\Omega_{u}\left(t,\infty\right)}\nonumber \\
 & = & e^{-\Omega_{u}\left(t,\infty\right)}\left(\beta+\mathcal{\tilde{O}}\left(t\right)+\tilde{\mathcal{E}}\left(t\right)-i\hat{\partial}_{t}\right)e^{\Omega_{u}\left(t,\infty\right)}\nonumber
\end{eqnarray}
As the first line in (\ref{eq:time-dependent  H_NW  I}) is manifestly
even, we convert, like before in the static case (\ref{sec:Perturbative H_NW Hamiltonian static}),
now the last line in (\ref{eq:time-dependent  H_NW  I}) to the identical guise
\begin{eqnarray}
\mathsf{\tilde{H}}^{\left(NW\right)}\left(t\right) & = & i\hat{\partial}_{t}+\cosh\left(\textrm{ad}_{\Omega_{u}\left(t,\infty\right)}\right)\circ\left(\beta+\tilde{\mathcal{E}}\left(t\right)
-i\hat{\partial}_{t}\right)-\sinh\left(\textrm{ad}_{\Omega_{u}\left(t,\infty\right)}\right)\circ\mathcal{\tilde{O}}\left(t\right)
\label{eq:time-dependent  H_NW  II}\\
 & = & \beta+\tilde{\mathcal{E}}\left(t\right)+\left(\cosh\left(\textrm{ad}_{\Omega_{u}\left(t,\infty\right)}\right)
 -\mathsf{1}\right)\circ\left(\beta+\tilde{\mathcal{E}}\left(t\right)-i\hat{\partial}_{t}\right)
 -\sinh\left(\textrm{ad}_{\Omega_{u}\left(t,\infty\right)}\right)\circ\mathcal{\tilde{O}}\left(t\right)\nonumber
\end{eqnarray}
So in \emph{all} commutators comprising in the
static case a term $\mathcal{\tilde{E}}$, the transition to the time-dependent case is enabled
making in the expansion (\ref{eq:representation H_NW   with Omega_u(infty) IV}),  respectively in (\ref{eq:h_2  II}), (\ref{eq:h_4  II}), (\ref{eq:h_6  II}) and so on, the substitution
\begin{equation}
\mathcal{\tilde{E}}\rightarrow\mathcal{\tilde{F}}\left(t\right)=\mathcal{\tilde{E}}\left(t\right)-i\hat{\partial}_{t}\label{eq:substitution E->F}
\end{equation}
Note that the substitution rule (\ref{eq:substitution E->F}) here arises out of the presented time-dependent generalization of the beta-flow striving to the limiting value (\ref{eq:properties of Z(t,s)}), whereby the latter reduces in the stationary case to the energy-sign operator of the Dirac Hamiltonian. The substitution (\ref{eq:substitution E->F}) has been established,  on an observational basis though, already in \cite{EriksenKolsrud}, as it facilitates as well the calculational effort with the tedious original step by step FW-method significantly \cite{Silenko2016}.

\section*{Conclusion}
 Proceeding from Brockett's approach to continuous unitary transformations via flow equations with quadratic nonlinearity with a purpose-built generator $\left[\Gamma,\mathsf{H\left(s\right)}\right]$ of the flow that strives for $s\to\infty$ towards zero \cite{Brockett91}, the Hamiltonian flow considered by Bylev and Pirner (BP) in \cite{Bylev1998} transforming the stationary Dirac Hamiltonian to a unitary equivalent \emph{even} form, came out known to the choice $\Gamma =\beta$.
Based on that perception the perturbative approach initiated in \cite{Bylev1998} for static external fields,  suitable to expand the limiting value of their Hamiltonian flow in powers of $\frac{v}{c}$, has been generalized in the above as well to apply to a relativistic fermion with minimal coupling to \emph{time-dependent} electromagnetic fields.
Different from \cite{Bylev1998} though,  in view of the initial data of the flow, the  electric potential term  $\mathcal{E}$ has been considered in order of magnitude being comparable to the kinetic energy term $\mathcal{OO}$  for reasons of consistency with the nonrelativistic limit.

At order $\frac{v^4}{c^4}$ the relativistic correction to the kinetic energy, the Darwin term and the spin-orbit interaction terms emerge, but taking into account coupling to time-dependent electromagnetic fields now the spin-orbit interaction term  manifestly couples to the longitudinal \emph{and} to the transversal electric field, this result being in full agreement with the result obtained early on by the step-by-step FW-transformation method \cite{FoldyW50}\footnote{There is a typo in the original article of Foldy and Wouthuysen regarding the  sign of the scalar potential term in the Hamiltonian, see their Eq. (34) in \cite{FoldyW50}. Consequently the sign of the Darwin term in their Eq.(35) is incorrect, whereas the sign of the spin-orbit term is correct. We caution the reader that in the article of Silenko \cite{Silenko2003}, see his Eq.(40), the sign of the spin-orbit term is incorrect, while then again the sign of the scalar potential term and that one of the Darwin term is correct.}

But at the next order $\frac{v^6}{c^6}$ the results obtained with the Hamiltonian flow equation approach reveal a discrepancy with results obtained by the (elaborate) step-by-step unitary transformation method of Foldy and Wouthuysen.
Comparing these results obtained for the special case of electrostatic and magnetostatic fields superposed, unfortunately also a discrepancy with the unambiguous energy-separating Eriksen transformation method is manifest.

So, in view of the Hamiltonian flow equation approach beyond order $\frac{v^4}{c^4}$ not being equivalent to the Eriksen transformation either, a purpose-built \emph{reverse} beta-flow equation approach was introduced striving for $s\to\infty$ to the energy-sign operator of the stationary Dirac-Hamiltonian. That beta-flow equation being a Riccati equation,
serendipitously the exact solution was found and a unitary transformation (\ref{eq:identification V(s)=U(s)*Beta}) was constructed in terms of that solution,
which turned out to coincide with the Eriksen transformation in the limit $s\to\infty$.

Based on this insight a link between the generators of the Hamiltonian flow and the generator of the reverse beta-flow was noticed, that finally enabled
to derive the central results  of this article, namely Eq.(\ref{eq:representation H_NW   with Omega_u(infty) IV}) and Eq. (\ref{eq:IGL  Omega_u_(s) I}), that way fully implementing a convenient energy-separating scheme to unambiguously reconstruct the expansion in powers of $\frac{v}{c}$ for the Newton-Wigner Hamiltonian of the relativistic fermion moving in the presence of  electrostatic and magnetostatic fields superposed.

Finally, the results obtained for static fields have been generalized to a Dirac Hamiltonian with coupling to weak amplitude and slowly varying \emph{time-dependent} electromagnetic fields, resulting in a series expansion of the NW-Hamiltonian in powers of $\frac{v}{c}$ fully coinciding with the correction scheme for the step-by-step FW-transformation method introduced by Silenko \cite{Silenko2016}.

So, the long standing problem of obtaining the \emph{explicit} series expansion in the parameter  $\frac{v}{c}$
of the unitary transformation of a \emph{general}  Dirac Hamiltonian to an
even \emph{and} energy-separating guise,  and concurrently being fully in accord with results obtainable with the unambiguous  Eriksen transformation \cite{Eriksen},  has been resolved.

In view of the step-by-step FW-transformation method not leading to an unambiguous energy-separating result, it would be interesting to apply the introduced flow equation approach as well to bosons carrying mass and charge with spin S=0 (Klein-Gordon) or spin S=1 (Proca), taking into account coupling to external electromagnetic fields, in this way checking the results obtained early on by Case \cite{Case54}.



\bibliography{MS_June_30_revtex_Literatur}

\begin{thebibliography}{42}%
\makeatletter
\providecommand \@ifxundefined [1]{%
 \@ifx{#1\undefined}
}%
\providecommand \@ifnum [1]{%
 \ifnum #1\expandafter \@firstoftwo
 \else \expandafter \@secondoftwo
 \fi
}%
\providecommand \@ifx [1]{%
 \ifx #1\expandafter \@firstoftwo
 \else \expandafter \@secondoftwo
 \fi
}%
\providecommand \natexlab [1]{#1}%
\providecommand \enquote  [1]{``#1''}%
\providecommand \bibnamefont  [1]{#1}%
\providecommand \bibfnamefont [1]{#1}%
\providecommand \citenamefont [1]{#1}%
\providecommand \href@noop [0]{\@secondoftwo}%
\providecommand \href [0]{\begingroup \@sanitize@url \@href}%
\providecommand \@href[1]{\@@startlink{#1}\@@href}%
\providecommand \@@href[1]{\endgroup#1\@@endlink}%
\providecommand \@sanitize@url [0]{\catcode `\\12\catcode `\$12\catcode
  `\&12\catcode `\#12\catcode `\^12\catcode `\_12\catcode `\%12\relax}%
\providecommand \@@startlink[1]{}%
\providecommand \@@endlink[0]{}%
\providecommand \url  [0]{\begingroup\@sanitize@url \@url }%
\providecommand \@url [1]{\endgroup\@href {#1}{\urlprefix }}%
\providecommand \urlprefix  [0]{URL }%
\providecommand \Eprint [0]{\href }%
\providecommand \doibase [0]{https://doi.org/}%
\providecommand \selectlanguage [0]{\@gobble}%
\providecommand \bibinfo  [0]{\@secondoftwo}%
\providecommand \bibfield  [0]{\@secondoftwo}%
\providecommand \translation [1]{[#1]}%
\providecommand \BibitemOpen [0]{}%
\providecommand \bibitemStop [0]{}%
\providecommand \bibitemNoStop [0]{.\EOS\space}%
\providecommand \EOS [0]{\spacefactor3000\relax}%
\providecommand \BibitemShut  [1]{\csname bibitem#1\endcsname}%
\let\auto@bib@innerbib\@empty
\bibitem [{\citenamefont {Newton}\ and\ \citenamefont
  {Wigner}(1949)}]{NewtonWigner}%
  \BibitemOpen
  \bibfield  {author} {\bibinfo {author} {\bibfnamefont {T.}~\bibnamefont
  {Newton}}\ and\ \bibinfo {author} {\bibfnamefont {E.}~\bibnamefont
  {Wigner}},\ }\href@noop {} {\bibfield  {journal} {\bibinfo  {journal}
  {Reviews of Modern Physics}\ }\textbf {\bibinfo {volume} {21}},\ \bibinfo
  {pages} {400} (\bibinfo {year} {1949})}\BibitemShut {NoStop}%
\bibitem [{\citenamefont {Foldy}\ and\ \citenamefont
  {Wouthuysen}(1950)}]{FoldyW50}%
  \BibitemOpen
  \bibfield  {author} {\bibinfo {author} {\bibfnamefont {L.}~\bibnamefont
  {Foldy}}\ and\ \bibinfo {author} {\bibfnamefont {S.}~\bibnamefont
  {Wouthuysen}},\ }\href@noop {} {\bibfield  {journal} {\bibinfo  {journal}
  {Physical Review}\ }\textbf {\bibinfo {volume} {78}},\ \bibinfo {pages} {29}
  (\bibinfo {year} {1950})}\BibitemShut {NoStop}%
\bibitem [{\citenamefont {Costella}\ and\ \citenamefont
  {McKellar}(1995)}]{CostellaMcKellar95}%
  \BibitemOpen
  \bibfield  {author} {\bibinfo {author} {\bibfnamefont {J.}~\bibnamefont
  {Costella}}\ and\ \bibinfo {author} {\bibfnamefont {B.}~\bibnamefont
  {McKellar}},\ }\href@noop {} {\bibfield  {journal} {\bibinfo  {journal}
  {American Journal of Physics}\ }\textbf {\bibinfo {volume} {63}} (\bibinfo
  {year} {1995})}\BibitemShut {NoStop}%
\bibitem [{\citenamefont {Case}(1954)}]{Case54}%
  \BibitemOpen
  \bibfield  {author} {\bibinfo {author} {\bibfnamefont {K.}~\bibnamefont
  {Case}},\ }\href@noop {} {\bibfield  {journal} {\bibinfo  {journal} {Physical
  Review}\ }\textbf {\bibinfo {volume} {95}},\ \bibinfo {pages} {1323}
  (\bibinfo {year} {1954})}\BibitemShut {NoStop}%
\bibitem [{\citenamefont {Eriksen}(1958)}]{Eriksen}%
  \BibitemOpen
  \bibfield  {author} {\bibinfo {author} {\bibfnamefont {E.}~\bibnamefont
  {Eriksen}},\ }\href@noop {} {\bibfield  {journal} {\bibinfo  {journal}
  {Physical Review}\ }\textbf {\bibinfo {volume} {111}},\ \bibinfo {pages}
  {1011} (\bibinfo {year} {1958})}\BibitemShut {NoStop}%
\bibitem [{\citenamefont {Wegner}(1994)}]{Wegner94}%
  \BibitemOpen
  \bibfield  {author} {\bibinfo {author} {\bibfnamefont {F.}~\bibnamefont
  {Wegner}},\ }\href@noop {} {\bibfield  {journal} {\bibinfo  {journal}
  {Annalen der Physik}\ }\textbf {\bibinfo {volume} {506}},\ \bibinfo {pages}
  {77} (\bibinfo {year} {1994})}\BibitemShut {NoStop}%
\bibitem [{\citenamefont {Wegner}(2001)}]{Wegner2001}%
  \BibitemOpen
  \bibfield  {author} {\bibinfo {author} {\bibfnamefont {F.~J.}\ \bibnamefont
  {Wegner}},\ }\href@noop {} {\bibfield  {journal} {\bibinfo  {journal}
  {Physics Reports}\ }\textbf {\bibinfo {volume} {348}},\ \bibinfo {pages} {77}
  (\bibinfo {year} {2001})}\BibitemShut {NoStop}%
\bibitem [{\citenamefont {Kehrein}(2006)}]{Kehrein}%
  \BibitemOpen
  \bibfield  {author} {\bibinfo {author} {\bibfnamefont {S.}~\bibnamefont
  {Kehrein}},\ }\href@noop {} {\bibfield  {journal} {\bibinfo  {journal}
  {Springer Tracts in Modern Physics}\ }\textbf {\bibinfo {volume} {217}}
  (\bibinfo {year} {2006})}\BibitemShut {NoStop}%
\bibitem [{\citenamefont {Brockett}(1991)}]{Brockett91}%
  \BibitemOpen
  \bibfield  {author} {\bibinfo {author} {\bibfnamefont {R.}~\bibnamefont
  {Brockett}},\ }\href@noop {} {\bibfield  {journal} {\bibinfo  {journal}
  {Linear Algebra and its Applications}\ }\textbf {\bibinfo {volume} {146}},\
  \bibinfo {pages} {79} (\bibinfo {year} {1991})}\BibitemShut {NoStop}%
\bibitem [{\citenamefont {Driessel}\ \emph {et~al.}(2001)\citenamefont
  {Driessel}, \citenamefont {Hentzel},\ and\ \citenamefont
  {Wasin}}]{DoubleBracketFlow}%
  \BibitemOpen
  \bibfield  {author} {\bibinfo {author} {\bibfnamefont {K.}~\bibnamefont
  {Driessel}}, \bibinfo {author} {\bibfnamefont {I.}~\bibnamefont {Hentzel}},\
  and\ \bibinfo {author} {\bibfnamefont {S.}~\bibnamefont {Wasin}},\
  }\href@noop {} {\bibfield  {journal} {\bibinfo  {journal} {JP Jour. Algebra,
  Number Theory \&Appl.}\ }\textbf {\bibinfo {volume} {1}},\ \bibinfo {pages}
  {87} (\bibinfo {year} {2001})}\BibitemShut {NoStop}%
\bibitem [{\citenamefont {Bylev}\ and\ \citenamefont
  {Pirner}(1998)}]{Bylev1998}%
  \BibitemOpen
  \bibfield  {author} {\bibinfo {author} {\bibfnamefont {A.}~\bibnamefont
  {Bylev}}\ and\ \bibinfo {author} {\bibfnamefont {H.}~\bibnamefont {Pirner}},\
  }\href@noop {} {\bibfield  {journal} {\bibinfo  {journal} {Physics Letters
  B}\ } (\bibinfo {year} {1998})}\BibitemShut {NoStop}%
\bibitem [{\citenamefont {Schopohl}\ and\ \citenamefont
  {Cetin}(2022)}]{Schopohl2022}%
  \BibitemOpen
  \bibfield  {author} {\bibinfo {author} {\bibfnamefont {N.}~\bibnamefont
  {Schopohl}}\ and\ \bibinfo {author} {\bibfnamefont {N.~S.}\ \bibnamefont
  {Cetin}},\ }\href@noop {} {\bibfield  {journal} {\bibinfo  {journal}
  {supplemental material}\ } (\bibinfo {year} {2022})}\BibitemShut {NoStop}%
\bibitem [{\citenamefont {Pursey}(1958)}]{Pursey}%
  \BibitemOpen
  \bibfield  {author} {\bibinfo {author} {\bibfnamefont {D.~L.}\ \bibnamefont
  {Pursey}},\ }\href@noop {} {\bibfield  {journal} {\bibinfo  {journal}
  {Nuclear Physics}\ }\textbf {\bibinfo {volume} {8}},\ \bibinfo {pages} {595}
  (\bibinfo {year} {1958})}\BibitemShut {NoStop}%
\bibitem [{\citenamefont {Eriksen}\ and\ \citenamefont
  {Kolsrud}(1960)}]{EriksenKolsrud}%
  \BibitemOpen
  \bibfield  {author} {\bibinfo {author} {\bibfnamefont {E.}~\bibnamefont
  {Eriksen}}\ and\ \bibinfo {author} {\bibfnamefont {M.}~\bibnamefont
  {Kolsrud}},\ }\href@noop {} {\bibfield  {journal} {\bibinfo  {journal} {Nuovo
  Cimento Suppl.}\ }\textbf {\bibinfo {volume} {18}},\ \bibinfo {pages} {1}
  (\bibinfo {year} {1960})}\BibitemShut {NoStop}%
\bibitem [{Note1()}]{Note1}%
  \BibitemOpen
  \bibinfo {note} {As of now the summation convention always applies, if not
  suspended explicitely.}\BibitemShut {Stop}%
\bibitem [{\citenamefont {Baym}(1969)}]{Baym}%
  \BibitemOpen
  \bibfield  {author} {\bibinfo {author} {\bibfnamefont {G.}~\bibnamefont
  {Baym}},\ }\href@noop {} {\bibfield  {journal} {\bibinfo  {journal} {Taylor
  and Francis Group}\ } (\bibinfo {year} {1969})}\BibitemShut {NoStop}%
\bibitem [{\citenamefont {Messiah}(1970)}]{Messiah}%
  \BibitemOpen
  \bibfield  {author} {\bibinfo {author} {\bibfnamefont {A.}~\bibnamefont
  {Messiah}},\ }\href@noop {} {\bibfield  {journal} {\bibinfo  {journal}
  {North-Holland}\ } (\bibinfo {year} {1970})}\BibitemShut {NoStop}%
\bibitem [{\citenamefont {Bjorken}\ and\ \citenamefont
  {Drell}(1964)}]{BjorkenDrell}%
  \BibitemOpen
  \bibfield  {author} {\bibinfo {author} {\bibfnamefont {J.}~\bibnamefont
  {Bjorken}}\ and\ \bibinfo {author} {\bibfnamefont {S.}~\bibnamefont
  {Drell}},\ }\href@noop {} {\bibfield  {journal} {\bibinfo  {journal}
  {McGraw-Hill Book Company}\ }\textbf {\bibinfo {volume} {ISBN 07-005493-2}}
  (\bibinfo {year} {1964})}\BibitemShut {NoStop}%
\bibitem [{\citenamefont {Itzykson}\ and\ \citenamefont
  {Zuber}(1980)}]{IzyksonZuber}%
  \BibitemOpen
  \bibfield  {author} {\bibinfo {author} {\bibfnamefont {C.}~\bibnamefont
  {Itzykson}}\ and\ \bibinfo {author} {\bibfnamefont {J.-B.}\ \bibnamefont
  {Zuber}},\ }\href@noop {} {\bibfield  {journal} {\bibinfo  {journal}
  {McGraw-Hill}\ } (\bibinfo {year} {1980})}\BibitemShut {NoStop}%
\bibitem [{\citenamefont {Grainer}(1990)}]{Grainer}%
  \BibitemOpen
  \bibfield  {author} {\bibinfo {author} {\bibfnamefont {W.}~\bibnamefont
  {Grainer}},\ }\href@noop {} {\bibfield  {journal} {\bibinfo  {journal}
  {Springer-Verlag Berlin Heidelberg}\ } (\bibinfo {year} {1990})}\BibitemShut
  {NoStop}%
\bibitem [{\citenamefont {Liping~Zou}\ and\ \citenamefont
  {Silenko}(2020)}]{LipingZou2020}%
  \BibitemOpen
  \bibfield  {author} {\bibinfo {author} {\bibfnamefont {P.~Z.}\ \bibnamefont
  {Liping~Zou}}\ and\ \bibinfo {author} {\bibfnamefont {A.~J.}\ \bibnamefont
  {Silenko}},\ }\href@noop {} {\bibfield  {journal} {\bibinfo  {journal}
  {Physical Review A}\ }\textbf {\bibinfo {volume} {101}} (\bibinfo {year}
  {2020})}\BibitemShut {NoStop}%
\bibitem [{\citenamefont {Sucher}(1963)}]{Sucher_on_KG}%
  \BibitemOpen
  \bibfield  {author} {\bibinfo {author} {\bibfnamefont {J.}~\bibnamefont
  {Sucher}},\ }\href@noop {} {\bibfield  {journal} {\bibinfo  {journal} {J.
  Math. Phys.}\ }\textbf {\bibinfo {volume} {4}},\ \bibinfo {pages} {17}
  (\bibinfo {year} {1963})}\BibitemShut {NoStop}%
\bibitem [{\citenamefont {H.-J.~Briegel}(1991)}]{Briegel_on_KG}%
  \BibitemOpen
  \bibfield  {author} {\bibinfo {author} {\bibfnamefont {G.~S.}\ \bibnamefont
  {H.-J.~Briegel}, \bibfnamefont {B.-G.~Englert}},\ }\href@noop {} {\bibfield
  {journal} {\bibinfo  {journal} {Z. Naturforsch.}\ }\textbf {\bibinfo {volume}
  {46a}},\ \bibinfo {pages} {933} (\bibinfo {year} {1991})}\BibitemShut
  {NoStop}%
\bibitem [{\citenamefont {Laemmerzahl}(1993)}]{Laemmerzahl1993}%
  \BibitemOpen
  \bibfield  {author} {\bibinfo {author} {\bibfnamefont {C.}~\bibnamefont
  {Laemmerzahl}},\ }\href@noop {} {\bibfield  {journal} {\bibinfo  {journal}
  {J. Math. Phys}\ }\textbf {\bibinfo {volume} {34}},\ \bibinfo {pages} {3918}
  (\bibinfo {year} {1993})}\BibitemShut {NoStop}%
\bibitem [{\citenamefont {Morpurgo}(1960)}]{Morpurgo}%
  \BibitemOpen
  \bibfield  {author} {\bibinfo {author} {\bibfnamefont {G.}~\bibnamefont
  {Morpurgo}},\ }\href@noop {} {\bibfield  {journal} {\bibinfo  {journal} {Il
  Nuovo Cimento}\ }\textbf {\bibinfo {volume} {XV}},\ \bibinfo {pages} {624}
  (\bibinfo {year} {1960})}\BibitemShut {NoStop}%
\bibitem [{\citenamefont {Osche}(1977)}]{Osche}%
  \BibitemOpen
  \bibfield  {author} {\bibinfo {author} {\bibfnamefont {G.}~\bibnamefont
  {Osche}},\ }\href@noop {} {\bibfield  {journal} {\bibinfo  {journal} {Phys.
  Rev. D}\ }\textbf {\bibinfo {volume} {15}},\ \bibinfo {pages} {2181}
  (\bibinfo {year} {1977})}\BibitemShut {NoStop}%
\bibitem [{\citenamefont {de~Vries}(1970)}]{deVries}%
  \BibitemOpen
  \bibfield  {author} {\bibinfo {author} {\bibfnamefont {E.}~\bibnamefont
  {de~Vries}},\ }\href@noop {} {\bibfield  {journal} {\bibinfo  {journal}
  {Fortschr. Phys.}\ }\textbf {\bibinfo {volume} {18}},\ \bibinfo {pages} {149}
  (\bibinfo {year} {1970})}\BibitemShut {NoStop}%
\bibitem [{\citenamefont {Barut}(1980)}]{Barut}%
  \BibitemOpen
  \bibfield  {author} {\bibinfo {author} {\bibfnamefont {A.}~\bibnamefont
  {Barut}},\ }\href@noop {} {\bibfield  {journal} {\bibinfo  {journal} {Dover
  Publications, Inc. New York}\ } (\bibinfo {year} {1980})}\BibitemShut
  {NoStop}%
\bibitem [{\citenamefont {Pauli}(1941)}]{Pauli}%
  \BibitemOpen
  \bibfield  {author} {\bibinfo {author} {\bibfnamefont {W.}~\bibnamefont
  {Pauli}},\ }\href@noop {} {\bibfield  {journal} {\bibinfo  {journal} {Rev.
  Mod. Phys.}\ }\textbf {\bibinfo {volume} {13}},\ \bibinfo {pages} {203}
  (\bibinfo {year} {1941})}\BibitemShut {NoStop}%
\bibitem [{\citenamefont {Silenko}(2003)}]{Silenko2003}%
  \BibitemOpen
  \bibfield  {author} {\bibinfo {author} {\bibfnamefont {A.~J.}\ \bibnamefont
  {Silenko}},\ }\href@noop {} {\bibfield  {journal} {\bibinfo  {journal}
  {Journal of Mathematical Physics}\ }\textbf {\bibinfo {volume} {44}}
  (\bibinfo {year} {2003})}\BibitemShut {NoStop}%
\bibitem [{Note2()}]{Note2}%
  \BibitemOpen
  \bibinfo {note} {Here $\protect \mathsf {\protect \tilde {H}}^{\left (D\right
  )}$ denotes a \protect \emph {scaled} Dirac Hamiltonian.}\BibitemShut {Stop}%
\bibitem [{\citenamefont {Levin}(1959)}]{matrixRiccati}%
  \BibitemOpen
  \bibfield  {author} {\bibinfo {author} {\bibfnamefont {J.~J.}\ \bibnamefont
  {Levin}},\ }\href@noop {} {\bibfield  {journal} {\bibinfo  {journal} {Trans.
  Am. Math. Soc.}\ }\textbf {\bibinfo {volume} {10}},\ \bibinfo {pages} {519}
  (\bibinfo {year} {1959})}\BibitemShut {NoStop}%
\bibitem [{\citenamefont {Furuta}(1983)}]{polar}%
  \BibitemOpen
  \bibfield  {author} {\bibinfo {author} {\bibfnamefont {T.}~\bibnamefont
  {Furuta}},\ }\href@noop {} {\bibfield  {journal} {\bibinfo  {journal} {Acta
  Sci. Math.}\ }\textbf {\bibinfo {volume} {46}},\ \bibinfo {pages} {261}
  (\bibinfo {year} {1983})}\BibitemShut {NoStop}%
\bibitem [{\citenamefont {de~Vries}\ and\ \citenamefont
  {Jonker}(1968)}]{Vries1968}%
  \BibitemOpen
  \bibfield  {author} {\bibinfo {author} {\bibfnamefont {E.}~\bibnamefont
  {de~Vries}}\ and\ \bibinfo {author} {\bibfnamefont {J.}~\bibnamefont
  {Jonker}},\ }\href@noop {} {\bibfield  {journal} {\bibinfo  {journal}
  {Nuclear Physics B}\ }\textbf {\bibinfo {volume} {6}},\ \bibinfo {pages}
  {213} (\bibinfo {year} {1968})}\BibitemShut {NoStop}%
\bibitem [{\citenamefont {Silenko}(2013)}]{Silenko2013}%
  \BibitemOpen
  \bibfield  {author} {\bibinfo {author} {\bibfnamefont {A.~J.}\ \bibnamefont
  {Silenko}},\ }\href@noop {} {\bibfield  {journal} {\bibinfo  {journal}
  {Physics of Particles and Nuclei Letters}\ }\textbf {\bibinfo {volume}
  {10}},\ \bibinfo {pages} {321} (\bibinfo {year} {2013})}\BibitemShut
  {NoStop}%
\bibitem [{\citenamefont {Silenko}(2016{\natexlab{a}})}]{Silenko2015}%
  \BibitemOpen
  \bibfield  {author} {\bibinfo {author} {\bibfnamefont {A.~J.}\ \bibnamefont
  {Silenko}},\ }\href@noop {} {\bibfield  {journal} {\bibinfo  {journal}
  {Physical Review A}\ }\textbf {\bibinfo {volume} {91}},\ \bibinfo {pages}
  {022103} (\bibinfo {year} {2016}{\natexlab{a}})}\BibitemShut {NoStop}%
\bibitem [{\citenamefont {Silenko}(2016{\natexlab{b}})}]{Silenko2016}%
  \BibitemOpen
  \bibfield  {author} {\bibinfo {author} {\bibfnamefont {A.~J.}\ \bibnamefont
  {Silenko}},\ }\href@noop {} {\bibfield  {journal} {\bibinfo  {journal}
  {Physical Review A}\ }\textbf {\bibinfo {volume} {93}},\ \bibinfo {pages}
  {022108} (\bibinfo {year} {2016}{\natexlab{b}})}\BibitemShut {NoStop}%
\bibitem [{\citenamefont {Zanna}(2004)}]{Zanna2004}%
  \BibitemOpen
  \bibfield  {author} {\bibinfo {author} {\bibfnamefont {A.}~\bibnamefont
  {Zanna}},\ }\href@noop {} {\bibfield  {journal} {\bibinfo  {journal}
  {Mathematics of Computation}\ }\textbf {\bibinfo {volume} {73}},\ \bibinfo
  {pages} {761} (\bibinfo {year} {2004})}\BibitemShut {NoStop}%
\bibitem [{\citenamefont {A.~Iserles}(2002)}]{A.Iserles2002}%
  \BibitemOpen
  \bibfield  {author} {\bibinfo {author} {\bibfnamefont {S.~N.}\ \bibnamefont
  {A.~Iserles}},\ }\href@noop {} {\bibfield  {journal} {\bibinfo  {journal}
  {Phil. Trans. R. Soc. Lond.}\ }\textbf {\bibinfo {volume} {A357 (1754)}},\
  \bibinfo {pages} {983} (\bibinfo {year} {2002})}\BibitemShut {NoStop}%
\bibitem [{\citenamefont {Wilcox}(1967)}]{Wilcox}%
  \BibitemOpen
  \bibfield  {author} {\bibinfo {author} {\bibfnamefont {R.~M.}\ \bibnamefont
  {Wilcox}},\ }\href@noop {} {\bibfield  {journal} {\bibinfo  {journal}
  {Journal of Mathematical Physics}\ }\textbf {\bibinfo {volume} {8}},\
  \bibinfo {pages} {962} (\bibinfo {year} {1967})}\BibitemShut {NoStop}%
\bibitem [{\citenamefont {Iserles}(2002)}]{Iserles}%
  \BibitemOpen
  \bibfield  {author} {\bibinfo {author} {\bibfnamefont {A.}~\bibnamefont
  {Iserles}},\ }\href@noop {} {\bibfield  {journal} {\bibinfo  {journal}
  {Notices of the AMS}\ }\textbf {\bibinfo {volume} {49}},\ \bibinfo {pages}
  {430} (\bibinfo {year} {2002})}\BibitemShut {NoStop}%
\bibitem [{Note3()}]{Note3}%
  \BibitemOpen
  \bibinfo {note} {There is a typo in the original article of Foldy and
  Wouthuysen regarding the sign of the scalar potential term in the
  Hamiltonian, see their Eq. (34) in \cite {FoldyW50}. Consequently the sign of
  the Darwin term in their Eq.(35) is incorrect, whereas the sign of the
  spin-orbit term is correct. We caution the reader that in the article of
  Silenko \cite {Silenko2003}, see his Eq.(40), the sign of the spin-orbit term
  is incorrect, while then again the sign of the scalar potential term and that
  one of the Darwin term is correct.}\BibitemShut {Stop}%
\end{thebibliography}%


\begin{thebibliography}{10}%
\makeatletter
\providecommand \@ifxundefined [1]{%
 \@ifx{#1\undefined}
}%
\providecommand \@ifnum [1]{%
 \ifnum #1\expandafter \@firstoftwo
 \else \expandafter \@secondoftwo
 \fi
}%
\providecommand \@ifx [1]{%
 \ifx #1\expandafter \@firstoftwo
 \else \expandafter \@secondoftwo
 \fi
}%
\providecommand \natexlab [1]{#1}%
\providecommand \enquote  [1]{``#1''}%
\providecommand \bibnamefont  [1]{#1}%
\providecommand \bibfnamefont [1]{#1}%
\providecommand \citenamefont [1]{#1}%
\providecommand \href@noop [0]{\@secondoftwo}%
\providecommand \href [0]{\begingroup \@sanitize@url \@href}%
\providecommand \@href[1]{\@@startlink{#1}\@@href}%
\providecommand \@@href[1]{\endgroup#1\@@endlink}%
\providecommand \@sanitize@url [0]{\catcode `\\12\catcode `\$12\catcode
  `\&12\catcode `\#12\catcode `\^12\catcode `\_12\catcode `\%12\relax}%
\providecommand \@@startlink[1]{}%
\providecommand \@@endlink[0]{}%
\providecommand \url  [0]{\begingroup\@sanitize@url \@url }%
\providecommand \@url [1]{\endgroup\@href {#1}{\urlprefix }}%
\providecommand \urlprefix  [0]{URL }%
\providecommand \Eprint [0]{\href }%
\providecommand \doibase [0]{https://doi.org/}%
\providecommand \selectlanguage [0]{\@gobble}%
\providecommand \bibinfo  [0]{\@secondoftwo}%
\providecommand \bibfield  [0]{\@secondoftwo}%
\providecommand \translation [1]{[#1]}%
\providecommand \BibitemOpen [0]{}%
\providecommand \bibitemStop [0]{}%
\providecommand \bibitemNoStop [0]{.\EOS\space}%
\providecommand \EOS [0]{\spacefactor3000\relax}%
\providecommand \BibitemShut  [1]{\csname bibitem#1\endcsname}%
\let\auto@bib@innerbib\@empty
\bibitem [{\citenamefont {Schopohl}\ and\ \citenamefont
  {Cetin}(2022)}]{Schopohl2022MS}%
  \BibitemOpen
  \bibfield  {author} {\bibinfo {author} {\bibfnamefont {N.}~\bibnamefont
  {Schopohl}}\ and\ \bibinfo {author} {\bibfnamefont {N.~S.}\ \bibnamefont
  {Cetin}},\ }\href@noop {} {\bibfield  {journal} {\bibinfo  {journal}
  {submitted}\ } (\bibinfo {year} {2022})}\BibitemShut {NoStop}%
\bibitem [{\citenamefont {Bruening}\ \emph {et~al.}(2011)\citenamefont
  {Bruening}, \citenamefont {Grushin}, \citenamefont {Dobrokhotov},\ and\
  \citenamefont {Tudorovskii}}]{Bruening2011}%
  \BibitemOpen
  \bibfield  {author} {\bibinfo {author} {\bibfnamefont {J.}~\bibnamefont
  {Bruening}}, \bibinfo {author} {\bibfnamefont {V.~V.}\ \bibnamefont
  {Grushin}}, \bibinfo {author} {\bibfnamefont {S.~Y.}\ \bibnamefont
  {Dobrokhotov}},\ and\ \bibinfo {author} {\bibfnamefont {T.~Y.}\ \bibnamefont
  {Tudorovskii}},\ }\href@noop {} {\bibfield  {journal} {\bibinfo  {journal}
  {Theoretical and Mathematical Physics}\ }\textbf {\bibinfo {volume} {167}},\
  \bibinfo {pages} {547} (\bibinfo {year} {2011})}\BibitemShut {NoStop}%
\bibitem [{\citenamefont {Eriksen}\ and\ \citenamefont
  {Kolsrud}(1960)}]{EriksenKolsrud}%
  \BibitemOpen
  \bibfield  {author} {\bibinfo {author} {\bibfnamefont {E.}~\bibnamefont
  {Eriksen}}\ and\ \bibinfo {author} {\bibfnamefont {M.}~\bibnamefont
  {Kolsrud}},\ }\href@noop {} {\bibfield  {journal} {\bibinfo  {journal} {Nuovo
  Cimento Suppl.}\ }\textbf {\bibinfo {volume} {18}},\ \bibinfo {pages} {1}
  (\bibinfo {year} {1960})}\BibitemShut {NoStop}%
\bibitem [{\citenamefont {Neznamov}\ and\ \citenamefont
  {Silenko}(2009)}]{NeznamovSilenko2009}%
  \BibitemOpen
  \bibfield  {author} {\bibinfo {author} {\bibfnamefont {V.}~\bibnamefont
  {Neznamov}}\ and\ \bibinfo {author} {\bibfnamefont {A.}~\bibnamefont
  {Silenko}},\ }\href@noop {} {\bibfield  {journal} {\bibinfo  {journal}
  {Journal of Mathematical Physics}\ }\textbf {\bibinfo {volume} {50}}
  (\bibinfo {year} {2009})}\BibitemShut {NoStop}%
\bibitem [{\citenamefont {de~Vries}\ and\ \citenamefont
  {Jonker}(1968)}]{Vries1968}%
  \BibitemOpen
  \bibfield  {author} {\bibinfo {author} {\bibfnamefont {E.}~\bibnamefont
  {de~Vries}}\ and\ \bibinfo {author} {\bibfnamefont {J.}~\bibnamefont
  {Jonker}},\ }\href@noop {} {\bibfield  {journal} {\bibinfo  {journal}
  {Nuclear Physics B}\ }\textbf {\bibinfo {volume} {6}},\ \bibinfo {pages}
  {213} (\bibinfo {year} {1968})}\BibitemShut {NoStop}%
\bibitem [{Note1()}]{Note1}%
  \BibitemOpen
  \bibinfo {note} {A particularly clear presentation of the
  Pauli-Achieser-Berestezki elimination method is due to Stephani \cite
  {Stephani}}\BibitemShut {NoStop}%
\bibitem [{\citenamefont {Silenko}(2016{\natexlab{a}})}]{Silenko2016}%
  \BibitemOpen
  \bibfield  {author} {\bibinfo {author} {\bibfnamefont {A.~J.}\ \bibnamefont
  {Silenko}},\ }\href@noop {} {\bibfield  {journal} {\bibinfo  {journal}
  {Physical Review A}\ }\textbf {\bibinfo {volume} {93}},\ \bibinfo {pages}
  {022108} (\bibinfo {year} {2016}{\natexlab{a}})}\BibitemShut {NoStop}%
\bibitem [{\citenamefont {Silenko}(2016{\natexlab{b}})}]{Silenko2015}%
  \BibitemOpen
  \bibfield  {author} {\bibinfo {author} {\bibfnamefont {A.~J.}\ \bibnamefont
  {Silenko}},\ }\href@noop {} {\bibfield  {journal} {\bibinfo  {journal}
  {Physical Review A}\ }\textbf {\bibinfo {volume} {91}},\ \bibinfo {pages}
  {022103} (\bibinfo {year} {2016}{\natexlab{b}})}\BibitemShut {NoStop}%
\bibitem [{\citenamefont {Foldy}\ and\ \citenamefont
  {Wouthuysen}(1950)}]{FoldyW50}%
  \BibitemOpen
  \bibfield  {author} {\bibinfo {author} {\bibfnamefont {L.}~\bibnamefont
  {Foldy}}\ and\ \bibinfo {author} {\bibfnamefont {S.}~\bibnamefont
  {Wouthuysen}},\ }\href@noop {} {\bibfield  {journal} {\bibinfo  {journal}
  {Physical Review}\ }\textbf {\bibinfo {volume} {78}},\ \bibinfo {pages} {29}
  (\bibinfo {year} {1950})}\BibitemShut {NoStop}%
\bibitem [{\citenamefont {Stephani}(1965)}]{Stephani}%
  \BibitemOpen
  \bibfield  {author} {\bibinfo {author} {\bibfnamefont {H.}~\bibnamefont
  {Stephani}},\ }\href@noop {} {\bibfield  {journal} {\bibinfo  {journal}
  {Annalen der Physik}\ }\textbf {\bibinfo {volume} {15}},\ \bibinfo {pages}
  {12} (\bibinfo {year} {1965})}\BibitemShut {NoStop}%
\end{thebibliography}%

\end{document}


%
%
\title{Supplemental Material for MS entitled \protect\\ "A Flow Equation Approach Striving towards an Energy-Separating Hamiltonian Unitary Equivalent to the Dirac Hamiltonian with Coupling to Electromagnetic Fields"}
\author{N. Schopohl}
\email{nils.schopohl@uni-tuebingen.de}
\author{N.S. Cetin}
\date{\today}
\affiliation{Eberhard-Karls-Universit{\"a}t T{\"u}bingen\\ Institute for Theoretical Physics and CQ Center for Collective Quantum
Phenomena\\  Auf der Morgenstelle 14 , D-72076 T{\"u}bingen, Germany}

\begin{abstract}
\begin{description}
\item[The ensuing comprises supplemental material for the following sections in \cite{Schopohl2022MS}]
\item[XIV] "Perturbation Series for Generator $\omega\left(s\right)$ of Beta-Flow".
\item[XV] "Perturbation Series for Operator $\Omega_{u}\left(s\right)$ .
\item[XVI] "The Relativistic Corrections to the Schrödinger-Pauli Hamiltonian $\mathsf{H}^{\left(SP\right)}$ \\as a Series Progressing in $\frac{v^2}{c^2}$.
\end{description}
\end{abstract}
\maketitle
\section*{Supplemental material for section XIV in \cite{Schopohl2022MS}}
Mind that Eq.(148) in \cite{Schopohl2022MS}
is a \emph{nonlinear}  ODE and therefore more than one solution may exist.
To find the (physical) correct solution the property
 $\mathsf{Z}\left(s\right)\mathsf{Z}\left(s\right)=\mathsf{1}_{4\times4}$ needs to be implemented into the ansatz Eq.(149) in \cite{Schopohl2022MS} , thus leading for
\begin{equation}
\mathsf{Q}\left(s\right)=\mathsf{Z}\left(s\right)\mathsf{\tilde{H}}^{\left(D\right)}\label{eq: def Q(s)}
\end{equation} to the constraint
\begin{equation}
\mathsf{Q}^{\dagger}\left(s\right)\mathsf{Q}\left(s\right)=\mathsf{\tilde{H}}^{\left(D\right)}\mathsf{\tilde{H}}^{\left(D\right)}\label{eq: constraint Q(s)^2=H^2}
\end{equation}
Insertion of the perturbation series ansatz
\begin{equation}
\mathsf{Q}\left(s\right)=\mathsf{1}_{4\times4}+\sum_{j=1}^{\infty}\kappa^{j}\left(\mathsf{Q}^{\left(j\right)}\left(s\right)\right)^{\dagger}\label{eq: ansatz Q(s)}
\end{equation}
and comparing coefficients of equal powers $\kappa^{n}$ for $n=1,2,3,...$ on either side of (\ref{eq: constraint Q(s)^2=H^2}) implies
\[
\left(\mathsf{1}_{4\times4}+\sum_{j=1}^{\infty}\kappa^{j}\left(\mathsf{Q}^{\left(j\right)}\left(s\right)\right)^{\dagger}\right)\left(\mathsf{1}_{4\times4}+\sum_{j'=1}^{\infty}\kappa^{j'}\mathsf{Q}^{\left(j'\right)}\left(s\right)\right)=\mathsf{\tilde{H}}^{\left(D\right)}\mathsf{\tilde{H}}^{\left(D\right)}=\mathsf{1}_{4\times4}+\sum_{n=2}^{4}\kappa^{n}\mathsf{R}^{\left(n\right)}
\]
, or equivalently
\begin{eqnarray}
 &  & \kappa\left(\left(\mathsf{Q}^{\left(1\right)}\left(s\right)\right)^{\dagger}+\mathsf{Q}^{\left(1\right)}\left(s\right)\right)+\sum_{n=2}^{\infty}\kappa^{n}\left(\left(\mathsf{Q}^{\left(n\right)}\left(s\right)\right)^{\dagger}+\mathsf{Q}^{\left(n\right)}\left(s\right)\right)\nonumber \\
 & = & \sum_{n=2}^{\infty}\kappa^{n}\mathsf{R}^{\left(n\right)}-\sum_{j=1}^{\infty}\kappa^{j}\left(\mathsf{Q}^{\left(j\right)}\left(s\right)\right)^{\dagger}\sum_{j'=1}^{\infty}\kappa^{j'}\mathsf{Q}^{\left(j'\right)}\left(s\right)\nonumber \\
 & = & \sum_{n=2}^{\infty}\kappa^{n}\left(\mathsf{R}^{\left(n\right)}-\sum_{j=1}^{n-1}\left(\mathsf{Q}^{\left(j\right)}\left(s\right)\right)^{\dagger}\mathsf{Q}^{\left(n-j\right)}\left(s\right)\right)\label{eq: constraint Q(s)^2=H^2 II}
\end{eqnarray}
, whereby  $\mathsf{R}^{\left(n\right)}$ is stated in Eq.(152) in \cite{Schopohl2022MS}.
So we conclude
\begin{eqnarray}
\left(\mathsf{Q}^{\left(1\right)}\left(s\right)\right)^{\dagger} & = & -\mathsf{Q}^{\left(1\right)}\left(s\right)\nonumber \\
 & n\geq2\nonumber \\
\left(\mathsf{Q}^{\left(n\right)}\left(s\right)\right)^{\dagger}+\mathsf{Q}^{\left(n\right)}\left(s\right) & = & \mathsf{R}^{\left(n\right)}-\sum_{j=1}^{n-1}\left(\mathsf{Q}^{\left(j\right)}\left(s\right)\right)^{\dagger}\mathsf{Q}^{\left(n-j\right)}\left(s\right)\label{eq: constraint Q(s)^2=H^2 III}
\end{eqnarray}
These relations apply for all terms $\mathsf{Q}^{\left(n\right)}\left(s\right)$ solving the recursion (154) in \cite{Schopohl2022MS}.

For $n=2$ then
\begin{eqnarray}
\left(\mathsf{Q}^{\left(2\right)}\left(s\right)\right)^{\dagger}+\mathsf{Q}^{\left(2\right)}\left(s\right) & = & \mathsf{R}^{\left(2\right)}-\left(\mathsf{Q}^{\left(1\right)}\left(s\right)\right)^{\dagger}\mathsf{Q}^{\left(1\right)}\left(s\right)\label{eq: constraint Q(s)^2=H^2 for  n=2}\\
 & = & \left(2\beta\mathcal{\tilde{E}}+\mathcal{\tilde{O}}^{2}\right)+\mathsf{Q}^{\left(1\right)}\left(s\right)\mathsf{Q}^{\left(1\right)}\left(s\right)\nonumber
\end{eqnarray}
, and in particular for $s=0$
\begin{eqnarray}
\left(\mathsf{Q}^{\left(2\right)}\left(0\right)\right)^{\dagger}+\mathsf{Q}^{\left(2\right)}\left(0\right) & = & \left(2\beta\mathcal{\tilde{E}}+\mathcal{\tilde{O}}^{2}\right)+\beta\mathcal{\tilde{O}}\beta\mathcal{\tilde{O}}\label{eq: constraint Q(s)^2=H^2 for  n=2  II}\\
 & = & \left(2\beta\tilde{\mathcal{E}}+\mathcal{\tilde{O}}^{2}\right)-\mathcal{\tilde{O}}^{2}\nonumber \\
 & = & 2\beta\mathcal{\tilde{E}}\nonumber
\end{eqnarray}
, this indeed being consistent with the posed initial values in Eq.(155) in \cite{Schopohl2022MS}.

Analysis of the recursion Eq.(157) in \cite{Schopohl2022MS} for
 $\omega^{\left(n\right)}\left(s\right)=\left(\mathsf{Q}^{\left(n\right)}\left(s\right)\right)^{\dagger}-\mathsf{Q}^{\left(n\right)}\left(s\right)$
leads to the following results
\begin{eqnarray}
\omega^{\left(1\right)}\left(s\right) & = & -2e^{-4s}\beta\tilde{\mathcal{O}}\label{eq: omega_n_(s) for n=1,2,3,4}\\
\omega^{\left(2\right)}\left(s\right) & = & \mathsf{0}_{4\times4}\nonumber \\
\omega^{\left(3\right)}\left(s\right) & = & -4e^{-4s}s\left[\tilde{\mathcal{O}},\tilde{\mathcal{E}}\right]+\left(e^{-4s}\left(-\frac{1}{2}+4s\right)+\frac{e^{-12s}}{2}\right)\beta\tilde{\mathcal{O}}^{3}\nonumber \\
\omega^{\left(4\right)}\left(s\right) & = & \left(e^{-4s}\left(-\frac{1}{2}+2s\right)+\frac{e^{-8s}}{2}\right)\beta\left[\mathcal{\tilde{O}},\left(\tilde{\mathcal{E}}\tilde{\mathcal{O}}+\tilde{\mathcal{O}}\tilde{\mathcal{E}}\right)\right]\nonumber
\end{eqnarray}
Inserting the elementary integrals resulting from the evaluation of the recursion Eq.(157) in \cite{Schopohl2022MS}
, one finds after some adjustments with help of the identities
\begin{eqnarray}
\tilde{\mathcal{O}}^{2}\left[\tilde{\mathcal{O}},\tilde{\mathcal{E}}\right]+\tilde{\mathcal{O}}\left[\mathcal{\tilde{O}},\tilde{\mathcal{E}}\right]\tilde{\mathcal{O}}+\left[\mathcal{\tilde{O}},\tilde{\mathcal{E}}\right]\tilde{\mathcal{O}}^{2} & = & \left[\mathcal{\tilde{O}}^{3},\tilde{\mathcal{E}}\right]\label{eq: commutator identities I}\\
\nonumber \\
\tilde{\mathcal{O}}^{2}\left[\mathcal{\tilde{O}},\tilde{\mathcal{E}}\right]+\left[\tilde{\mathcal{O}},\tilde{\mathcal{E}}\right]\mathcal{\tilde{O}}^{2} & = & \left[\tilde{\mathcal{O}},\left(\tilde{\mathcal{E}}\mathcal{\tilde{O}}^{2}+\tilde{\mathcal{O}}^{2}\tilde{\mathcal{E}}\right)\right]\nonumber \\
\nonumber \\
\left[\tilde{\mathcal{O}},\left[\tilde{\mathcal{O}},\left[\mathcal{\tilde{O}},\tilde{\mathcal{E}}\right]\right]\right] & = & \tilde{\mathcal{O}}^{2}\left[\mathcal{\tilde{O}},\tilde{\mathcal{E}}\right]+\left[\tilde{\mathcal{O}},\tilde{\mathcal{E}}\right]\mathcal{\tilde{O}}^{2}-2\tilde{\mathcal{O}}\left[\mathcal{\tilde{O}},\tilde{\mathcal{E}}\right]\tilde{\mathcal{O}}\nonumber \\
 & = & \left[\mathcal{\tilde{O}}^{3},\tilde{\mathcal{E}}\right]-3\tilde{\mathcal{O}}\left[\mathcal{\tilde{O}},\tilde{\mathcal{E}}\right]\tilde{\mathcal{O}}\nonumber \\
\nonumber \\
\tilde{\mathcal{O}}\left[\tilde{\mathcal{O}},\tilde{\mathcal{E}}\right]\tilde{\mathcal{O}} & = & \frac{1}{3}\left[\tilde{\mathcal{O}}^{3},\tilde{\mathcal{E}}\right]-\frac{1}{3}\left[\tilde{\mathcal{O}},\left[\tilde{\mathcal{O}},\left[\tilde{\mathcal{O}},\tilde{\mathcal{E}}\right]\right]\right]\nonumber \\
\nonumber \\
\tilde{\mathcal{O}}^{2}\left[\tilde{\mathcal{O}},\tilde{\mathcal{E}}\right]+\left[\tilde{\mathcal{O}},\tilde{\mathcal{E}}\right]\mathcal{\tilde{O}}^{2} & = & \frac{2}{3}\left[\tilde{\mathcal{O}}^{3},\tilde{\mathcal{E}}\right]+\frac{1}{3}\left[\tilde{\mathcal{O}},\left[\tilde{\mathcal{O}},\left[\tilde{\mathcal{O}},\tilde{\mathcal{E}}\right]\right]\right]\nonumber
\end{eqnarray}
the result
\begin{eqnarray}
\label{eq:omega_5(s)}\\
\omega^{\left(5\right)}\left(s\right) & = & \left(\begin{array}{c}
-4e^{-4s}s^{2}\beta\left[\left[\tilde{\mathcal{O}},\tilde{\mathcal{E}}\right],\tilde{\mathcal{E}}\right]\\
\\
+\left(\begin{array}{c}
e^{-4s}\left(-\frac{1}{12}+\frac{1}{3}s+\frac{2}{3}s^{2}\right)-\frac{1}{12}e^{-8s}+e^{-12s}\left(\frac{1}{6}+\frac{2}{3}s\right)\end{array}\right)\left[\tilde{\mathcal{O}},\left[\mathcal{\tilde{O}},\left[\mathcal{\tilde{O}},\mathcal{\tilde{E}}\right]\right]\right]\\
\\
+\left(\begin{array}{c}
e^{-4s}\left(-\frac{1}{6}-\frac{1}{3}s+\frac{16}{3}s^{2}\right)+\frac{1}{3}e^{-8s}+e^{-12s}\left(-\frac{1}{6}+\frac{1}{3}s\right)\end{array}\right)\left[\mathcal{\tilde{O}}^{3},\tilde{\mathcal{E}}\right]\\
\\
+\left(\begin{array}{c}
e^{-4s}\left(\frac{1}{4}-4s^{2}\right)-e^{-12s}\left(\frac{1}{8}+3s\right)-\frac{1}{8}e^{-20s}\end{array}\right)\beta\mathcal{\tilde{O}}^{5}
\end{array}\right)\nonumber
\end{eqnarray}

\section*{Supplemental material for section XV in \cite{Schopohl2022MS}}
Insertion of the perturbation series
\begin{equation}
\Omega_{u}\left(s\right)=\sum_{n=0}^{\infty}\kappa^{2n+1}\Omega^{\left(2n+1\right)}\left(s\right)\label{eq:perturbation series Omega_u_(s)}
\end{equation}
into Eq.(168 in \cite{Schopohl2022MS} leads to
\begin{equation}
\Omega^{\left(1\right)}\left(s\right)=\int_{0}^{s}ds'\omega^{\left(1\right)}\left(s'\right)=\left(-\frac{1}{2}+\frac{e^{-4s}}{2}\right)\beta\tilde{\mathcal{O}}\label{eq: Omega_1_(s)}
\end{equation}
Now, because
\begin{equation}
\left[\Omega^{\left(1\right)}\left(s'\right),\omega^{\left(1\right)}\left(s'\right)\right]=\mathsf{0}_{4\times4}\label{eq: [Omega_1 , omega_1]=0}
\end{equation}
, all the commutator term  $\left(\mathsf{ad}_{\Omega_{u}\left(s'\right)}\right)^{2n}\circ\omega_{u}\left(s'\right)$
 in Eq. (168) in\cite{Schopohl2022MS} are actually not small of order
 $O\left(\kappa^{2n+1}\right)$  but small of order  $O\left(\kappa^{2n+3}\right)$ :
\begin{eqnarray}
 &  & \left(\mathsf{ad}_{\Omega_{u}\left(s'\right)}\right)^{2}\circ\omega_{u}\left(s'\right)\label{eq: O(kapa^5)}\\
 & = & \left[\Omega_{u}\left(s'\right),\left[\Omega_{u}\left(s'\right),\omega_{u}\left(s'\right)\right]\right]\nonumber \\
 & = & \left[\sum_{j''=0}^{\infty}\kappa^{2j''+1}\Omega^{\left(2j''+1\right)}\left(s'\right),\left[\sum_{j'=0}^{\infty}\kappa^{2j'+1}\Omega^{\left(2j'+1\right)}\left(s'\right),\sum_{j=0}^{\infty}\kappa^{2j+1}\omega^{\left(2j+1\right)}\left(s'\right)\right]\right]\nonumber \\
 & = & \sum_{j''=0}^{\infty}\sum_{j'=0}^{\infty}\sum_{j=0}^{\infty}\left[\Omega^{\left(2j''+1\right)}\left(s'\right),\left[\Omega^{\left(2j'+1\right)}\left(s'\right),\omega^{\left(2j+1\right)}\left(s'\right)\right]\right]\kappa^{2j''+1}\kappa^{2j'+1}\kappa^{2j+1}\nonumber \\
 & = & \sum_{j''=0}^{\infty}\sum_{j'=0}^{\infty}\sum_{j=0}^{\infty}\left[\Omega^{\left(2j''+1\right)}\left(s'\right),\left[\Omega^{\left(2j'+1\right)}\left(s'\right),\omega^{\left(2j+1\right)}\left(s'\right)\right]\right]\sum_{n=0}^{\infty}\delta_{2n+3,\left(2j+1\right)+\left(2j'+1\right)+\left(2j''+1\right)}\kappa^{2n+3}\nonumber \\
 & = & \sum_{n=0}^{\infty}\kappa^{2n+3}\sum_{j''=0}^{\infty}\sum_{j'=0}^{\infty}\Theta\left(n-j'-j''\right)\left[\Omega^{\left(2j''+1\right)}\left(s'\right),\left[\Omega^{\left(2j'+1\right)}\left(s'\right),\omega^{\left(2\left(n-j'-j''\right)+1\right)}\left(s'\right)\right]\right]\nonumber \\
 & = & \sum_{n=1}^{\infty}\kappa^{2n+3}\sum_{j''=0}^{\infty}\sum_{j'=0}^{\infty}\Theta\left(n-j'-j''\right)\left[\Omega^{\left(2j''+1\right)}\left(s'\right),\left[\Omega^{\left(2j'+1\right)}\left(s'\right),\omega^{\left(2\left(n-j'-j''\right)+1\right)}\left(s'\right)\right]\right]\nonumber \\
 & = & O\left(\kappa^{5}\right)\nonumber
\end{eqnarray}
, and for that same reason
\begin{eqnarray}
\left(\mathsf{ad}_{\Omega_{u}\left(s'\right)}\right)^{2n}\circ\omega_{u}\left(s'\right) & = & O\left(\kappa^{2n+3}\right)\label{eq: O(kappa^(2n+3)}\\
n & = & 1,2,3,....\nonumber
\end{eqnarray}

If we are only interested in an expansion up to and including terms of order  $\kappa^{6}$  it is adequate to ignore in Eq.(168) in \cite{Schopohl2022MS}
all commutator terms but the first one, leading to
\begin{eqnarray}
\Omega^{\left(3\right)}\left(s\right) & = & \int_{0}^{s}ds'\omega^{\left(3\right)}\left(s'\right)\label{eq: Omega_3_(s)}\\
 & = & \begin{cases}
\left(-\frac{1}{4}+e^{-4s}\left(\frac{1}{4}+s\right)\right)\left[\tilde{\mathcal{O}},\tilde{\mathcal{E}}\right]\\
+\left(\frac{1}{6}-e^{-4s}\left(\frac{1}{8}+s\right)-\frac{e^{-12s}}{24}\right)\beta\mathcal{\tilde{O}}^{3}
\end{cases}\nonumber
\end{eqnarray}

The determination of the fifth-order term $\Omega^{\left(5\right)}\left(s\right)$ requires to evaluate two double commutator terms:
\begin{eqnarray}
\label{eq: Omega_5_(s)}\\
\Omega^{\left(5\right)}\left(s\right) & = & \begin{cases}
\int_{0}^{s}ds'\omega^{\left(5\right)}\left(s'\right)\\
\\
-\frac{2}{3}\int_{0}^{s}ds'\left(\left[\Omega^{\left(1\right)}\left(s'\right),\left[\Omega^{\left(3\right)}\left(s'\right),\omega^{\left(1\right)}\left(s'\right)\right]\right]+\left[\Omega^{\left(1\right)}\left(s'\right),\left[\Omega^{\left(1\right)}\left(s'\right),\omega^{\left(3\right)}\left(s'\right)\right]\right]\right)
\end{cases}\nonumber
\end{eqnarray}
Noting
\begin{eqnarray*}
\left[\beta\tilde{\mathcal{O}}^{3},\beta\tilde{\mathcal{O}}\right] & = & \beta\tilde{\mathcal{O}}^{3}\beta\tilde{\mathcal{O}}-\beta\tilde{\mathcal{O}}\beta\mathcal{\tilde{O}}^{3}\\
 & = & -\mathcal{\tilde{O}}^{3}\tilde{\mathcal{O}}+\tilde{\mathcal{O}}\tilde{\mathcal{O}}^{3}\\
 & \equiv & 0_{4\times4}
\end{eqnarray*}
then
\begin{eqnarray}
 &  & \int_{0}^{s}ds'\left[\Omega^{\left(1\right)}\left(s'\right),\left[\Omega^{\left(1\right)}\left(s'\right),\omega^{\left(3\right)}\left(s'\right)\right]\right]\label{eq: C1}\\
 & = & \int_{0}^{s}ds'\left(-\frac{1-e^{-4s'}}{2}\right)^{2}\left(-4e^{-4s}s'\right)\left[\beta\tilde{\mathcal{O}},\left[\beta\tilde{\mathcal{O}},\left[\tilde{\mathcal{O}},\tilde{\mathcal{E}}\right]\right]\right]\nonumber
\end{eqnarray}
and
\begin{eqnarray}
 &  & \int_{0}^{s}ds'\left[\Omega^{\left(1\right)}\left(s'\right),\left[\Omega^{\left(3\right)}\left(s'\right),\omega^{\left(1\right)}\left(s'\right)\right]\right]\label{eq: C2}\\
 & = & \int_{0}^{s}ds'\left(-\frac{1-e^{-4s'}}{2}\right)\left(-\frac{1}{4}+e^{-4s'}\left(\frac{1}{4}+s'\right)\right)\left(-2e^{-4s'}\right)\left(-\left[\beta\tilde{\mathcal{O}},\left[\beta\tilde{\mathcal{O}},\left[\tilde{\mathcal{O}},\tilde{\mathcal{E}}\right]\right]\right]\right)\nonumber
\end{eqnarray}
Taking into account the identities (\ref{eq: commutator identities I})  the commutator term may be rewritten as
\begin{eqnarray}
 &  & \left[\beta\tilde{\mathcal{O}},\left[\beta\tilde{\mathcal{O}},\left[\tilde{\mathcal{O}},\tilde{\mathcal{E}}\right]\right]\right]\label{eq: commutator identities II}\\
 & = & -\mathcal{\tilde{O}}^{2}\left[\mathcal{\tilde{O}},\tilde{\mathcal{E}}\right]-\left[\mathcal{\tilde{O}},\tilde{\mathcal{E}}\right]\mathcal{\tilde{O}}^{2}-2\tilde{\mathcal{O}}\left[\mathcal{\tilde{O}},\tilde{\mathcal{E}}\right]\tilde{\mathcal{O}}\nonumber \\
 & = & -\frac{2}{3}\left[\mathcal{\tilde{O}}^{3},\tilde{\mathcal{E}}\right]-\frac{1}{3}\left[\mathcal{\tilde{O}},\left[\tilde{\mathcal{O}},\left[\mathcal{\tilde{O}},\tilde{\mathcal{E}}\right]\right]\right]-2\left(\frac{1}{3}\left[\mathcal{\tilde{O}}^{3},\tilde{\mathcal{E}}\right]-\frac{1}{3}\left[\mathcal{\tilde{O}},\left[\mathcal{\tilde{O}},\left[\tilde{\mathcal{O}},\tilde{\mathcal{E}}\right]\right]\right]\right)\nonumber \\
 & = & -\frac{4}{3}\left[\mathcal{\tilde{O}}^{3},\mathcal{\tilde{E}}\right]+\frac{1}{3}\left[\tilde{\mathcal{O}},\left[\mathcal{\tilde{O}},\left[\mathcal{\tilde{O}},\tilde{\mathcal{E}}\right]\right]\right]\nonumber
\end{eqnarray}
That way
\begin{eqnarray}
 &  & \int_{0}^{s}ds'\left(\left[\Omega^{\left(1\right)}\left(s'\right),\left[\Omega^{\left(3\right)}\left(s'\right),\omega^{\left(1\right)}\left(s'\right)\right]\right]
 +\left[\Omega^{\left(1\right)}\left(s'\right),\left[\Omega^{\left(1\right)}\left(s'\right),\omega^{\left(3\right)}\left(s'\right)\right]\right]\right)\label{eq: C1+C2}\\
 & = & \left(-\frac{5}{192}+\frac{1}{4}e^{-4s}s+e^{-8s}\left(\frac{3}{64}-\frac{1}{8}s\right)-\frac{1}{48}e^{-12s}\right)\left(-\frac{4}{3}\left[\mathcal{\tilde{O}}^{3},\mathcal{\tilde{E}}\right]
 +\frac{1}{3}\left[\tilde{\mathcal{O}},\left[\mathcal{\tilde{O}},\left[\mathcal{\tilde{O}},\tilde{\mathcal{E}}\right]\right]\right]\right)\nonumber
\end{eqnarray}
, and upon insertion into (\ref{eq: Omega_5_(s)})  finally
\begin{eqnarray}
\label{eq:Omega_5_(s)  II}\\
\Omega^{\left(5\right)}\left(s\right) & = & \left(\begin{array}{c}
\left(-\frac{1}{8}+e^{-4s}\left(\frac{1}{8}+\frac{1}{2}s+s^{2}\right)\right)\beta\left[\left[\tilde{\mathcal{O}},\tilde{\mathcal{E}}\right],\tilde{\mathcal{E}}\right]\\
\\
+\left(\begin{array}{c}
\frac{25}{864}-e^{-4s}\left(\frac{1}{48}+\frac{1}{6}s+\frac{1}{6}s^{2}\right)
+\frac{1}{96}e^{-8s}-e^{-12s}\left(\frac{1}{54}+\frac{1}{18}s\right)\end{array}\right)\left[\tilde{\mathcal{O}},\left[\tilde{\mathcal{O}},\left[\tilde{\mathcal{O}},\tilde{\mathcal{E}}\right]\right]\right]\\
\\
+\left(\begin{array}{c}
\frac{29}{216}\end{array}-e^{-4s}\left(\frac{5}{48}+\frac{7}{12}s+\frac{4}{3}s^{2}\right)-\frac{1}{24}e^{-8s}+e^{-12s}\left(\frac{5}{432}-\frac{1}{36}s\right)\right)\left[\mathcal{\tilde{O}}^{3},\tilde{\mathcal{E}}\right]\\
\\
+\left(\begin{array}{c}
-\frac{1}{10}+e^{-4s}\left(\frac{1}{16}+\frac{1}{2}s+s^{2}\right)+e^{-12s}\left(\frac{1}{32}+\frac{1}{4}s\right)+\frac{1}{160}e^{-20s}\end{array}\right)\beta\mathcal{\tilde{O}}^{5}\\
\\
\\
-\frac{2}{3}\left(-\frac{5}{192}+\frac{1}{4}e^{-4s}s
+e^{-8s}\left(\frac{3}{64}-\frac{1}{8}s\right)-\frac{1}{48}e^{-12s}\right)\left(-\frac{4}{3}\left[\mathcal{\tilde{O}}^{3},\mathcal{\tilde{E}}\right]
+\frac{1}{3}\left[\tilde{\mathcal{O}},\left[\mathcal{\tilde{O}},\left[\mathcal{\tilde{O}},\tilde{\mathcal{E}}\right]\right]\right]\right)
\end{array}\right)
\nonumber
\end{eqnarray}
\[
=\left(\begin{array}{c}
\left(-\frac{1}{8}+e^{-4s}\left(\frac{1}{8}+\frac{1}{2}s+s^{2}\right)\right)\beta\left[\left[\tilde{\mathcal{O}},\tilde{\mathcal{E}}\right],\tilde{\mathcal{E}}\right]\\
\\
+\left(\frac{5}{144}-e^{-4s}\left(\frac{1}{48}+\frac{2}{9}s+\frac{1}{6}s^{2}\right)+e^{-8s}\frac{1}{36}s-e^{-12s}\left(\frac{1}{72}+\frac{1}{18}s\right)\right)\left[\tilde{\mathcal{O}},\left[\tilde{\mathcal{O}},\left[\tilde{\mathcal{O}},\tilde{\mathcal{E}}\right]\right]\right]\\
\\
+\left(\frac{1}{9}-e^{-4s}\left(\frac{5}{48}+\frac{13}{36}s+\frac{4}{3}s^{2}\right)-e^{-8s}\frac{1}{9}s-e^{-12s}\left(\frac{1}{144}+\frac{1}{36}s\right)\right)\left[\mathcal{\tilde{O}}^{3},\tilde{\mathcal{E}}\right]\\
\\
+\left(\begin{array}{c}
-\frac{1}{10}+e^{-4s}\left(\frac{1}{16}+\frac{1}{2}s+s^{2}\right)+e^{-12s}\left(\frac{1}{32}+\frac{1}{4}s\right)+\frac{1}{160}e^{-20s}\end{array}\right)\beta\mathcal{\tilde{O}}^{5}
\end{array}\right)
\]
The searched for Magnus type series expansion in the limit $s\rightarrow\infty$  thus reads
\begin{eqnarray}
\Omega_{u}\left(\infty\right) & = & \kappa\Omega^{\left(1\right)}\left(\infty\right)+\kappa^{3}\Omega^{\left(3\right)}\left(\infty\right)+\kappa^{5}\Omega^{\left(5\right)}\left(\infty\right)+...\label{eq: perturbation series Omega_u_(infty)}\\
\nonumber \\
 & = & \begin{cases}
\kappa\left(-\frac{1}{2}\beta\mathcal{\tilde{O}}\right)+\kappa^{3}\left(-\frac{1}{4}\left[\tilde{\mathcal{O}},\tilde{\mathcal{E}}\right]+\frac{1}{6}\beta\tilde{\mathcal{O}}^{3}\right)\\
\\
+\kappa^{5}\left(-\frac{1}{10}\beta\mathcal{\tilde{O}}^{5}+\frac{1}{9}\left[\tilde{\mathcal{O}}^{3},\tilde{\mathcal{E}}\right]
+\frac{5}{144}\left[\tilde{\mathcal{O}},\left[\tilde{\mathcal{O}},\left[\tilde{\mathcal{O}},\tilde{\mathcal{E}}\right]\right]\right]
-\frac{1}{8}\beta\left[\left[\tilde{\mathcal{O}},\tilde{\mathcal{E}}\right],\tilde{\mathcal{E}}\right]\right)+O\left(\kappa^{7}\right)
\end{cases}\nonumber
\end{eqnarray}

\section*{Supplemental material for section XVI in \cite{Schopohl2022MS}}
 Restricting to accuracy $O\left(\kappa^{8}\right)$ then
\begin{eqnarray}
\mathsf{H}^{\left(NW\right)} & = & \begin{cases}
\left(\beta+\kappa^{2}\tilde{\mathcal{E}}\right)\\
\\
+\frac{1}{2}\left(\textrm{ad}_{\Omega_{u}}\right)^{2}\circ\left(\beta+\kappa^{2}\tilde{\mathcal{E}}\right)+\frac{1}{24}\left(\textrm{ad}_{\Omega_{u}}\right)^{4}\circ\left(\beta+\kappa^{2}\tilde{\mathcal{E}}\right)+\frac{1}{720}\left(\textrm{ad}_{\Omega_{u}}\right)^{6}\circ\left(\beta+\kappa^{2}\tilde{\mathcal{E}}\right)\\
\\
-\left(\textrm{ad}_{\Omega_{u}}\right)\circ\kappa\mathcal{\tilde{O}}-\frac{1}{6}\left(\textrm{ad}_{\Omega_{u}}\right)^{3}\circ\kappa\mathcal{\tilde{O}}-\frac{1}{120}\left(\textrm{ad}_{\Omega_{u}}\right)^{5}\circ\kappa\mathcal{\tilde{O}}\\
\\
+O\left(\kappa^{8}\right)
\end{cases}\label{eq: series expansion  H_NW  I}\\
 & \equiv & \beta+\kappa^{2}\tilde{\mathsf{h}}^{\left(2\right)}+\kappa^{4}\tilde{\mathsf{h}}^{\left(4\right)}+\kappa^{6}\tilde{\mathsf{h}}^{\left(6\right)}+O\left(\kappa^{8}\right)\nonumber
\end{eqnarray}
At first sight the indicated accuracy $O\left(\kappa^{8}\right)$
holds indeed with $\Omega_{u}\left(\infty\right)$ being expanded according to (\ref{eq:perturbation series Omega_u_(s)}) up to and including the fifth order term  $\Omega^{\left(5\right)}$
, because upon insertion of  (\ref{eq:perturbation series Omega_u_(s)}) there holds
\begin{equation}
\textrm{ad}_{\Omega_{u}}\circ\mathcal{F}=\sum_{j=0}^{\infty}\kappa^{2j+1}\textrm{ad}_{\Omega^{\left(2j+1\right)}}\circ\mathcal{F}\label{eq: perturbation series ad_Omega_u}
\end{equation}
Therefore
\begin{eqnarray*}
\left(\textrm{ad}_{\Omega_{u}}\right)^{6}\circ\beta & = & \kappa^{6}\left(\textrm{ad}_{\Omega^{\left(1\right)}}\right)^{6}\circ\beta+O\left(\kappa^{8}\right)\\
\left(\textrm{ad}_{\Omega_{u}\left(\infty\right)}\right)^{5}\circ\left(\kappa\mathcal{\tilde{O}}\right) & = & \kappa^{6}\left(\textrm{ad}_{\Omega^{\left(1\right)}}\right)^{5}\circ\tilde{O}+O\left(\kappa^{8}\right)\\
\left(\textrm{ad}_{\Omega_{u}}\right)^{4}\circ\left(\kappa^{2}\tilde{\mathcal{E}}\right) & = & \kappa^{6}\left(\textrm{ad}_{\Omega^{\left(1\right)}}\right)^{4}\circ\tilde{\mathcal{E}}+O\left(\kappa^{8}\right)
\end{eqnarray*}
We readily evaluate
\begin{eqnarray*}
\textrm{ad}_{\Omega^{\left(1\right)}}\circ\beta=\left[\Omega^{\left(1\right)},\beta\right] & = & \left[\left(-\frac{1}{2}\beta\mathcal{\mathcal{\tilde{O}}}\right),\beta\right]=\mathcal{\tilde{O}}\\
\left(\textrm{ad}_{\Omega^{\left(1\right)}}\right)^{2}\circ\beta\equiv\left[\Omega^{\left(1\right)},\textrm{ad}_{\Omega^{\left(1\right)}}\circ\beta\right] & = & \left[\left(-\frac{1}{2}\beta\mathcal{\tilde{O}}\right),\mathcal{\tilde{O}}\right]=-\beta\mathcal{\mathcal{\tilde{O}}}^{2}\\
\left(\textrm{ad}_{\Omega^{\left(1\right)}}\right)^{3}\circ\beta=\left[\Omega^{\left(1\right)},\left(\textrm{ad}_{\Omega^{\left(1\right)}}\right)^{2}\circ\beta\right] & = & \left[\left(-\frac{1}{2}\beta\mathcal{\tilde{O}}\right),-\beta\mathcal{\mathcal{\tilde{O}}}^{2}\right]=-\mathcal{\tilde{O}}^{3}\\
\left(\textrm{ad}_{\Omega^{\left(1\right)}}\right)^{4}\circ\beta=\left[\Omega^{\left(1\right)},\left(\textrm{ad}_{\Omega^{\left(1\right)}}\right)^{3}\circ\beta\right] & = & \left[\left(-\frac{1}{2}\beta\mathcal{\tilde{O}}\right),-\mathcal{\mathcal{\tilde{O}}}^{3}\right]=\beta\mathcal{\tilde{O}}^{4}\\
\left(\textrm{ad}_{\Omega^{\left(1\right)}}\right)^{5}\circ\beta=\left[\Omega^{\left(1\right)},\left(\textrm{ad}_{\Omega^{\left(1\right)}}\right)^{4}\circ\beta\right] & = & \left[\left(-\frac{1}{2}\beta\mathcal{\tilde{O}}\right),\beta\mathcal{\tilde{O}}^{4}\right]=\mathcal{\tilde{O}}^{5}\\
\left(\textrm{ad}_{\Omega^{\left(1\right)}}\right)^{6}\circ\beta=\left[\Omega^{\left(1\right)},\left(\textrm{ad}_{\Omega^{\left(1\right)}}\right)^{5}\circ\beta\right] & = & \left[\left(-\frac{1}{2}\beta\mathcal{\tilde{O}}\right),\mathcal{\tilde{O}}^{5}\right]=-\beta\mathcal{\tilde{O}}^{6}
\end{eqnarray*}
and
\begin{eqnarray*}
\textrm{ad}_{\Omega^{\left(1\right)}}\circ\mathcal{\tilde{O}}=\left[\Omega^{\left(1\right)},\mathcal{\tilde{O}}\right] & = & \left[\left(-\frac{1}{2}\beta\mathcal{\tilde{O}}\right),\mathcal{\tilde{O}}\right]=-\beta\mathcal{\mathcal{\tilde{O}}}^{2}\\
\left(\textrm{ad}_{\Omega^{\left(1\right)}}\right)^{2}\circ\mathcal{\tilde{O}}=\left[\Omega^{\left(1\right)},\textrm{ad}_{\Omega^{\left(1\right)}}\circ\mathcal{\tilde{O}}\right] & = & \left[\left(-\frac{1}{2}\beta\mathcal{\tilde{O}}\right),-\beta\mathcal{\mathcal{\tilde{O}}}^{2}\right]=-\mathcal{\mathcal{\tilde{O}}}^{3}\\
\left(\textrm{ad}_{\Omega^{\left(1\right)}}\right)^{3}\circ\mathcal{\tilde{O}}=\left[\Omega^{\left(1\right)},\left(\textrm{ad}_{\Omega^{\left(1\right)}}\right)^{2}\circ\mathcal{\tilde{O}}\right] & = & \left[\left(-\frac{1}{2}\beta\mathcal{\tilde{O}}\right),-\mathcal{\mathcal{\tilde{O}}}^{3}\right]=\beta\mathcal{\tilde{O}}^{4}\\
\left(\textrm{ad}_{\Omega^{\left(1\right)}}\right)^{4}\circ\mathcal{\tilde{O}}=\left[\Omega^{\left(1\right)},\left(\textrm{ad}_{\Omega^{\left(1\right)}}\right)^{3}\circ\mathcal{\tilde{O}}\right] & = & \left[\left(-\frac{1}{2}\beta\mathcal{\tilde{O}}\right),\beta\mathcal{\tilde{O}}^{4}\right]=\mathcal{\tilde{O}}^{5}\\
\left(\textrm{ad}_{\Omega^{\left(1\right)}}\right)^{5}\circ\mathcal{\tilde{O}}=\left[\Omega^{\left(1\right)},\left(\textrm{ad}_{\Omega^{\left(1\right)}}\right)^{4}\circ\mathcal{\tilde{O}}\right] & = & \left[\left(-\frac{1}{2}\beta\mathcal{\tilde{O}}\right),\mathcal{\tilde{O}}^{5}\right]=-\beta\mathcal{\tilde{O}}^{6}
\end{eqnarray*}
This implies it being adequate to terminate the expansion of
$\sinh\left(\textrm{ad}_{\Omega_{u}\left(\infty\right)}\right)$
and the expansion of
$\cosh\left(\textrm{ad}_{\Omega_{u}\left(\infty\right)}\right)$
in Eq.(126) in \cite{Schopohl2022MS}
in the manner implemented, as we are here only interested in the terms up to and including the sixth order $\kappa^{6}$.

Observation: for the commutator terms of order  $\kappa^{2n}$ with $n\geq2$ holds
\begin{eqnarray}
\frac{1}{2}\textrm{ad}_{\Omega^{\left(1\right)}}\circ\textrm{ad}_{\Omega^{\left(2n-1\right)}}\circ\beta+\frac{1}{2}\textrm{ad}_{\Omega^{\left(2n-1\right)}}\circ\textrm{ad}_{\Omega^{\left(1\right)}}\circ\beta & = & \textrm{ad}_{\Omega^{\left(2n-1\right)}}\circ\mathcal{\tilde{O}}\label{eq: identity  ad-Operators}
\end{eqnarray}
Demonstration:
\begin{eqnarray*}
 &  & \frac{1}{2}\textrm{ad}_{\Omega^{\left(1\right)}}\circ\textrm{ad}_{\Omega^{\left(2n-1\right)}}\circ\beta+\frac{1}{2}\textrm{ad}_{\Omega^{\left(2n-1\right)}}\circ\textrm{ad}_{\Omega^{\left(1\right)}}\circ\beta\\
 & = & \frac{1}{2}\left[\Omega^{\left(1\right)},\left[\Omega^{\left(2n-1\right)},\beta\right]\right]+\frac{1}{2}\left[\Omega^{\left(2n-1\right)},\left[\Omega^{\left(1\right)},\beta\right]\right]\\
 & = & -\left[\Omega^{\left(1\right)},\beta\Omega^{\left(2n-1\right)}\right]-\left[\Omega^{\left(2n-1\right)},\beta\Omega^{\left(1\right)}\right]\\
 & = & -\left[\Omega^{\left(1\right)},\beta\right]\Omega^{\left(2n-1\right)}-\beta\left[\Omega^{\left(1\right)},\Omega^{\left(2n-1\right)}\right]-\left[\Omega^{\left(2n-1\right)},\beta\right]\Omega^{\left(1\right)}-\beta\left[\Omega^{\left(2n-1\right)},\Omega^{\left(1\right)}\right]\\
 & = & 2\beta\Omega^{\left(1\right)}\Omega^{\left(2n-1\right)}+2\beta\Omega^{\left(2n-1\right)}\Omega^{\left(1\right)}\\
 & = & 2\beta\left(-\frac{1}{2}\beta\mathcal{\mathcal{\tilde{O}}}\right)\Omega^{\left(2n-1\right)}+2\beta\Omega^{\left(2n-1\right)}\left(-\frac{1}{2}\beta\mathcal{\mathcal{\tilde{O}}}\right)\\
 & = & -\mathcal{\mathcal{\tilde{O}}}\Omega^{\left(2n-1\right)}+\Omega^{\left(2n-1\right)}\mathcal{\mathcal{\tilde{O}}}\\
 & = & \left[\Omega^{\left(2n-1\right)},\mathcal{\mathcal{\tilde{O}}}\right]\\
 & = & \textrm{ad}_{\Omega^{\left(2n-1\right)}}\circ\mathcal{\tilde{O}}
\end{eqnarray*}
This is a happy coincidence, as it implies to achieve accuracy  $O\left(\kappa^{2n}\right)$ for $n\geq2$ now only the terms
 $\Omega^{\left(1\right)},\Omega^{\left(3\right)},...,\Omega^{\left(2n-3\right)}$
are required, because the commutator terms involving  $\Omega^{\left(2n-1\right)}$  cancel !

Trivially
\begin{equation}
\mathsf{\tilde{\mathsf{h}}}^{\left(0\right)}=\beta\label{eq: h_0}
\end{equation}
, while the second order term, identical to the nonrelativistic Schrödinger-Pauli Hamiltonian, is readily identified as
\begin{equation}
\tilde{\mathsf{h}}^{\left(2\right)}=\tilde{\mathcal{E}}+\frac{1}{2}\left[\Omega^{\left(1\right)},\left[\Omega^{\left(1\right)},\beta\right]\right]-\left[\Omega^{\left(1\right)},\mathcal{\tilde{O}}\right]=\tilde{\mathcal{E}}+\frac{1}{2}\beta\mathcal{\tilde{O}}^{2}\label{eq: h_2}
\end{equation}
Up next the fourth order term
\begin{eqnarray}
\tilde{\mathsf{h}}^{\left(4\right)} & = & \begin{cases}
+\frac{1}{2}\left[\Omega^{\left(1\right)},\left[\Omega^{\left(3\right)},\beta\right]\right]+\frac{1}{2}\left[\Omega^{\left(3\right)},\left[\Omega^{\left(1\right)},\beta\right]\right]+\frac{1}{24}\left[\Omega^{\left(1\right)},\left[\Omega^{\left(1\right)},\left[\Omega^{\left(1\right)},\left[\Omega^{\left(1\right)},\mathcal{\beta}\right]\right]\right]\right]\\
+\frac{1}{2}\left[\Omega^{\left(1\right)},\left[\Omega^{\left(1\right)},\tilde{\mathcal{E}}\right]\right]\\
-\left[\Omega^{\left(3\right)},\mathcal{\tilde{O}}\right]-\frac{1}{6}\left[\Omega^{\left(1\right)},\left[\Omega^{\left(1\right)},\left[\Omega^{\left(1\right)},\mathcal{\tilde{O}}\right]\right]\right]
\end{cases}\nonumber \\
 & = & -\frac{1}{8}\left[\mathcal{\mathcal{\tilde{O}}},\left[\mathcal{\tilde{O}},\tilde{\mathcal{E}}\right]\right]-\frac{1}{8}\beta\mathcal{\tilde{O}}^{4}\label{eq: h_4}
\end{eqnarray}
The term  $\tilde{\mathsf{h}}^{\left(4\right)}$ indeed represents all the leading order relativistic corrections to the Schrödinger-Pauli Hamiltonian in a
 manifestly gauge invariant manner, namely the spin-orbit interaction together with the Darwin term and also comprising the relativistic correction to the kinetic energy term (with the corresponding corrections to the Zeeman-term
 included in $\mathcal{\tilde{O}}^{4}$.

The sixth order term we may write, because of the cancellation of the terms involving $\Omega^{\left(5\right)}$, now in the guise
\begin{eqnarray}
\label{eq: h_6  I}\\
\tilde{\mathsf{h}}^{\left(6\right)} & = & \begin{cases}
+\frac{1}{2}\left[\Omega^{\left(3\right)},\left[\Omega^{\left(3\right)},\beta\right]\right]+\frac{1}{2}\left[\Omega^{\left(1\right)},\left[\Omega^{\left(3\right)},\tilde{\mathcal{E}}\right]\right]+\frac{1}{2}\left[\Omega^{\left(3\right)},\left[\Omega^{\left(1\right)},\tilde{\mathcal{E}}\right]\right]\\
\\
+\frac{1}{24}\left[\Omega^{\left(1\right)},\left[\Omega^{\left(1\right)},\left[\Omega^{\left(1\right)},\left[\Omega^{\left(3\right)},\beta\right]\right]\right]\right]+\frac{1}{24}\left[\Omega^{\left(1\right)},\left[\Omega^{\left(1\right)},\left[\Omega^{\left(3\right)},\left[\Omega^{\left(1\right)},\beta\right]\right]\right]\right]\\
+\frac{1}{24}\left[\Omega^{\left(3\right)},\left[\Omega^{\left(1\right)},\left[\Omega^{\left(1\right)},\left[\Omega^{\left(1\right)},\beta\right]\right]\right]\right]+\frac{1}{24}\left[\Omega^{\left(1\right)},\left[\Omega^{\left(3\right)},\left[\Omega^{\left(1\right)},\left[\Omega^{\left(1\right)},\beta\right]\right]\right]\right]\\
+\frac{1}{24}\left[\Omega^{\left(1\right)},\left[\Omega^{\left(1\right)},\left[\Omega^{\left(1\right)},\left[\Omega^{\left(1\right)},\tilde{\mathcal{E}}\right]\right]\right]\right]\\
\\
+\frac{1}{720}\left[\Omega^{\left(1\right)},\left[\Omega^{\left(1\right)},\left[\Omega^{\left(1\right)},\left[\Omega^{\left(1\right)},\left[\Omega^{\left(1\right)},\left[\Omega^{\left(1\right)},\beta\right]\right]\right]\right]\right]\right]\\
-\frac{1}{6}\left[\Omega^{\left(1\right)},\left[\Omega^{\left(1\right)},\left[\Omega^{\left(3\right)},\mathcal{\tilde{O}}\right]\right]\right]-\frac{1}{6}\left[\Omega^{\left(1\right)},\left[\Omega^{\left(3\right)},\left[\Omega^{\left(1\right)},\mathcal{\tilde{O}}\right]\right]\right]-\frac{1}{6}\left[\Omega^{\left(3\right)},\left[\Omega^{\left(1\right)},\left[\Omega^{\left(1\right)},\mathcal{\tilde{O}}\right]\right]\right]\\
\\
-\frac{1}{120}\left[\Omega^{\left(1\right)},\left[\Omega^{\left(1\right)},\left[\Omega^{\left(1\right)},\left[\Omega^{\left(1\right)},\left[\Omega^{\left(1\right)},\mathcal{\tilde{O}}\right]\right]\right]\right]\right]
\end{cases}\nonumber
\end{eqnarray}
Making use of
\begin{eqnarray*}
\left[\Omega^{\left(1\right)},\beta\right] & = & \mathcal{\tilde{O}}\\
\\
\left[\Omega^{\left(1\right)},\left[\Omega^{\left(1\right)},\left[\Omega^{\left(1\right)},\left[\Omega^{\left(1\right)},\left[\Omega^{\left(1\right)},\left[\Omega^{\left(1\right)},\beta\right]\right]\right]\right]\right]\right] & = & -\beta\mathcal{\tilde{O}}^{6}\\
\left[\Omega^{\left(1\right)},\left[\Omega^{\left(1\right)},\left[\Omega^{\left(1\right)},\left[\Omega^{\left(1\right)},\left[\Omega^{\left(1\right)},\mathcal{\tilde{O}}\right]\right]\right]\right]\right] & = & -\beta\mathcal{\tilde{O}}^{6}\\
\\
\left[\Omega^{\left(1\right)},\left[\Omega^{\left(1\right)},\left[\Omega^{\left(1\right)},\beta\right]\right]\right] & = & -\mathcal{\tilde{O}}^{3}\\
\left[\Omega^{\left(1\right)},\left[\Omega^{\left(1\right)},\mathcal{\tilde{O}}\right]\right] & = & -\mathcal{\tilde{O}}^{3}\\
\\
\left[\Omega^{\left(1\right)},\left[\Omega^{\left(1\right)},\beta\right]\right] & = & -\beta\mathcal{\mathcal{\tilde{O}}}^{2}\\
\left[\Omega^{\left(1\right)},\mathcal{\tilde{O}}\right] & = & -\beta\mathcal{\mathcal{\tilde{O}}}^{2}
\end{eqnarray*}
then
\begin{eqnarray}
\label{eq: h_6  II}\\
\tilde{\mathsf{h}}^{\left(6\right)} & = & \begin{cases}
\frac{1}{2}\left[\Omega^{\left(3\right)},\left[\Omega^{\left(3\right)},\beta\right]\right]+\frac{1}{2}\left[\Omega^{\left(1\right)},\left[\Omega^{\left(3\right)},\tilde{\mathcal{E}}\right]\right]+\frac{1}{2}\left[\Omega^{\left(3\right)},\left[\Omega^{\left(1\right)},\tilde{\mathcal{E}}\right]\right]\\
\\
+\frac{1}{24}\left[\Omega^{\left(1\right)},\left[\Omega^{\left(1\right)},\left[\Omega^{\left(1\right)},\left[\Omega^{\left(3\right)},\beta\right]\right]\right]\right]\\
\\
+\frac{1}{24}\left[\Omega^{\left(1\right)},\left[\Omega^{\left(1\right)},\left[\Omega^{\left(1\right)},\left[\Omega^{\left(1\right)},\tilde{\mathcal{E}}\right]\right]\right]\right]\\
\\
+\frac{5}{720}\beta\mathcal{\tilde{O}}^{6}\\
\\
-\frac{1}{8}\left[\Omega^{\left(1\right)},\left[\Omega^{\left(1\right)},\left[\Omega^{\left(3\right)},\mathcal{\tilde{O}}\right]\right]\right]-\frac{1}{8}\left[\Omega^{\left(1\right)},\left[\Omega^{\left(3\right)},\left(-\beta\mathcal{O}^{2}\right)\right]\right]-\frac{1}{8}\left[\Omega^{\left(3\right)},\left(-\mathcal{\mathcal{\tilde{O}}}^{3}\right)\right]
\end{cases}\nonumber
\end{eqnarray}

Let us evaluate
\[
\Omega^{\left(3\right)}=\frac{1}{6}\beta\mathcal{\tilde{O}}^{3}-\frac{1}{4}\left[\mathcal{\tilde{O}},\tilde{\mathcal{E}}\right]
\]

\[
\left[\Omega^{\left(3\right)},\beta\right]=-\frac{1}{3}\mathcal{\tilde{O}}^{3}+\frac{1}{2}\beta\left[\tilde{\mathcal{O}},\tilde{\mathcal{E}}\right]
\]

\begin{eqnarray*}
\left[\Omega^{\left(1\right)},\left[\Omega^{\left(3\right)},\beta\right]\right] & = & \left[\left(-\frac{1}{2}\beta\mathcal{\mathcal{\tilde{O}}}\right),\left(-\frac{1}{3}\mathcal{\tilde{O}}^{3}+\frac{1}{2}\beta\left[\tilde{\mathcal{O}},\tilde{\mathcal{E}}\right]\right)\right]\\
 & = & \frac{1}{3}\beta\mathcal{\tilde{O}}^{4}+\frac{1}{4}\left[\tilde{\mathcal{O}},\left[\tilde{\mathcal{O}},\tilde{\mathcal{E}}\right]\right]
\end{eqnarray*}

\begin{eqnarray*}
\left[\Omega^{\left(1\right)},\left[\Omega^{\left(1\right)},\left[\Omega^{\left(3\right)},\beta\right]\right]\right] & = & \left[\left(-\frac{1}{2}\beta\mathcal{\mathcal{\tilde{O}}}\right),\left(\frac{1}{3}\beta\mathcal{\tilde{O}}^{4}+\frac{1}{4}\left[\tilde{\mathcal{O}},\left[\tilde{\mathcal{O}},\tilde{\mathcal{E}}\right]\right]\right)\right]\\
 & = & \frac{1}{3}\mathcal{\tilde{O}}^{5}-\frac{1}{8}\beta\left[\tilde{\mathcal{O}},\left[\tilde{\mathcal{O}},\left[\tilde{\mathcal{O}},\tilde{\mathcal{E}}\right]\right]\right]
\end{eqnarray*}

\begin{eqnarray*}
\left[\Omega^{\left(1\right)},\left[\Omega^{\left(1\right)},\left[\Omega^{\left(1\right)},\left[\Omega^{\left(3\right)},\beta\right]\right]\right]\right] & = & \left[\left(-\frac{1}{2}\beta\mathcal{\mathcal{\tilde{O}}}\right),\left(\frac{1}{3}\mathcal{\tilde{O}}^{5}-\frac{1}{8}\beta\left[\tilde{\mathcal{O}},\left[\tilde{\mathcal{O}},\left[\tilde{\mathcal{O}},\tilde{\mathcal{E}}\right]\right]\right]\right)\right]\\
 & = & -\frac{1}{3}\beta\mathcal{\tilde{O}}^{6}-\frac{1}{16}\left[\mathcal{\tilde{O}},\left[\mathcal{\tilde{O}},\left[\mathcal{\tilde{O}},\left[\mathcal{\tilde{O}},\tilde{\mathcal{E}}\right]\right]\right]\right]
\end{eqnarray*}

In addition
\begin{eqnarray*}
\left[\Omega^{\left(3\right)},\mathcal{\tilde{O}}\right] & = & \left[\left(-\frac{1}{4}\left[\mathcal{O},\mathcal{E}\right]+\frac{1}{6}\beta\mathcal{O}^{3}\right),\mathcal{\tilde{O}}\right]\\
 & = & \frac{1}{4}\left[\mathcal{\tilde{O}},\left[\mathcal{\tilde{O}},\tilde{\mathcal{E}}\right]\right]+\frac{1}{3}\beta\mathcal{\tilde{O}}^{4}
\end{eqnarray*}

\begin{eqnarray*}
\left[\Omega^{\left(1\right)},\left[\Omega^{\left(3\right)},\mathcal{\tilde{O}}\right]\right] & = & \left[\left(-\frac{1}{2}\beta\mathcal{\mathcal{\tilde{O}}}\right),\left(\frac{1}{4}\left[\mathcal{\tilde{O}},\left[\mathcal{\tilde{O}},\tilde{\mathcal{E}}\right]\right]+\frac{1}{3}\beta\mathcal{\tilde{O}}^{4}\right)\right]\\
 & = & -\frac{1}{8}\beta\left[\mathcal{\tilde{O}},\left[\mathcal{\tilde{O}},\left[\mathcal{\tilde{O}},\tilde{\mathcal{E}}\right]\right]\right]+\frac{1}{3}\mathcal{\tilde{O}}^{5}
\end{eqnarray*}

\begin{eqnarray*}
\left[\Omega^{\left(1\right)},\left[\Omega^{\left(1\right)},\left[\Omega^{\left(3\right)},\mathcal{\tilde{O}}\right]\right]\right] & = & \left[\left(-\frac{1}{2}\beta\mathcal{\mathcal{\tilde{O}}}\right),\left(-\frac{1}{8}\beta\left[\mathcal{\tilde{O}},\left[\mathcal{\tilde{O}},\left[\mathcal{\tilde{O}},\tilde{\mathcal{E}}\right]\right]\right]+\frac{1}{3}\mathcal{\tilde{O}}^{5}\right)\right]\\
 & = & -\frac{1}{3}\beta\mathcal{\tilde{O}}^{6}-\frac{1}{16}\left[\mathcal{\tilde{O}},\left[\mathcal{\tilde{O}},\left[\mathcal{\tilde{O}},\left[\mathcal{\tilde{O}},\tilde{\mathcal{E}}\right]\right]\right]\right]
\end{eqnarray*}

, in addition
\begin{eqnarray*}
\left[\Omega^{\left(3\right)},\left(-\beta\mathcal{O}^{2}\right)\right] & = & \left[\left(\frac{1}{6}\beta\mathcal{\tilde{O}}^{3}-\frac{1}{4}\left[\mathcal{\tilde{O}},\tilde{\mathcal{E}}\right]\right),\left(-\beta\mathcal{O}^{2}\right)\right]\\
 & = & \frac{1}{3}\mathcal{\mathcal{\tilde{O}}}^{5}-\frac{1}{4}\beta\left(\mathcal{O}^{2}\left[\tilde{\mathcal{O}},\mathcal{\tilde{E}}\right]+\left[\tilde{\mathcal{O}},\mathcal{\tilde{E}}\right]\mathcal{O}^{2}\right)
\end{eqnarray*}

\begin{eqnarray*}
\left[\Omega^{\left(1\right)},\left[\Omega^{\left(3\right)},\left(-\beta\mathcal{O}^{2}\right)\right]\right] & = & \left[\left(-\frac{1}{2}\beta\mathcal{\mathcal{\tilde{O}}}\right),\left(\frac{1}{3}\mathcal{\mathcal{\tilde{O}}}^{5}-\frac{1}{4}\beta\left(\mathcal{O}^{2}\left[\tilde{\mathcal{O}},\mathcal{\tilde{E}}\right]+\left[\tilde{\mathcal{O}},\mathcal{\tilde{E}}\right]\mathcal{O}^{2}\right)\right)\right]\\
 & = & -\frac{1}{3}\beta\mathcal{\tilde{O}}^{6}-\frac{1}{8}\left[\mathcal{\mathcal{\tilde{O}}},\left(\mathcal{O}^{2}\left[\tilde{\mathcal{O}},\mathcal{\tilde{E}}\right]+\left[\tilde{\mathcal{O}},\mathcal{\tilde{E}}\right]\mathcal{O}^{2}\right)\right]
\end{eqnarray*}

, in addition
\begin{eqnarray*}
\left[\Omega^{\left(3\right)},\left(-\mathcal{\mathcal{\tilde{O}}}^{3}\right)\right] & = & \left[\left(\frac{1}{6}\beta\mathcal{\tilde{O}}^{3}-\frac{1}{4}\left[\mathcal{\tilde{O}},\tilde{\mathcal{E}}\right]\right),\left(-\mathcal{\mathcal{\tilde{O}}}^{3}\right)\right]\\
 & = & -\frac{1}{3}\beta\mathcal{O}^{6}-\frac{1}{4}\left[\mathcal{\mathcal{\tilde{O}}}^{3},\left[\tilde{\mathcal{O}},\mathcal{\tilde{E}}\right]\right]
\end{eqnarray*}

, in addition
\begin{eqnarray*}
\frac{1}{2}\left[\Omega^{\left(3\right)},\left[\Omega^{\left(3\right)},\beta\right]\right] & = & \frac{1}{2}\left[\Omega^{\left(3\right)},\left(-2\beta\Omega^{\left(3\right)}\right)\right]\\
 & = & -\left[\Omega^{\left(3\right)},\beta\Omega^{\left(3\right)}\right]\\
 & = & -\left[\Omega^{\left(3\right)},\beta\right]\Omega^{\left(3\right)}-\beta\underset{\equiv\mathsf{0}}{\underbrace{\left[\Omega^{\left(3\right)},\Omega^{\left(3\right)}\right]}}\\
 & = & 2\beta\,\Omega^{\left(3\right)}\Omega^{\left(3\right)}\\
 & = & 2\beta\left(\frac{1}{6}\beta\mathcal{\tilde{O}}^{3}-\frac{1}{4}\left[\mathcal{\tilde{O}},\tilde{\mathcal{E}}\right]\right)\left(\frac{1}{6}\beta\mathcal{\tilde{O}}^{3}-\frac{1}{4}\left[\mathcal{\tilde{O}},\tilde{\mathcal{E}}\right]\right)\\
 & = & -\frac{1}{18}\beta\mathcal{\tilde{O}}^{6}-\frac{1}{12}\left[\mathcal{\tilde{O}}^{3},\left[\mathcal{\tilde{O}},\tilde{\mathcal{E}}\right]\right]+\frac{1}{8}\beta\left(\left[\mathcal{\tilde{O}},\tilde{\mathcal{E}}\right]\right)^{2}
\end{eqnarray*}

, in addition
\begin{eqnarray*}
\left[\Omega^{\left(1\right)},\tilde{\mathcal{E}}\right] & = & \left[\left(-\frac{1}{2}\beta\mathcal{\mathcal{\tilde{O}}}\right),\tilde{\mathcal{E}}\right]=-\frac{1}{2}\beta\left[\tilde{\mathcal{O}},\mathcal{\tilde{E}}\right]\\
\left[\Omega^{\left(1\right)},\left[\Omega^{\left(1\right)},\tilde{\mathcal{E}}\right]\right] & = & \left[\left(-\frac{1}{2}\beta\mathcal{\mathcal{\tilde{O}}}\right),-\frac{1}{2}\beta\left[\tilde{\mathcal{O}},\mathcal{\tilde{E}}\right]\right]\\
 & = & -\frac{1}{4}\left[\mathcal{\mathcal{\tilde{O}}},\left[\tilde{\mathcal{O}},\mathcal{\tilde{E}}\right]\right]
\end{eqnarray*}
\begin{eqnarray*}
\left[\Omega^{\left(1\right)},\left[\Omega^{\left(1\right)},\left[\Omega^{\left(1\right)},\tilde{\mathcal{E}}\right]\right]\right] & = & \left[\left(-\frac{1}{2}\beta\mathcal{\mathcal{\tilde{O}}}\right),\left(-\frac{1}{4}\left[\mathcal{\mathcal{\tilde{O}}},\left[\tilde{\mathcal{O}},\mathcal{\tilde{E}}\right]\right]\right)\right]\\
 & = & \frac{1}{8}\beta\left[\mathcal{\mathcal{\tilde{O}}},\left[\mathcal{\mathcal{\tilde{O}}},\left[\tilde{\mathcal{O}},\mathcal{\tilde{E}}\right]\right]\right]
\end{eqnarray*}
\begin{eqnarray*}
\left[\Omega^{\left(1\right)},\left[\Omega^{\left(1\right)},\left[\Omega^{\left(1\right)},\left[\Omega^{\left(1\right)},\tilde{\mathcal{E}}\right]\right]\right]\right] & = & \left[\left(-\frac{1}{2}\beta\mathcal{\mathcal{\tilde{O}}}\right),\frac{1}{8}\beta\left[\mathcal{\mathcal{\tilde{O}}},\left[\mathcal{\mathcal{\tilde{O}}},\left[\tilde{\mathcal{O}},\mathcal{\tilde{E}}\right]\right]\right]\right]\\
 & = & \frac{1}{16}\left[\mathcal{\mathcal{\tilde{O}}},\left[\mathcal{\mathcal{\tilde{O}}},\left[\mathcal{\mathcal{\tilde{O}}},\left[\tilde{\mathcal{O}},\mathcal{\tilde{E}}\right]\right]\right]\right]
\end{eqnarray*}
, in addition
\begin{eqnarray*}
\left[\Omega^{\left(3\right)},\left[\Omega^{\left(1\right)},\tilde{\mathcal{E}}\right]\right]
 & = & +\frac{1}{12}\left[\mathcal{O}^{3},\left[\tilde{\mathcal{O}},\mathcal{\tilde{E}}\right]\right]-\frac{1}{4}\beta\left(\left[\tilde{\mathcal{O}},\mathcal{\tilde{E}}\right]\right)^{2}
\end{eqnarray*}
\begin{eqnarray*}
\left[\Omega^{\left(3\right)},\tilde{\mathcal{E}}\right]
& = & -\frac{1}{4}\left[\left[\tilde{\mathcal{O}},\mathcal{\tilde{E}}\right],\tilde{\mathcal{E}}\right]+\frac{1}{6}\beta\left[\mathcal{\tilde{O}}^{3},\tilde{\mathcal{E}}\right]
\end{eqnarray*}
\begin{eqnarray*}
\left[\Omega^{\left(1\right)},\left[\Omega^{\left(3\right)},\tilde{\mathcal{E}}\right]\right]
 & = & \frac{1}{8}\beta\left(\mathcal{\mathcal{\tilde{O}}}\left[\left[\tilde{\mathcal{O}},\mathcal{\tilde{E}}\right],\tilde{\mathcal{E}}\right]+\left[\left[\tilde{\mathcal{O}},\mathcal{\tilde{E}}\right],\tilde{\mathcal{E}}\right]\mathcal{\tilde{O}}\right)+\frac{1}{12}\left[\mathcal{\mathcal{\tilde{O}}}^{3},\left[\mathcal{\tilde{O}},\tilde{\mathcal{E}}\right]\right]
\end{eqnarray*}


Insertion of the afore calculated commutator terms into the expression (\ref{eq: h_6  II}) leads to
\begin{eqnarray*}
\tilde{\mathsf{h}}^{\left(6\right)} & = & \begin{cases}
\underset{=\frac{1}{16}}{\underbrace{\left(-\frac{1}{18}-\frac{1}{24\cdot3}+\frac{1}{8}\left(\frac{1}{3}+\frac{1}{3}+\frac{1}{3}\right)+\frac{5}{720}\right)}}\beta\mathcal{\tilde{O}}^{6}\\
\\
\underset{=\frac{1}{32}}{\underbrace{\left(\frac{1}{24}+\frac{1}{8\cdot4}-\frac{1}{12}+\frac{1}{24}\right)}}\left[\mathcal{\tilde{O}}^{3},\left[\mathcal{\tilde{O}},\tilde{\mathcal{E}}\right]\right]\\
\\
+\underset{=\frac{1}{128}}{\underbrace{\left(\frac{1}{24}+\frac{1}{8}-\frac{1}{24}\right)\frac{1}{16}}}\left[\mathcal{\mathcal{\tilde{O}}},\left[\mathcal{\mathcal{\tilde{O}}},\left[\mathcal{\mathcal{\tilde{O}}},\left[\tilde{\mathcal{O}},\mathcal{\tilde{E}}\right]\right]\right]\right]\\
\\
+\frac{1}{8\cdot8}\left[\mathcal{\mathcal{\tilde{O}}},\left(\mathcal{O}^{2}\left[\tilde{\mathcal{O}},\mathcal{\tilde{E}}\right]+\left[\tilde{\mathcal{O}},\mathcal{\tilde{E}}\right]\mathcal{O}^{2}\right)\right]\\
\\
+\frac{1}{16}\beta\left(\mathcal{\mathcal{\tilde{O}}}\left[\left[\tilde{\mathcal{O}},\mathcal{\tilde{E}}\right],\tilde{\mathcal{E}}\right]+\left[\left[\tilde{\mathcal{O}},\mathcal{\tilde{E}}\right],\tilde{\mathcal{E}}\right]\mathcal{\tilde{O}}\right)\\
\\
+\underset{=0}{\underbrace{\left(\frac{1}{8}-\frac{1}{8}\right)}}\beta\left(\left[\mathcal{\tilde{O}},\tilde{\mathcal{E}}\right]\right)^{2}
\end{cases}
\end{eqnarray*}
and finally
\begin{equation}
\tilde{\mathsf{h}}^{\left(6\right)}=\begin{cases}
\frac{1}{16}\beta\mathcal{\tilde{O}}^{6}\\
\\
\frac{1}{32}\left[\mathcal{\tilde{O}}^{3},\left[\mathcal{\tilde{O}},\tilde{\mathcal{E}}\right]\right]\\
\\
+\frac{1}{128}\left[\mathcal{\mathcal{\tilde{O}}},\left[\mathcal{\mathcal{\tilde{O}}},\left[\mathcal{\mathcal{\tilde{O}}},\left[\tilde{\mathcal{O}},\mathcal{\tilde{E}}\right]\right]\right]\right]\\
\\
+\frac{1}{64}\left[\mathcal{\mathcal{\tilde{O}}},\left(\mathcal{O}^{2}\left[\tilde{\mathcal{O}},\mathcal{\tilde{E}}\right]+\left[\tilde{\mathcal{O}},\mathcal{\tilde{E}}\right]\mathcal{O}^{2}\right)\right]\\
\\
+\frac{1}{16}\beta\left(\mathcal{\mathcal{\tilde{O}}}\left[\left[\tilde{\mathcal{O}},\mathcal{\tilde{E}}\right],\tilde{\mathcal{E}}\right]
+\left[\left[\tilde{\mathcal{O}},\mathcal{\tilde{E}}\right],\tilde{\mathcal{E}}\right]\mathcal{\tilde{O}}\right)
\end{cases}
\label{eq: h_6  III}
\end{equation}

Note that the obtained sixth order term  $\tilde{\mathsf{h}}^{\left(6\right)}$  in (\ref{eq: h_6  III}) comprises no term being proportional to the square of the electric field.
As we have seen, such a term will be represented by $\beta\left(\left[\mathcal{\tilde{O}},\tilde{\mathcal{E}}\right]\right)^{2}$
, but it's weight cancels to zero. This is in contrast to the expansion found by Brüning et al. \cite{Bruening2011}
and indicates the result of these authors, regarding their higher order terms, being not in accord with the criterion of energy energy-separation,  as required for the unitary transformation of the Dirac Hamiltonian to a ${2\times2}$-blockdiagonal guise being determined unambigously \cite{EriksenKolsrud}\cite{NeznamovSilenko2009}\cite{Schopohl2022MS}. 

Last not least, for comparison of our result with results of De Vries and   Jonker \cite{Vries1968}, actually  obtained by computer algebra with help of the Pauli-Achieser-Berestezki
 elimination method \footnote{ A particularly clear presentation of the Pauli-Achieser-Berestezki elimination method is due to Stephani \cite{Stephani} }, let us expand
\begin{eqnarray*}
\left[\mathcal{\tilde{O}}^{3},\left[\mathcal{\tilde{O}},\tilde{\mathcal{E}}\right]\right] & = & \mathcal{\mathcal{\tilde{O}}}^{4}\mathcal{\tilde{E}}-\mathcal{\mathcal{\tilde{O}}}^{3}\mathcal{\tilde{E}}\mathcal{\mathcal{\tilde{O}}}-\mathcal{\mathcal{\tilde{O}}}\mathcal{\tilde{E}}\mathcal{\mathcal{\tilde{O}}}^{3}+\mathcal{\tilde{E}}\mathcal{\mathcal{\tilde{O}}}^{4}\\
\\
\left[\mathcal{\mathcal{\tilde{O}}},\left[\mathcal{\mathcal{\tilde{O}}},\left[\mathcal{\mathcal{\tilde{O}}},\left[\tilde{\mathcal{O}},\mathcal{\tilde{E}}\right]\right]\right]\right] & = & \tilde{\mathcal{O}}^{4}\mathcal{\tilde{E}}-4\mathcal{\mathcal{\tilde{O}}}^{3}\mathcal{\tilde{E}}\mathcal{\mathcal{\tilde{O}}}+6\mathcal{\mathcal{\tilde{O}}}^{2}\mathcal{\tilde{E}}\mathcal{\mathcal{\tilde{O}}}^{2}-4\mathcal{\mathcal{\tilde{O}}}\mathcal{\tilde{E}}\mathcal{\mathcal{\tilde{O}}}^{3}+\mathcal{\tilde{E}}\mathcal{\mathcal{\tilde{O}}}^{4}\\
\\
\left[\mathcal{\mathcal{\tilde{O}}},\left(\mathcal{O}^{2}\left[\tilde{\mathcal{O}},\mathcal{\tilde{E}}\right]+\left[\tilde{\mathcal{O}},\mathcal{\tilde{E}}\right]\mathcal{O}^{2}\right)\right] & = & \mathcal{\mathcal{\tilde{O}}}^{4}\mathcal{\tilde{E}}-2\mathcal{\mathcal{\tilde{O}}}^{3}\mathcal{\tilde{E}}\mathcal{\mathcal{\tilde{O}}}+2\mathcal{\mathcal{\tilde{O}}}^{2}\mathcal{\tilde{E}}\mathcal{\mathcal{\tilde{O}}}^{2}-2\mathcal{\mathcal{\tilde{O}}}\mathcal{\tilde{E}}\mathcal{\mathcal{\tilde{O}}}^{3}+\mathcal{\tilde{E}}\mathcal{\mathcal{\tilde{O}}}^{4}\\
\end{eqnarray*}
Then
\begin{eqnarray*}
 &  & \frac{1}{32}\left[\mathcal{\tilde{O}}^{3},\left[\mathcal{\tilde{O}},\tilde{\mathcal{E}}\right]\right]+\frac{1}{128}\left[\mathcal{\mathcal{\tilde{O}}},\left[\mathcal{\mathcal{\tilde{O}}},\left[\mathcal{\mathcal{\tilde{O}}},\left[\tilde{\mathcal{O}},\mathcal{\tilde{E}}\right]\right]\right]\right]+\frac{1}{64}\left[\mathcal{\mathcal{\tilde{O}}},\left(\mathcal{O}^{2}\left[\tilde{\mathcal{O}},\mathcal{\tilde{E}}\right]+\left[\tilde{\mathcal{O}},\mathcal{\tilde{E}}\right]\mathcal{O}^{2}\right)\right]\\
\\
 & = & \begin{cases}
\frac{1}{32}\left(\mathcal{\mathcal{\tilde{O}}}^{4}\mathcal{\tilde{E}}-\mathcal{\mathcal{\tilde{O}}}^{3}\mathcal{\tilde{E}}\mathcal{\mathcal{\tilde{O}}}-\mathcal{\mathcal{\tilde{O}}}\mathcal{\tilde{E}}\mathcal{\mathcal{\tilde{O}}}^{3}+\mathcal{\tilde{E}}\mathcal{\mathcal{\tilde{O}}}^{4}\right)\\
+\frac{1}{128}\left(\tilde{\mathcal{O}}^{4}\mathcal{\tilde{E}}-4\mathcal{\mathcal{\tilde{O}}}^{3}\mathcal{\tilde{E}}\mathcal{\mathcal{\tilde{O}}}+6\mathcal{\mathcal{\tilde{O}}}^{2}\mathcal{\tilde{E}}\mathcal{\mathcal{\tilde{O}}}^{2}-4\mathcal{\mathcal{\tilde{O}}}\mathcal{\tilde{E}}\mathcal{\mathcal{\tilde{O}}}^{3}+\mathcal{\tilde{E}}\mathcal{\mathcal{\tilde{O}}}^{4}\right)\\
+\frac{1}{64}\left(\mathcal{\mathcal{\tilde{O}}}^{4}\mathcal{\tilde{E}}-2\mathcal{\mathcal{\tilde{O}}}^{3}\mathcal{\tilde{E}}\mathcal{\mathcal{\tilde{O}}}+2\mathcal{\mathcal{\tilde{O}}}^{2}\mathcal{\tilde{E}}\mathcal{\mathcal{\tilde{O}}}^{2}-2\mathcal{\mathcal{\tilde{O}}}\mathcal{\tilde{E}}\mathcal{\mathcal{\tilde{O}}}^{3}+\mathcal{\tilde{E}}\mathcal{\mathcal{\tilde{O}}}^{4}\right)
\end{cases}\\
\\
 & = & \begin{cases}
\underset{=\frac{7}{128}}{\underbrace{\left(\frac{1}{32}+\frac{1}{128}+\frac{1}{64}\right)}}\left(\mathcal{\mathcal{\tilde{O}}}^{4}\mathcal{\tilde{E}}+\mathcal{\tilde{E}}\mathcal{\mathcal{\tilde{O}}}^{4}\right)\\
\\
-\underset{=\frac{3}{32}}{\underbrace{\left(\frac{1}{32}+\frac{4}{128}+\frac{2}{64}\right)}}\left(\mathcal{\mathcal{\tilde{O}}}^{3}\mathcal{\tilde{E}}\mathcal{\mathcal{\tilde{O}}}+\mathcal{\mathcal{\tilde{O}}}\mathcal{\tilde{E}}\mathcal{\mathcal{\tilde{O}}}^{3}\right)\\
\\
+\underset{\frac{5}{64}}{\underbrace{\left(\frac{6}{128}+\frac{2}{64}\right)}}\left(\frac{6}{128}+\frac{2}{64}\right)\left(\mathcal{\mathcal{\tilde{O}}}^{2}\mathcal{\tilde{E}}\mathcal{\mathcal{\tilde{O}}}^{2}\right)
\end{cases}\\
\\
 & = & \frac{7}{128}\left(\mathcal{\mathcal{\tilde{O}}}^{4}\mathcal{\tilde{E}}+\mathcal{\tilde{E}}\mathcal{\mathcal{\tilde{O}}}^{4}\right)-\frac{3}{32}\left(\mathcal{\mathcal{\tilde{O}}}^{3}\mathcal{\tilde{E}}\mathcal{\mathcal{\tilde{O}}}+\mathcal{\mathcal{\tilde{O}}}\mathcal{\tilde{E}}\mathcal{\mathcal{\tilde{O}}}^{3}\right)+\frac{5}{64}\mathcal{\mathcal{\tilde{O}}}^{2}\mathcal{\tilde{E}}\mathcal{\mathcal{\tilde{O}}}^{2}
\end{eqnarray*}
Furthermore
\begin{eqnarray*}
 &  & \mathcal{\mathcal{\tilde{O}}}\left[\left[\tilde{\mathcal{O}},\mathcal{\tilde{E}}\right],\tilde{\mathcal{E}}\right]+\left[\left[\tilde{\mathcal{O}},\mathcal{\tilde{E}}\right],\tilde{\mathcal{E}}\right]\mathcal{\tilde{O}}\\
\\
 & = & \begin{cases}
\mathcal{\mathcal{\tilde{O}}}\left(\mathcal{\mathcal{\tilde{O}}}\mathcal{\tilde{E}}-\mathcal{\tilde{E}}\tilde{\mathcal{O}}\right)\tilde{\mathcal{E}}-\mathcal{\mathcal{\tilde{O}}}\tilde{\mathcal{E}}\left(\mathcal{\mathcal{\tilde{O}}}\mathcal{\tilde{E}}-\mathcal{\tilde{E}}\tilde{\mathcal{O}}\right)\\
+\left(\mathcal{\mathcal{\tilde{O}}}\mathcal{\tilde{E}}-\mathcal{\tilde{E}}\tilde{\mathcal{O}}\right)\tilde{\mathcal{E}}\mathcal{\mathcal{\tilde{O}}}-\tilde{\mathcal{E}}\left(\mathcal{\mathcal{\tilde{O}}}\mathcal{\tilde{E}}-\mathcal{\tilde{E}}\tilde{\mathcal{O}}\right)\tilde{\mathcal{O}}
\end{cases}\\
\\
 & = & \begin{cases}
\mathcal{\tilde{O}}^{2}\mathcal{\tilde{E}}^{2}-2\mathcal{\tilde{O}}\mathcal{\tilde{E}}\mathcal{\tilde{O}}\mathcal{\tilde{E}}+\mathcal{\tilde{O}}\mathcal{\tilde{E}}^{2}\mathcal{\tilde{O}}\\
+\mathcal{\tilde{O}}\mathcal{\tilde{E}}^{2}\mathcal{\tilde{O}}-2\mathcal{\tilde{E}}\mathcal{\tilde{O}}\mathcal{\tilde{E}}\mathcal{\tilde{O}}+\mathcal{\tilde{E}}^{2}\mathcal{\tilde{O}}^{2}
\end{cases}\\
\\
 & = & \mathcal{\tilde{O}}^{2}\mathcal{\tilde{E}}^{2}+\mathcal{\tilde{E}}^{2}\mathcal{\tilde{O}}^{2}-2\mathcal{\tilde{O}}\mathcal{\tilde{E}}\mathcal{\tilde{O}}\mathcal{\tilde{E}}-2\mathcal{\tilde{E}}\mathcal{\tilde{O}}\mathcal{\tilde{E}}\mathcal{\tilde{O}}+2\mathcal{\tilde{O}}\mathcal{\tilde{E}}^{2}\mathcal{\tilde{O}}
\end{eqnarray*}
So our expression for $\tilde{\mathsf{h}}^{\left(6\right)}$  agrees with the results listed in the table of  DeVries and Jonker \cite{Vries1968}.
 It also agrees with the result obtained by Silenko with his correction scheme \cite{Silenko2016,Silenko2015} for the original step- by-step transformation of Foldy and Wouthuysen \cite{FoldyW50}.
 Note, that the flow equation based approach leading to Eqs.(126), (144) in \cite{Schopohl2022MS} is by construction energy-separating and therefore doesn't need any correction scheme.

\bibliography{Supplemental_MS_June_30_revtex_Literatur}